\documentclass{article}
\usepackage{multirow}
\usepackage{amsmath,amssymb}
\usepackage[retainorgcmds]{IEEEtrantools}
\usepackage{feynmp}
\usepackage{slashed}
\usepackage[dvips]{graphicx}

\author{Jamil Hetzel}
\title{Probing the Supersymmetry Breaking Mechanism Using Renormalisation Group Invariants}
\date{}

\numberwithin{equation}{section}

\begin{document}
\setlength{\unitlength}{1mm}

\renewcommand{\d}{\mathrm{d}}
\newcommand{\R}{\mathbb{R}}
\newcommand{\ord}[1]{\mathcal{O}\left(#1\right)}
\newcommand{\lag}{\mathcal{L}}
\newcommand{\scatmat}{\mathcal{M}}
\newcommand{\msq}{m_{\widetilde{Q}}^2}
\newcommand{\msqa}{m_{\widetilde{Q}_1}^2}
\newcommand{\msqc}{m_{\widetilde{Q}_3}^2}
\newcommand{\msu}{m_{\widetilde{\bar{u}}}^2}
\newcommand{\msua}{m_{\widetilde{\bar{u}}_1}^2}
\newcommand{\msuc}{m_{\widetilde{\bar{u}}_3}^2}
\newcommand{\msd}{m_{\widetilde{\bar{d}}}^2}
\newcommand{\msda}{m_{\widetilde{\bar{d}}_1}^2}
\newcommand{\msdc}{m_{\widetilde{\bar{d}}_3}^2}
\newcommand{\msl}{m_{\widetilde{L}}^2}
\newcommand{\msla}{m_{\widetilde{L}_1}^2}
\newcommand{\mslc}{m_{\widetilde{L}_3}^2}
\newcommand{\mse}{m_{\widetilde{\bar{e}}}^2}
\newcommand{\msea}{m_{\widetilde{\bar{e}}_1}^2}
\newcommand{\msec}{m_{\widetilde{\bar{e}}_3}^2}
\newcommand{\mhu}{m_{H_u}^2}
\newcommand{\mhd}{m_{H_d}^2}
\newcommand{\reusediagram}[2]{\parbox{#1mm}{\fmfreuse{#2}}}

\maketitle
\begin{abstract}
If supersymmetric particles are discovered, an important problem will be to determine how supersymmetry has been broken.
At collider energies, supersymmetry breaking can be parameterised by soft supersymmetry breaking parameters.
Several mechanisms for supersymmetry breaking have been proposed, which are all characterised by patterns in the high scale values of these parameters.
Therefore, looking for such patterns will give us important clues about the way supersymmetry has been broken in Nature.
In this master thesis, we study an approach to find these patterns using Renormalisation Group invariants.
We construct sum rules that test properties of the spectrum at the scale of supersymmetry breaking, provided that the Minimal Supersymmetric Standard Model is a good description of Nature at collider energies and all soft mass parameters and gauge couplings have been determined.
Subsequently, we examine to what extent these sum rules can distinguish between different supersymmetry breaking scenarios.
It is found that our sum rules provide unambiguous checks in almost all cases.
\end{abstract}
\vskip7cm
\begin{center}
\textit{Supervisor: Wim Beenakker}
\end{center}
\thispagestyle{empty}

\pagenumbering{roman}
\section*{Acknowledgements}
A master's thesis like this is a bit like a movie: although the leading actor gets his face put in neon letters everywhere and receives a lot of the credit, he or she would have been nothing without all those people appearing in the closing credits.
Seeing only my name on the title page of this theses, I cannot help but wonder how many people have actually contributed to finishing this thesis.
Therefore, I would like to seize the opportunity to thank my `crew' for their contributions, be it directly or indirectly.

First of all, every movie has a director: his guidance is necessary to correct the main actor when necessary and ensure that the movie conveys to the audience what it was meant to.
I would like to thank my supervisor Wim Beenakker for the essential guidance, especially during my writing of this thesis.
Your countless corrections and suggestions were necessary to arrive at a satisfactory end result.
I would also like to thank Irene Niessen and Jari Laamanen for all useful discussions we have had on the way.

Then there are the background actors, without whom a movie would look rather dull.
They provide the atmosphere that the main actor needs to flourish.
I would like to thank everybody at our high energy physics department for providing a great environment to work in.

Finally, there is the technical crew, who are never seen by the majority of the public but whose constant support is vital to success.
I would like to thank my friends and family for all support, especially my parents.
Without your unconditional support, I would never have gotten this far.
Last but far from the least, I would like to thank my girlfriend Maaike.
Your presence and support were indispensable to me, be it during periods of stress or relaxation, progress or stagnation.

\tableofcontents

\pagenumbering{arabic}
\setcounter{page}{1}
\section{Introduction}
The Standard Model (SM) is one of the most succesful theories in physics.
It provides an excellent description of the experimental data collected so far at particle colliders, such as the Large Hadron Collider (LHC).
All of its particles have been detected except one: the famous Higgs boson.
It is hoped that we will soon find it at the LHC, where protons will be smashed into each other with total center-of-mass energies up to 14 TeV.

Despite the success of the Standard Model, it has a number of theoretical shortcomings that make us believe that it does not provide a complete description of Nature.
At some energy scale new physics should arise, and around this scale the Standard Model is expected to break down.
Apart from searching for the Higgs boson, the LHC will also look for new physics.

There are various proposals for what this new physics should be.
The most popular extension of the Standard Model involves supersymmetry, a symmetry relating bosons and fermions.
If we extend the Standard Model to include supersymmetry, it turns out that we get a lot of new particles: for each bosonic (fermionic) SM particle, there must be a fermionic (bosonic) counterpart.
Furthermore, these fermionic and bosonic `twin brothers' should have equal masses.
However, in that case we would already have observed these new particles.
Hence, if supersymmetry is realised in Nature, it must be broken somehow.

There is no consensus on how supersymmetry should be broken: the literature offers a plethora of models that propose some mechanism to break supersymmetry.
Unfortunately, in these models the supersymmetry breakdown usually takes place at energies far beyond experimental access.
This leaves us without a way to test these models directly.
For phenomenological purposes, however, we can just parameterise our ignorance of how supersymmetry has been broken: all possible supersymmetry breaking interactions can be described in terms of a limited set of `soft' supersymmetry breaking parameters.
The patterns between these parameters at high energies provide an important clue about the way supersymmetry has been broken.
Hence, by looking for such patterns, we might be able to identify the mechanism that is responsible for supersymmetry breaking.

In this thesis, we examine a strategy for testing what patterns might occur at high energies, using measurements at low energies only.
In section \ref{s:standardmodel}, we summarise the concepts behind the Standard Model that we will need in the rest of this thesis.
We also need to know the relation between parameters at low energies and those at high energies; the necessary background is given in section \ref{s:renormalisationgroup} (Renormalisation Group) and \ref{s:EFTs} (effective field theories).
Section \ref{s:susy} contains a brief introduction to supersymmetry; we will give reasons for studying supersymmetry, discuss ingredients of supersymmetric theories and describe an important example of such a theory, the Minimal Supersymmetric Standard Model (MSSM).
In section \ref{s:susybreaking} we discuss how supersymmetry can be broken and list the most common breaking mechanisms.
Sections \ref{s:probinghighscales} and \ref{s:results} are the core of this work: in section \ref{s:probinghighscales} we discuss several methods to determine the supersymmetry breaking mechanism and propose a new strategy to do this.
In section \ref{s:results} we list our results and discuss them.
We conclude in section \ref{s:conclusion}.

\subsection{Notation and conventions}
In this thesis, we use natural units, in which $\hbar=c=1$.
In these units, mass is the only dimension that is left:
\begin{equation}
[\text{mass}] = [\text{energy}] = [\text{length}]^{-1} = [\text{time}]^{-1}
\end{equation}
Lorentz indices are represented by letters from the middle of the Greek alphabet $\mu, \nu, \ldots$ and take the values $0, 1, 2, 3$.
Letters from the beginning of the Roman alphabet $a, b, \ldots$ label either the gauge group ($a=1$ for $U(1)_Y$, $a=2$ for $SU(2)_L$ and $a=3$ for $SU(3)_C$) or the Lie algebra generators of the gauge group ($a=1, 2, 3$ for $SU(2)_L$ and $a=1, 2, \ldots, 8$ for $SU(3)_C$).
It should always be clear from the context which one of them is meant.
Repeated indices are always summed over unless explicitly stated otherwise.

We use the following signature of the Minkowskian spacetime metric:
\begin{equation}
g_{\mu\nu} = \text{diag}(+1, -1, -1, -1)
\end{equation}
The Dirac gamma matrices are $4\times4$ matrices, given by:
\begin{equation}
\gamma^0 = \left(\begin{array}{cc} 0 & I \\ I & 0 \end{array}\right) \qquad,\qquad \gamma^i = \left(\begin{array}{cc} 0 & \sigma_i \\ -\sigma_i & 0 \end{array}\right) \qquad (i=1,2,3)
\end{equation}
Here $I$ is the $2\times2$ identity matrix and $\sigma_i$ are the $2\times2$ Pauli matrices:
\begin{equation}
\sigma_1 = \left(\begin{array}{cc} 0 & 1 \\ 1 & 0 \end{array}\right) \quad,\quad \sigma_2 = \left(\begin{array}{cc} 0 & -i \\ i & 0 \end{array}\right) \quad,\quad \sigma_3 = \left(\begin{array}{cc} 1 & 0 \\ 0 & -1 \end{array}\right)
\end{equation}
We also define an additional gamma matrix:
\begin{equation}
\gamma^5 \equiv i\gamma^0\gamma^1\gamma^2\gamma^3 = \left(\begin{array}{cc} -I & 0 \\ 0 & I \end{array}\right)
\end{equation}
The abbreviation `h.c.' in equations stands for Hermitian conjugate.
The $\log$ function denotes the natural logarithm, rather than the logarithm to base 10.

\section{Standard Model}\label{s:standardmodel}
In this section, we summarise the ideas behind the Standard Model (SM).
We start with the concepts of quantum gauge theories that are used in this thesis; for an extensive treatment, see e.g.\ \cite{PeskinSchroeder}.
Then we discuss the particle content of the SM and its group theoretical aspects.
Subsequently we discuss the Higgs mechanism.
We conclude with some comments on naturalness, which will be important to our discussion of supersymmetry in section \ref{s:susy}.

\subsection{Quantum field theory}
The fundamental quantity of a quantum field theory is the action $S$.
This is the spacetime integral of a Lagrangian\footnote{Technically speaking, the term \emph{Lagrangian density} should be used here. The Lagrangian $L$ is the spatial integral of the Lagrangian density: $L=\int \d^3x\lag$. But since $\lag$ is the only quantity that is used in practice, it is alway called the Lagrangian.} $\lag$, a function of one or more fields $\phi(x)$ and their derivatives $\partial_\mu\phi(x)$ which depend on the spacetime coordinate $x$:
\begin{equation}
S = \int\d^4x\lag(\phi,\partial_\mu\phi)
\end{equation}
In the Standard Model, three types of fields appear in the Lagrangian:
\begin{itemize}
\item Spin-0 scalar fields $\phi$. A complex scalar field has two real degrees of freedom. A scalar field has mass dimension 1.
\item Spin-1/2 Dirac fields, which are represented by spinors $\psi$ with four degrees of freedom. They can be split into a left-handed part $\psi_L$ and a right-handed part $\psi_R$ using the chirality operator (see section~\ref{s:SMfieldcontent}). For massless fermions, these chirality eigenstates are also eigenstates of helicity, which is the projection of the spin onto the momentum. The two-component objects $\psi_L$, $\psi_R$ are called \emph{Weyl spinors}. A spinor has mass dimension 3/2.
\item Spin-1 vector boson fields $A_\mu$. Massive vector boson fields have three degrees of freedom, whereas massless ones have only two. A vector boson field has mass dimension 1.
\end{itemize}

The principle of least action states that a system always evolves from one configuration to another along the path in configuration space for which $S$ is an extremum.
This condition leads to the Euler-Lagrange equation of motion for a field:
\begin{equation}
\partial_\mu\left(\frac{\partial\lag}{\partial(\partial_\mu\phi)}\right) - \frac{\partial\lag}{\partial\phi} = 0 \label{eq:eulerlagrange}
\end{equation}
However, experiments measure cross sections, not equations of motion.
The transition amplitude for a given process can be calculated as a perturbation expansion.
In turn, this expansion can be represented as a set of graphs (called \emph{Feynman diagrams}) with \emph{propagators}, i.e.\ lines that denote propagating fields, and \emph{vertices}, which denote interactions.
Each propagator and each vertex corresponds to a mathematical expression, which is given by the \emph{Feynman rules}.

As an example, consider the toy model $\phi^3$ theory, which describes a real scalar field $\phi$ with mass $m_0$ by the Lagrangian:
\begin{equation}
\lag = \frac12\left(\partial_\mu\phi\right)\left(\partial^\mu\phi\right) - \frac12m_0^2\phi^2 - \frac{1}{3!}g_0\phi^3 \label{eq:phithirdlagrangian}
\end{equation}
where $g_0$ is a coupling constant.
The momentum space Feynman rules for $\phi^3$ theory are (see also appendix \ref{a:feynmanrules}):
\begin{fmffile}{phithirdfeynmanrules}
\begin{align}
\parbox{20mm}{
  \begin{fmfgraph*}(20,10)
    \fmfleft{in}
    \fmfright{out}
    \fmf{scalar,label=$p$}{in,out}
  \end{fmfgraph*}}
&= \frac{i}{p^2-m_0^2+i\epsilon} \label{eq:phithirdpropagator} \\
\parbox{20mm}{
  \begin{fmfgraph*}(20,10)
    \fmfleft{in}
    \fmfright{out1,out2}
    \fmf{dashes}{out2,v,out1}
    \fmf{dashes,tension=1.5}{v,in}
  \end{fmfgraph*}}
&= -ig_0 \label{eq:phithirdvertex}
\end{align}
\end{fmffile}
Here $p$ is the four-momentum of the propagating field; it is conserved at every vertex.
The propagator \eqref{eq:phithirdpropagator} contains an infinitesimal $\epsilon>0$ that becomes relevant when we perform integrations over four-momenta: if we have to integrate over poles, it tells us how to move the integration contour around it in the complex plane.

Once the Feynman rules are given, we can calculate any cross section by drawing all possible arrangements of propagators and vertices with the given initial and final states.
Then we can retrieve a mathematical expression using the Feynman rules.
There are additional rules for Feynman diagrams as a whole: we should multiply by combinatorial factors for interchangeable lines, integrate over internal momenta in loops, add minus signs for fermion loops and insert the appropriate polarisation vectors for the initial and final states.
For a complete overview, see e.g.\ \cite{PeskinSchroeder}.

\subsection{Gauge theories}\label{s:gaugetheories}
The Standard Model is a relativistic gauge theory.
The possible terms in the Lagrangian of such a theory obey the principle of gauge invariance.
To see how this works, let us review the derivation of Quantum Electrodynamics (QED).

We start with the free Dirac Lagrangian, which describes a free, massive spin-$1/2$ particle:
\begin{equation}
\lag_\text{Dirac} = i\bar\psi(x)\gamma^\mu\partial_\mu\psi(x) - m\bar{\psi}(x)\psi(x) \label{eq:diraclagrangian}
\end{equation}
Here $\psi(x)$ is a Dirac spinor, $\gamma^\mu$ are the Dirac matrices and $\bar{\psi}\equiv \psi^\dagger\gamma^0$.
Note that $\lag_\text{Dirac}$ is invariant under a global (i.e.\ the same in each spacetime point $x$) $U(1)$ gauge transformation:
\begin{equation}
\psi(x)\rightarrow e^{i\alpha}\psi(x) \qquad (\alpha\in\R)
\end{equation}
QED can be derived by imposing invariance of \eqref{eq:diraclagrangian} under a \emph{local} (i.e.\ possibly different in each spacetime point $x$) $U(1)$ gauge transformation:
\begin{equation}
\psi(x) \rightarrow e^{i\alpha(x)}\psi(x) \qquad (\alpha(x)\in\R) \label{eq:localgaugetransformation}
\end{equation}
The mass term $-m\bar{\psi}\psi$ is clearly invariant under \eqref{eq:localgaugetransformation}, but in the kinetic term $i\bar{\psi}(x)\gamma^\mu\partial_\mu\psi(x)$ the partial derivative spoils local gauge invariance.
In order to make the theory locally gauge invariant, we need to replace the ordinary derivative $\partial_\mu$ by a \emph{gauge covariant derivative} $D_\mu$ such that $D_\mu\psi(x)\rightarrow e^{i\alpha(x)}D_\mu\psi(x)$ under \eqref{eq:localgaugetransformation}.
This is achieved by the replacement:
\begin{equation}
\partial_\mu \rightarrow D_\mu = \partial_\mu + ieA_\mu(x) \label{eq:covariantderivative}
\end{equation}
Here we have introduced a covariant vector field $A_\mu(x)$; a constant $e$ has been extracted for future convenience.
We define the vector field to transform under \eqref{eq:localgaugetransformation} as:
\begin{equation}
A_\mu(x) \rightarrow A_\mu(x) - \frac1e \partial_\mu\alpha(x) \label{eq:Atransformation}
\end{equation}
In the end, $A_\mu(x)$ will turn out to be the electromagnetic field.
Then the transformation \eqref{eq:Atransformation} corresponds to the gauge freedom of the electromagnetic field.
Note that since $i\partial_\mu$ is the momentum operator, \eqref{eq:covariantderivative} corresponds to a trick we know in electrodynamics as \emph{minimal substitution}, which is used to describe the effects of the electromagnetic field on charged particles.
Now the modified Lagrangian:
\begin{equation}
\lag = i\bar\psi(x)\gamma^\mu\big(\partial_\mu + ieA_\mu(x)\big)\psi(x) - m\bar{\psi}(x)\psi(x)
\end{equation}
is invariant under the gauge transformation \eqref{eq:localgaugetransformation}+\eqref{eq:Atransformation}.
As a last step, we should include a kinetic energy term for the new field $A_\mu(x)$.
The QED Lagrangian is defined as the most general one, including only operators up to mass dimension four,\footnote{This is to ensure renormalisability, see section \ref{s:renormalisationprocedure}.} that is consistent with local gauge invariance and invariance under parity and time reversal:
\begin{equation}
\lag_\text{QED} = i\bar{\psi}\gamma^\mu \big(\partial_\mu + ieA_\mu\big)\psi - m\bar{\psi}\psi - \frac14F_{\mu\nu}F^{\mu\nu} \label{eq:QEDLagrangian}
\end{equation}
Here we have suppressed the spacetime dependence of all fields.
We have also introduced the electromagnetic field tensor:
\begin{equation}
F_{\mu\nu} \equiv \partial_\mu A_\nu - \partial_\nu A_\mu
\end{equation}
The Euler-Lagrange equation \eqref{eq:eulerlagrange} for the field $A_\mu$ yields the Maxwell equations, with conserved four-current $j^\mu=e\bar{\psi}\gamma^\mu\psi$.
So indeed $A_\mu$ is the electromagnetic field and $e$ can be interpreted as the electric charge.
Hence, we can derive QED simply by postulating a local $U(1)$ symmetry of the Lagrangian!

\subsubsection*{Non-Abelian gauge theories}
If one simple principle allows us to reproduce the Maxwell equations, it is natural to ask whether we could reconstruct other theories by using other gauge groups.
It turns out that we can: the Lagrangian \eqref{eq:QEDLagrangian} can be generalised to other continuous gauge groups that describe unitary transformations, such as $SU(2)$.
Let $T^a$ denote the generators of the gauge group.
Their commutation relations can be written as:
\begin{equation}
[T^a,T^b] = if^{abc}T^c
\end{equation}
where the $f^{abc}$ are numbers called \emph{structure constants} of the gauge group.
We can always choose a basis for the generators such that the $f^{abc}$ are totally antisymmetric.

If the generators do not commute, the procedure for constructing the gauge theory becomes more complicated.
This feature is important enough to give the resulting theories a special name: if the $T^a$ do not commute, the gauge group is called \emph{non-Abelian}, so we refer to the corresponding theories as \emph{non-Abelian gauge theories}.
QED is an \emph{Abelian gauge theory}.

We will only list the results for non-Abelian gauge theories.
We demand invariance of the Lagrangian under the gauge transformation:
\begin{equation}
\psi(x)\rightarrow e^{i\alpha^a(x)T^a}\psi(x) \qquad (\alpha^a(x)\in\R)
\end{equation}
where $\psi(x)$ is a multiplet in the fundamental representation of the gauge group, i.e.\ it is a column vector with spinors as components.
This requires us to introduce the covariant derivative:
\begin{equation}
D_\mu = \partial_\mu - igA_\mu^aT^a \label{eq:covariantderivative2}
\end{equation}
For each generator $T^a$, we have introduced a vector field $A_\mu^a$; a constant $g$ has been extracted for future convenience.
The vector fields transform as:
\begin{equation}
A_\mu^a \rightarrow A_\mu^a + \frac1g\partial_\mu\alpha^a + f^{abc}A_\mu^b\alpha^c
\end{equation}
The field tensor is given by:
\begin{equation}
F_{\mu\nu}^a = \partial_\mu A_\nu^a - \partial_\nu A_\mu^a + gf^{abc}A_\mu^bA_\nu^c
\end{equation}
The most general Lagrangian that is locally gauge invariant, has operators up to mass dimension four and conserves $P$ and $T$, is given by:
\begin{equation}
\lag_\text{non-Abelian} = i\bar{\psi}\gamma^\mu(\partial_\mu-igA_\mu^aT^a)\psi - m\bar{\psi}\psi - \frac14F_{\mu\nu}^aF^{\mu\nu,a}
\end{equation}
The vector field $A_\mu^a$ couples to the conserved current $-g\bar{\psi}\gamma^\mu T^a\psi$.
Hence, there are as many conserved currents as there are generators of the group.
By looking for conserved currents, experiment will guide us towards the correct gauge group.

Note that the above expressions reduce to the Abelian case when we put $f^{abc}=0$.
Here we can also see what is so special about non-Abelian theories: because of the additional term $gf^{abc}A_\mu^bA_\nu^c$ in the field tensor, the Lagrangian contains self-interactions among the gauge fields.
These are not present in the Abelian case.

\subsection{Standard Model field content}\label{s:SMfieldcontent}
The Standard Model (SM) is derived by requiring gauge invariance under the gauge group $SU(3)_C\times SU(2)_L\times U(1)_Y$.
The first factor is the gauge group of quantum chromodynamics (QCD); the $C$ stands for colour charge.
The second factor is the gauge group of the weak interactions; the $L$ refers to the fact that it only affects the left-handed fermions.
Left-handed and right-handed fermions are eigenstates of the chirality operator $\gamma^5$ with eigenvalues $-1$ and $+1$ respectively.
A spinor $\psi$ can be split into a left-handed part $\psi_L\equiv \frac12(1-\gamma^5)\psi$ and a right-handed part $\psi_R\equiv \frac12(1+\gamma^5)\psi$.
The $U(1)$ factor in the SM gauge group is that of the weak hypercharge $Y$, which is related to the electromagnetic charge $Q$ and the third component of weak $(SU(2))$ isospin $I_3$ by the formula $Q=I_3+Y$.
The gauge couplings for the SM gauge groups are $g_3$ for QCD, $g_2$ for the weak interactions and $g'$ for hypercharge.

The fields that appear in the SM Lagrangian transform according to a finite-dimensional unitary representation of the SM gauge group.
These representations are characterised by their dimension.
The Standard Model fermions that are charged under $SU(N)$ (with $N=2,3$) are in the \emph{fundamental representation} of $SU(N)$; this means we can view them as $N$-dimensional complex vectors with spinors as components.
We denote the fundamental representations of $SU(N)$ as $\mathbf{N}$.
If a particle is in the fundamental representation $\mathbf{2}$ of $SU(2)_L$, we say it is an $SU(2)_L$ \emph{doublet}.
If it is in the fundamental representation $\mathbf{3}$ of $SU(3)_C$, we say it is an $SU(3)_C$ \emph{triplet}.
Fermions that are uncharged under $SU(N)$ are in the \emph{trivial representation} $\mathbf{1}$; we call them \emph{singlets} (e.g.\ $\psi_R$ is a singlet under $SU(2)_L$).
For $SU(3)$, the fundamental representation is complex,\footnote{In general, the fundamental representation of $SU(N)$ is complex for $N>2$.} so there is a second, inequivalent representation $\mathbf{\bar{3}}$, called the \emph{conjugate representation} of $\mathbf{3}$.
Quarks are in the representation $\mathbf{3}$ whereas antiquarks are in $\mathbf{\bar{3}}$.

The Standard Model gauge fields are in the \emph{adjoint representation} of the corresponding gauge group; this means we can view them as matrices.
The dimension of the adjoint representation of $SU(N)$ is $N^2-1$.
Thus gluons are in the \emph{octet} representation $\mathbf{8}$ of $SU(3)_C$ and the weak bosons are in the triplet representation $\mathbf{3}$ of $SU(2)_L$.

The representations of $U(1)_Y$ are denoted by the eigenvalues of the hypercharge generator $Y$.
All SM particles and their corresponding representations of $SU(3)_C\times SU(2)_L\times U(1)_Y$ are listed in table \ref{t:SMparticlecontent}.

\begin{table}[t]
\begin{center}
{\renewcommand{\arraystretch}{1.2}
\begin{tabular}{|l|c|c|}
\hline
Name	&	Symbol	&	Gauge group representation	\\\hline
\multicolumn{3}{|l|}{Quarks (3 generations)}	\\\hline
Left-handed doublet	&	$Q=(u_L \; d_L)$	&	$(\mathbf{3}, \mathbf{2}, \frac16)$	\\
Right-handed up-type singlet	&	$u_R$	&	$(\mathbf{3}, \mathbf{1}, \frac23)$	\\
Right-handed down-type singlet	&	$d_R$	&	$(\mathbf{3}, \mathbf{1}, -\frac13)$	\\\hline
\multicolumn{3}{|l|}{Leptons (3 generations)}	\\\hline
Left-handed leptons	&	$L=(\nu \; e_L)$	&	$(\mathbf{1}, \mathbf{2}, -\frac12)$	\\
Right-handed charged leptons	&	$e_R$	&	$(\mathbf{1}, \mathbf{1}, -1)$	\\\hline
\multicolumn{3}{|l|}{Gauge bosons}	\\\hline
Gluons	&	$g$	&	$(\mathbf{8}, \mathbf{1}, 0)$	\\
$W$ bosons	&	$W^1, W^2, W^3$	&	$(\mathbf{1}, \mathbf{3}, 0)$	\\
$B$ boson	&	$B$	&	$(\mathbf{1}, \mathbf{1}, 0)$	\\\hline
\multicolumn{3}{|l|}{Higgs sector (see section \ref{s:higgsmechanism})}	\\\hline
Higgs boson	&	$H$	&	$(\mathbf{1}, \mathbf{2}, \frac12)$	\\\hline
\end{tabular}}
\end{center}
\caption{Standard Model particles and their representations of the SM gauge group $SU(3)_C\times SU(2)_L\times U(1)_Y$. Of the three generations of quarks and leptons, only the first is shown. Note that the $W$ and $B$ bosons are interaction eigenstates of the gauge bosons. After electroweak symmetry breaking (see section \ref{s:higgsmechanism}), mixtures of them become the mass eigenstates $W^\pm, Z^0$ (weak interaction) and the photon $\gamma$ (electromagnetism). There are no right-handed neutrinos, but these may be included in extensions of the Standard Model.}\label{t:SMparticlecontent}
\end{table}

\subsection{Higgs mechanism}\label{s:higgsmechanism}
It is pretty nice that we can derive the Standard Model from group theory arguments.
However, there is a problem: we know that the weak interaction bosons, quarks and charged leptons are massive, but if we add mass terms to the Lagrangian, we get in trouble.
A mass term for fermions is of the form $m\bar{\psi}\psi = m(\bar{\psi}_L\psi_R + \bar{\psi}_R\psi_L)$.
This couples an $SU(2)$ doublet $\psi_L$ to a singlet $\psi_R$, so the mass term is an $SU(2)$ doublet.
Hence, the Lagrangian would not be gauge-invariant.
Since gauge-invariance is the cornerstone of our theory, this is unacceptable.
A mass term for the gauge bosons would be of the form $\frac12m^2A_\mu A^\mu$ and is not gauge-invariant either.
Clearly, we need a new mechanism to generate mass.

Let us start with the fermion mass term.
We want the Lagrangian to be a gauge singlet (i.e.\ it must be gauge-invariant), but a term proportional to $\bar{\psi}\psi$ is an $SU(2)$ doublet.
This suggests that we should combine $\bar{\psi}\psi$ with another $SU(2)$ doublet into an $SU(2)$ singlet.
Since we are restricting ourselves to operators up to mass dimension four (see section \ref{s:renormalisationprocedure}), we can only couple $\bar{\psi}\psi$ to a scalar field.
Therefore we introduce a new $SU(2)$ doublet complex scalar field $\phi$ and couple it to the left-handed doublet $\psi_L$ and the right-handed singlet $\psi_R$ with a Yukawa interaction:
\begin{align}
\lag_\text{Yukawa} &= -y\bar{\psi}_L\phi\psi_R + \text{h.c.} = -y(\bar{\psi}_u, \bar{\psi}_d)_L \left(\begin{array}{c} \phi_u \\ \phi_d \end{array}\right) \psi_R + \text{h.c.} \nonumber\\
&= -y\phi_u\bar{\psi}_{u,L}\psi_R - y\phi_d\bar{\psi}_{d,L}\psi_R + \text{h.c.} \label{eq:yukawaLagrangian}
\end{align}
Here $u$ and $d$ denote the up and down components of the $SU(2)$ doublets; they are not indices to be summed over.
The coupling constant $y$ is different for each fermion.

Now suppose that $\phi$ has a nonzero vacuum expectation value (VEV), i.e.\ $\phi=0$ is not a minimum of the scalar potential $V(\phi)$.
We should separate $\phi$ into its classical minimum $\langle\phi\rangle$ and the quantum fluctuation (i.e.\ particle) part $\eta$:
\begin{equation}
\phi = \langle\phi\rangle + \eta
\end{equation}
We can simplify the form of $\langle\phi\rangle$ by fixing the gauge: using $SU(2)$ gauge transformations we can bring it to the form
\begin{equation}
\langle\phi\rangle = \frac{1}{\sqrt{2}}\left(\begin{array}{c} 0 \\ v \end{array}\right)
\end{equation}
Then we can make $v$ real and positive with a $U(1)_Y$ phase transformation.
If we insert this into \eqref{eq:yukawaLagrangian}, we get:
\begin{equation}
\lag_\text{Yukawa} = -\frac{yv}{\sqrt{2}}\bar{\psi}_{d,L}\psi_R - y\eta_u\bar{\psi}_{u,L}\psi_R - y\eta_d\bar{\psi}_{d,L}\psi_R + \text{h.c.}
\end{equation}
This looks familiar: the first term is an effective mass term, and the other terms are Yukawa interactions with the new particles described by $\eta_{u,d}$.
Note that this only yields a mass term for the down-type fermions.
Mass terms for the up-type fermions are generated by the charge conjugate field $\phi^c=i\sigma_2\phi^*$.

Mass terms for the $W$ bosons are also generated: if we expand the gauge-invariant scalar kinetic term $(D_\mu\phi)^\dagger(D^\mu\phi)$ around the VEV, we end up with mass terms for the gauge bosons.
Thus we have found a gauge-invariant way to generate masses!

Now we see another reason why we had to introduce a scalar field: the fermions and gauge bosons cannot acquire a VEV without breaking either Lorentz or gauge invariance.
The only remaining task is to make sure that $\phi$ acquires a VEV.
The most general Lagrangian for a complex scalar, compatible with gauge invariance and renormalisability, is:
\begin{equation}
\lag_\phi = (D_\mu\phi)^\dagger(D^\mu\phi) - V(\phi) = (D_\mu\phi)^\dagger(D^\mu\phi) - \mu_H^2\phi^\dagger\phi - \frac{\lambda}{4}(\phi^\dagger\phi)^2
\end{equation}
where $\mu_H^2, \lambda\in\R$.
The covariant derivative follows from $SU(2)$ and $U(1)_Y$ gauge invariance; the analog of \eqref{eq:covariantderivative2} is:
\begin{equation}
D_\mu = \partial_\mu - \frac{i}{2}g'B_\mu - ig_2T^aA_\mu^a
\end{equation}
Here $B_\mu$ is the $U(1)_Y$ gauge boson field, $A_\mu^a$ are the $SU(2)_L$ gauge boson fields and $T^a$ are the $SU(2)_L$ generators.
The scalar potential:
\begin{equation}
V(\phi) = \mu_H^2\phi^\dagger\phi + \frac{\lambda}{4}(\phi^\dagger\phi)^2
\end{equation}
must be bounded from below, so $\lambda>0$.
For $\mu_H^2$, there are two different scenarios:
\begin{itemize}
\item If $\mu_H^2\geq0$, the minimum of the scalar potential is at $\phi=0$ and no mass generation occurs.
\item If $\mu_H^2<0$, the scalar potential has a minimum at $|\phi|=\sqrt{-2\mu_H^2/\lambda}$. In that case, $\phi$ acquires a VEV and we have electroweak symmetry breaking.
\end{itemize}

The latter case is an example of spontaneous symmetry breaking: the Lagrangian is still invariant under local gauge transformations, but the vacuum state is not.
According to Goldstone's theorem, each broken symmetry generator gives rise to a massless Goldstone boson.
Since $SU(2)$ has three generators (namely the three Pauli matrices, up to a normalisation factor), this would give three Goldstone bosons.
However, they have been turned into the longitudinal polarisation states of the gauge bosons:\footnote{In the literature, this fact is often expressed by saying that the gauge bosons have `eaten' three degrees of freedom of the Higgs doublet.} recall that a massive vector boson has three degrees of freedom, whereas a massless one has only two.
This is known as the \emph{Higgs mechanism}.

The fourth degree of freedom of the Higgs doublet has become the scalar field $\eta$.
This corresponds to the famous Higgs boson that we are hoping to find at the LHC.

\subsection{Naturalness}\label{s:naturalness}
As we will see in section \ref{s:renormalisationgroup}, the parameters in the Lagrangian are not the ones we measure in experiment.
This is because the original parameters receive corrections from Feynman diagrams with particle loops (quantum corrections), which involve contributions from all energy scales.
It is remarkable that all masses of the Standard Model particles are so small compared to (say) the Planck scale $M_\text{pl}=2.4\cdot10^{18}$ GeV, where gravity starts to become important.
The particle masses are free parameters of the Standard Model (their values are not predicted by the theory, but have to be determined by experiment), and we could say that they all just happen to be small.
But why would all masses be concentrated in a low energy regime?
We would expect them to be spread over the whole range between 0 GeV and the Planck scale, unless we tweaked all initial values of the Lagrangian parameters such that the observable masses (including the quantum corrections) are small.
But that is considered as unnatural.

However, there is a reason why most particle masses can remain naturally small if the Lagrangian parameters are small.
Consider for example the fermion masses.
In the limit $m\rightarrow0$, the Dirac Lagrangian \eqref{eq:diraclagrangian} is invariant under the transformation:
\begin{equation}
\psi \rightarrow e^{i\alpha\gamma^5}\psi \quad,\quad \bar\psi \rightarrow \bar\psi e^{i\alpha\gamma^5}
\end{equation}
with $\alpha\in\R$.
This is called \emph{chiral symmetry}, which is broken by the mass term in \eqref{eq:diraclagrangian}.
However, the fermion mass $m$ is protected against large corrections by the original symmetry: in the limit $m\rightarrow0$ chiral symmetry must be restored.
Hence, by dimensional analysis the loop corrections must be proportional to $m$ (and not to other relevant mass scales, such as the Planck scale).
We say that the fermion mass is \emph{protected by a symmetry}, in this case chiral symmetry.

The gauge bosons are also protected, namely by gauge symmetry.
Even if the symmetry is broken spontaneously, as happens in the Higgs mechanism, the masses stay naturally small.
The smallness of a mass is called \emph{natural} if setting the Lagrangian parameter to zero enhances the symmetry of the Lagrangian.
All Standard Model fermions and gauge bosons are protected against large corrections this way.
Only the Higgs boson is not protected by a symmetry; this will be important in our motivation for supersymmetry in section \ref{s:susy}.

\section{Renormalisation Group}\label{s:renormalisationgroup}
As we will see in section~\ref{s:susybreaking}, a study of supersymmetry breaking requires us to examine physics at very high energy scales.
These energies are currently far beyond experimental access.
Hence we need to use techniques from the Renormalisation Group (RG).
In this section, the concepts and most important lessons of the RG are reviewed.
A more extensive treatment of the RG can be found in any textbook on quantum field theory, e.g.\ \cite{PeskinSchroeder}.
For a conceptually clear treatment of RG, disentangled from any cumbersome quantum-field-theoretical calculations, see \cite{delamotte}.

\subsection{Example: mass renormalisation in $\phi^3$ theory}\label{s:renormalisationprocedure}
If we perform calculations beyond tree level, we often encounter divergent integrals.
Here we will consider an example of such an integral and recall how we can deal with it.
Thereafter we will examine the origin of these divergences and justify the steps involved in the `taming' of infinities.

Let us consider $\phi^3$ theory; recall that it is described by the Lagrangian \eqref{eq:phithirdlagrangian}:
\begin{equation}
\lag = \frac12\left(\partial_\mu\phi\right)\left(\partial^\mu\phi\right) - \frac12m_0^2\phi^2 - \frac{1}{3!}g_0\phi^3
\end{equation}
The Feynman rules for $\phi^3$ theory are given by \eqref{eq:phithirdpropagator}-\eqref{eq:phithirdvertex} (see also appendix \ref{a:feynmanrules}).

The propagator \eqref{eq:phithirdpropagator} is not the one we measure, because we can only measure the sum of all possible diagrams.
At one-loop order in perturbation theory, we have contributions from the one-loop self-energy, which is defined by:
\begin{fmffile}{phithirdloop}
\begin{align}
\Sigma_1(p,m_0) &\equiv i\times
\parbox{35mm}{
  \begin{fmfgraph*}(35,15)\fmfkeep{oneloopselfenergy}
    \fmfleft{in}
    \fmfright{out}
    \fmf{scalar,left,label=$k$}{v1,v2}
    \fmf{scalar,left,label=$k-p$}{v2,v1}
    \fmf{scalar,label=$p$}{in,v1}
    \fmf{scalar,label=$p$}{v2,out}
  \end{fmfgraph*}} \nonumber\\
&= \frac{ig_0^2}{2}\int\frac{\d^4k}{(2\pi)^4}\frac{1}{\left((k-p)^2-m_0^2+i\epsilon\right)\left(k^2-m_0^2+i\epsilon\right)} \label{eq:selfenergyintegral}
\end{align}
\end{fmffile}
This integral requires a toolbox of tricks (see appendix \ref{a:oneloopselfenergy}) to solve it, but we can already see a divergence appearing: after transforming the integral over Minkowskian four-momentum $k$ into an integral over Euclidean four-momentum $k_E$, there are four powers of momentum in the denominator and three in the numerator (from the volume element $\d^4k_E$ in spherical coordinates).
Hence the integral will behave as:
\begin{equation}
\int^\infty \d k_E\frac{k_E^3}{k_E^4} = \ln{k_E}|^\infty \rightarrow \infty
\end{equation}
We can regulate (i.e.\ `tame') this divergence by introducing a momentum cutoff $\Lambda$ in the radial component of $k_E$.
Then we will decide how to deal with the divergence.
In the end, our results for physical observables should be well-defined in the limit $\Lambda\rightarrow\infty$.

We continue our calculations with the regulated one-loop self energy:
\begin{equation}
\Sigma_{1,\Lambda}(p,m_0) = \frac{ig_0^2}{2}\int_{\Lambda}\frac{\d^4k}{(2\pi)^4}\frac{1}{\left((k-p)^2-m_0^2+i\epsilon\right)\left(k^2-m_0^2+i\epsilon\right)}
\end{equation}
where $\int_\Lambda$ means that we integrate the radial component of $k_E$ from $0$ to $\Lambda$.
Then the effective propagator has the following form at the one-particle-irreducible one-loop level:
\begin{fmffile}{phithirdseries}
\begin{align}
\parbox{20mm}{
  \begin{fmfgraph*}(20,15)\fmfkeep{effectivepropagator}
    \fmfleft{in}
    \fmfright{out}
    \fmfblob{15}{v1}
    \fmf{dashes}{in,v1}
    \fmf{dashes}{v1,out}
  \end{fmfgraph*}}
&=
\hskip2mm\parbox{12mm}{
  \begin{fmfgraph*}(12,15)
    \fmfleft{in}
    \fmfright{out}
    \fmf{dashes}{in,out}
  \end{fmfgraph*}}
+
\hskip2mm\parbox{15mm}{
  \begin{fmfgraph*}(15,15)
    \fmfleft{in}
    \fmfright{out}
    \fmf{dashes,left,tension=.5}{v1,v2}
    \fmf{dashes,left,tension=.5}{v2,v1}
    \fmf{dashes}{in,v1}
    \fmf{dashes}{v2,out}
  \end{fmfgraph*}}
+
\hskip2mm\parbox{25mm}{
  \begin{fmfgraph*}(25,15)
    \fmfleft{in}
    \fmfright{out}
    \fmf{dashes,left,tension=.5}{v1,v2}
    \fmf{dashes,left,tension=.5}{v2,v1}
    \fmf{dashes,left,tension=.5}{v3,v4}
    \fmf{dashes,left,tension=.5}{v4,v3}
    \fmf{dashes}{in,v1}
    \fmf{dashes}{v2,v3}
    \fmf{dashes}{v4,out}
  \end{fmfgraph*}}
+\ldots \nonumber \\
&= \frac{i}{p^2-m_0^2+i\epsilon} + \frac{i}{p^2-m_0^2+i\epsilon}\big(-i\Sigma_{1,\Lambda}(p,m_0)\big)\frac{i}{p^2-m_0^2+i\epsilon} \nonumber \\
&\quad+ \frac{i}{p^2-m_0^2+i\epsilon}\left(\big(-i\Sigma_{1,\Lambda}(p,m_0)\big)\frac{i}{p^2-m_0^2+i\epsilon}\right)^2 + \ldots \nonumber \\
&= \frac{i}{p^2-m_0^2+i\epsilon} \cdot \frac{1}{1-\Sigma_{1,\Lambda}(p,m_0)/(p^2-m_0^2+i\epsilon)} \nonumber \\
&= \frac{i}{p^2-m_0^2-\Sigma_{1,\Lambda}(p,m_0)+i\epsilon}
\end{align}
\end{fmffile}
Thus the effective propagator has the same form as the tree-level propagator \eqref{eq:phithirdpropagator}, but with an effective mass $m_\text{eff}^2(p) = m_0^2 + \Sigma_{1,\Lambda}(p,m_0)$ that is energy-dependent.
This already suggests that we can redefine the theory in terms of physical (i.e.\ measurable) parameters.
If we do so, the theory should be free of divergences, because it would relate only measurable quantities.

Thus we deal with the divergence as follows: we eliminate the parameter $m_0$ in favour of the physical  mass $m_\text{phys}(\mu)$ at a reference scale $\mu$.
To this end we write:
\begin{equation}
m_0^2 = m_\text{phys}^2(\mu) + \delta m^2
\end{equation}
The emerging term in the Lagrangian proportional to $\delta m^2$ is called a \emph{coun\-ter\-term}, since its purpose is to cancel the divergences in our theory.

At this point, let us isolate the diverging part of the one-loop self energy as follows:
\begin{IEEEeqnarray}{rCl}
\Sigma_{1,\Lambda}(p,m_0) &=& \frac{ig_0^2}{2}\int_{\Lambda}\frac{\d^4k}{(2\pi)^4}\left(\frac{1}{\left((k-p)^2-m_0^2+i\epsilon\right)\left(k^2-m_0^2+i\epsilon\right)}\right. \nonumber \\
&&\left.- \frac{1}{(k^2-\mu^2+i\epsilon)^2}\right) + \frac{ig_0^2}{2}\int_{\Lambda}\frac{\d^4k}{(2\pi)^4}\frac{1}{(k^2-\mu^2+i\epsilon)^2} \nonumber \\
&\equiv& \Sigma_{1,\Lambda}^\text{fin}(p,m_0,\mu) + \Sigma_{1,\Lambda}^\text{inf}(\mu)
\end{IEEEeqnarray}
This way, the finite part $\Sigma_{1,\Lambda}^\text{fin}(p,m_0,\mu)$ contains all information of the physical process (i.e.\ $p,m_0$) while the divergent part $\Sigma_{1,\Lambda}^\text{inf}(\mu)$ only contains our arbitrary parameters (i.e.\ $\mu,\Lambda$).
Now we define the renormalised one-loop self energy as:
\begin{align}
\Sigma_1^\text{R}(p,m_\text{phys}(\mu)) &\equiv \lim_{\Lambda\rightarrow\infty}\left(\Sigma_{1,\Lambda}(p,m_\text{phys}(\mu))+\delta m^2\right) \nonumber \\
&= \Sigma_1^\text{fin}(p,m_\text{phys}(\mu),\mu) + \left(\delta m^2 + \Sigma_1^\text{inf}(\mu)\right)
\end{align}
where $\delta m^2$ is chosen such that the above expression is finite; for a renormalisable theory, this is always possible.
Then the effective propagator has the form:
\begin{equation}
\reusediagram{20}{effectivepropagator} = \frac{i}{p^2-m_\text{phys}^2(\mu) - \Sigma_1^\text{R}(p,m_\text{phys}(\mu)) + i\epsilon}
\end{equation}
which is the propagator \eqref{eq:phithirdpropagator} with the \emph{renormalised mass} $m_R^2(p) = m_\text{phys}^2(\mu)+\Sigma_1^\text{R}(p,m_\text{phys}(\mu))$.

In a general renormalisable theory, all divergences can be removed by a redefinition of the masses and couplings.
Suppose a theory is described by a set of masses $m_j$ and couplings $g_i$.
Then similarly to the above calculation we should regularise the divergent diagrams, split the $m_j$ and $g_i$ into physical parameters and counterterms and tune the counterterms to absorb the divergences.

Whether or not a theory is renormalisable can be checked easily by power counting.
A Lagrangian has dimension 4, so an operator with dimension $d$ has a coupling constant with dimension $4-d$.
If the Lagrangian contains couplings with negative dimension, the theory is nonrenormalisable: an infinite number of counterterms is needed to deal with all divergences.
Then the theory would have an infinite number of free parameters and would have no predictive power.

\subsection{On the origin of divergences}
The above procedure seems odd: we calculated something that turned out to be infinite, then subtracted infinity from our original mass in an arbitrary way and ended up with something finite.
Moreover, we have added a divergent term to our Lagrangian and the mass we started with has suddenly been replaced by an energy-dependent mass.
Why would a procedure consisting of such ill-defined mathematical tricks be legitimate?
To see what has really happened, let us closely examine the starting point of our calculation.

In general, we start with a Lagrangian containing all possible terms that are compatible with basic assumptions such as relativity, causality, locality and gauge invariance.
It still contains a few parameters such as $m$ and $e$ in the case of QED.
It is tempting to call them `mass' and `charge', as they turn out to be just that in the classical (i.e.\ tree-level) theory.
But up to this point, they are just free parameters.
In order to make the theory predictive, the parameters need to be fixed by a set of measurements: we should calculate a set of cross-sections at a given order in perturbation theory, measure their values and then fit the parameters so that they reproduce the experimental data.
After this procedure, the theory is completely determined and becomes predictive.

Note that since the so-called `bare parameters' $m,e$ (which are only useful in intermediate calculations) will be replaced by physical (i.e.\ measured) quantities in the end anyway, we might as well parametrise the theory in terms of the latter.
The \emph{renormalisability hypothesis} is that this reparametrisation of the theory is enough to turn the perturbation expansion into a well-defined expansion.
The divergence problem then has nothing to do with the perturbation expansion itself: we have just chosen unsuitable parameters to perform it.

Also, the fact that our physical masses and couplings are scale-dependent should not surprise us.
The physical reason for this `running' is the existence of quantum fluctuations, which were not there in the classical theory.
These fluctuations correspond to intermediate particle states: at sufficiently high (i.e.\ relativistic) energies, new particles can be created and annihilated.
As the available energy increases, more particles can be created.
This effectively changes the couplings.

Having traded the bare parameters $m,e$ for renormalised parameters $m_R,e_R$, let us take a closer look at the internal consistency of the renormalisation procedure.
We have introduced the physical parameters at a reference scale $\mu$, but we could equally well have chosen an energy scale $\mu'$ with corresponding parameters $m_R'=m_R(\mu'), e_R'=e_R(\mu')$.
Physical processes should not depend on our choice of $\mu$, hence the masses and couplings should be related in such a way that for any observable\footnote{That is, any cross section we can think of.} $\Gamma(p)$ we have $\Gamma(p)=\Gamma(p,\mu,m_R,e_R)=\Gamma(p,\mu',m_R',e_R')$.
In other words, there should exist an equivalence class of parametrisations of the theory and it should not matter which element of the class we choose.
This observation clarifies where the divergences came from: our initial perturbation expansion consisted of taking $\Lambda\rightarrow\infty$ while keeping $m,e$ \emph{finite}.
From the renormalisation group viewpoint, however, the set $(\Lambda=\infty,m<\infty,e<\infty)$ does not belong to any equivalence class of a sensible theory.

\subsection{Renormalisation Group equations}
The fact that our reference scale $\mu$ is arbitrary, implies that anything observable should be independent of it.
This simple observation can be used to determine the $\mu$-dependence of the running couplings and masses.
Consider for example any observable $\Gamma$, which is a function of some couplings $\{g_i(\mu)\}$ and masses $\{m_j(\mu)\}$.
The above observation implies that
\begin{equation}
0 = \mu\frac{\d}{\d\mu}\Gamma = \left(\mu\frac{\partial}{\partial\mu} + \mu\frac{\d g_i(\mu)}{\d\mu}\frac{\partial}{\partial g_i(\mu)} + \mu\frac{\d m_j^2(\mu)}{\d\mu}\frac{\partial}{\partial m_j^2(\mu)}\right)\Gamma
\end{equation}
By explicitly calculating a set of observables,\footnote{In practice, one chooses the simplest observables that contain the necessary information.} we get coupled differential equations that govern the $\mu$-dependence of the masses and couplings.
These are called \emph{Renormalisation Group equations}.
In practice, one calculates an observable up to a certain loop order; hence the RG equations will also depend on the loop order at which we calculate them.
They also depend on the particle content of the theory, since this determines which particles can appear in loops.

One may wonder what we have gained from the Renormalisation Group; after all, we have introduced additional loop calculations to determine the running of the masses and couplings.
The essential advantage of using renormalised quantities is the fact that we can afford to calculate observables up to a much lower loop order.
This fact is often expressed by saying that `the renormalisation scheme partially resums the perturbation expansion', which can be visualised as follows.
Let us consider the scattering amplitude for the process $e^-e^-\rightarrow e^-e^-$ in QED.
We can arrange its perturbation expansion according to loop order:

\begin{fmffile}{eescatter}
\begin{eqnarray}
i\scatmat &=&
\parbox{25mm}{
  \begin{fmfgraph*}(40,10)
    \fmfleft{i1,i2}
    \fmfright{o1,o2}
    \fmf{fermion}{i1,v1}
    \fmf{plain}{v1,v2,v3}
    \fmf{fermion}{v3,o1}
    \fmf{fermion}{i2,v4}
    \fmf{plain}{v4,v5,v6}
    \fmf{fermion}{v6,o2}
    \fmffreeze
    \fmfblob{45}{b}
    \fmf{phantom}{v2,b,v5}
  \end{fmfgraph*}}
\nonumber\\[5ex]
&=&
\parbox{25mm}{
  \begin{fmfgraph*}(25,10)
    \fmfleft{i1,i2}
    \fmfright{o1,o2}
    \fmf{fermion}{i1,v1,o1}
    \fmf{fermion}{i2,v2,o2}
    \fmf{photon,tension=0}{v1,v2}
  \end{fmfgraph*}}
+
\parbox{25mm}{
  \begin{fmfgraph*}(25,10)
    \fmfleft{i1,i2}
    \fmfright{o1,o2}
    \fmf{phantom}{i1,v1,o1}
    \fmf{phantom}{i2,v2,o2}
    \fmf{photon,tension=0}{v1,v2}
    \fmffreeze
    \fmf{fermion}{i1,v1}
    \fmf{plain}{v1,v3}
    \fmf{fermion}{v3,o2}
    \fmf{fermion}{i2,v2}
    \fmf{plain}{v2,v3}
    \fmf{fermion}{v3,o1}
  \end{fmfgraph*}}
+
\parbox{25mm}{
  \begin{fmfgraph*}(25,10)
    \fmfleft{i1,i2}
    \fmfright{o1,o2}
    \fmf{fermion}{i2,v1}
    \fmf{plain}{v1,v2,v3}
    \fmf{fermion}{v3,o2}
    \fmf{fermion}{i1,v4,o1}
    \fmf{photon,tension=0}{v2,v4}
    \fmf{photon,left,tension=.8}{v1,v3}
  \end{fmfgraph*}}
+ \nonumber\\[3ex]
&&\parbox{25mm}{
  \begin{fmfgraph*}(25,10)
    \fmfleft{i1,i2}
    \fmfright{o1,o2}
    \fmf{fermion}{i1,v1}
    \fmf{plain}{v1,v2,v3}
    \fmf{fermion}{v3,o1}
    \fmf{fermion}{i2,v4,o2}
    \fmf{photon,tension=0}{v2,v4}
    \fmf{photon,right,tension=.8}{v1,v3}
  \end{fmfgraph*}}
+
\parbox{25mm}{
  \begin{fmfgraph*}(25,10)
    \fmfleft{i1,i2}
    \fmfright{o1,o2}
    \fmf{fermion}{i1,v1,o1}
    \fmf{fermion}{i2,v2,o2}
    \fmf{phantom,tension=0}{v1,v2}
    \fmffreeze
    \fmf{photon}{v1,v3}
    \fmf{photon}{v2,v4}
    \fmf{fermion,left,tension=.3}{v3,v4,v3}
  \end{fmfgraph*}}
+
\parbox{25mm}{
  \begin{fmfgraph*}(25,10)
    \fmfleft{i1,i2}
    \fmfright{o1,o2}
    \fmf{fermion}{i1,v1}
    \fmf{plain}{v1,v2}
    \fmf{fermion}{v2,o1}
    \fmf{fermion}{i2,v3}
    \fmf{plain}{v3,v4}
    \fmf{fermion}{v4,o2}
    \fmf{photon,tension=0}{v1,v3}
    \fmf{photon,tension=0}{v2,v4}
  \end{fmfgraph*}}
+ \nonumber\\[3ex]
&&\parbox{25mm}{
  \begin{fmfgraph*}(25,10)
    \fmfleft{i1,i2}
    \fmfright{o1,o2}
    \fmf{fermion}{i1,v1}
    \fmf{plain}{v1,v2}
    \fmf{fermion}{v2,o1}
    \fmf{fermion}{i2,v3}
    \fmf{plain}{v3,v4}
    \fmf{fermion}{v4,o2}
    \fmf{photon,tension=0}{v1,v4}
    \fmf{photon,tension=0}{v2,v3}
  \end{fmfgraph*}}
+\left(\begin{array}{c}\text{same one-loop diagrams with}\\ \text{final electron states interchanged}\end{array}\right)+ \nonumber\\[5ex]
&&\parbox{25mm}{
  \begin{fmfgraph*}(25,10)
    \fmfleft{i1,i2}
    \fmfright{o1,o2}
    \fmf{fermion}{i1,v1,o1}
    \fmf{fermion}{i2,v2}
    \fmf{plain}{v2,v3,v4,v5,v6}
    \fmf{fermion}{v6,o2}
    \fmf{photon,tension=0}{v1,v4}
    \fmf{photon,left,tension=.7}{v3,v5}
    \fmf{photon,left,tension=.3}{v2,v6}
  \end{fmfgraph*}}
+
\parbox{25mm}{
  \begin{fmfgraph*}(25,10)
    \fmfleft{i1,i2}
    \fmfright{o1,o2}
    \fmf{fermion}{i1,v1}
    \fmf{plain}{v1,v2,v3}
    \fmf{fermion}{v3,o1}
    \fmf{fermion}{i2,v4}
    \fmf{plain}{v4,v5,v6}
    \fmf{fermion}{v6,o2}
    \fmf{photon,tension=0}{v2,v5}
    \fmf{photon,left,tension=.8}{v4,v6}
    \fmf{photon,right,tension=.8}{v1,v3}
  \end{fmfgraph*}}
+\ldots \label{eq:eeperturbationseries}\\\nonumber
\end{eqnarray}

\noindent where the solid lines are electrons, the wavy lines are photons and all external lines are understood to be amputated.
Now let us define an \emph{effective} vertex, photon propagator and electron propagator:

\begin{IEEEeqnarray}{rCCCCl}
\hskip-5mm\parbox{25mm}{
  \begin{fmfgraph*}(25,10)\fmfkeep{effvert1}
    \fmfsurround{v1,v2,v3}
    \fmfblob{10}{v}
    \fmf{photon,tension=2}{v1,v}
    \fmf{fermion}{v2,v,v3}
  \end{fmfgraph*}}
&\hskip2mm=\hskip2mm&
\parbox{25mm}{
  \begin{fmfgraph*}(25,10)\fmfkeep{effvert2}
    \fmfsurround{v1,v2,v3}
    \fmf{photon,tension=2}{v1,v}
    \fmf{fermion}{v2,v,v3}
  \end{fmfgraph*}}
&\hskip2mm+\hskip2mm&
\parbox{25mm}{
  \begin{fmfgraph*}(25,10)\fmfkeep{effvert3}
    \fmfsurround{v1,v2,v3}
    \fmf{photon,tension=2}{v1,v}
    \fmf{fermion}{v2,u1}
    \fmf{plain,tension=2}{u1,v,u2}
    \fmf{fermion}{u2,v3}
    \fmffreeze
    \fmf{photon,right}{u1,u2}
  \end{fmfgraph*}}
&\hskip2mm+\ldots \IEEEyessubnumber\label{eq:eeeffvertex}\\
\hskip-5mm\parbox{25mm}{
  \begin{fmfgraph*}(25,10)\fmfkeep{effphot1}
    \fmfleft{i}
    \fmfright{o}
    \fmfblob{10}{v}
    \fmf{photon}{i,v,o}
  \end{fmfgraph*}}
&\hskip2mm=\hskip2mm&
\parbox{25mm}{
  \begin{fmfgraph*}(25,10)\fmfkeep{effphot2}
    \fmfleft{i}
    \fmfright{o}
    \fmf{photon}{i,o}
  \end{fmfgraph*}}
&\hskip2mm+\hskip2mm&
\parbox{25mm}{
  \begin{fmfgraph*}(25,10)\fmfkeep{effphot3}
    \fmfleft{i}
    \fmfright{o}
    \fmf{photon}{i,v1}
    \fmf{photon}{v2,o}
    \fmf{fermion,left,tension=.7}{v1,v2,v1}
  \end{fmfgraph*}}
&\hskip2mm+\ldots \IEEEyessubnumber\label{eq:eeeffphoton}\\
\hskip-5mm\parbox{25mm}{
  \begin{fmfgraph*}(25,10)\fmfkeep{effelec1}
    \fmfleft{i}
    \fmfright{o}
    \fmfblob{10}{v}
    \fmf{fermion}{i,v,o}
  \end{fmfgraph*}}
&\hskip2mm=\hskip2mm&
\parbox{25mm}{
  \begin{fmfgraph*}(25,10)\fmfkeep{effelec2}
    \fmfleft{i}
    \fmfright{o}
    \fmf{fermion}{i,o}
  \end{fmfgraph*}}
&\hskip2mm+\hskip2mm&
\parbox{25mm}{
  \begin{fmfgraph*}(25,10)\fmfkeep{effelec3}
    \fmfleft{i}
    \fmfright{o}
    \fmf{fermion}{i,v1}
    \fmf{plain}{v1,v2}
    \fmf{fermion}{v2,o}
    \fmf{photon,left,tension=.5}{v1,v2}
  \end{fmfgraph*}}
&\hskip2mm+\ldots \IEEEyessubnumber\label{eq:eeeffelectron}
\end{IEEEeqnarray}

\noindent We are going to express the scattering amplitude in terms of these effective quantities. To this end we rearrange and then partially resum the perturbation series \eqref{eq:eeperturbationseries} as follows:

\begin{align}
i\scatmat =&
\parbox{25mm}{
  \begin{fmfgraph*}(25,10)
    \fmfleft{i1,i2}
    \fmfright{o1,o2}
    \fmf{fermion}{i1,v1,o1}
    \fmf{fermion}{i2,v2,o2}
    \fmf{photon,tension=0}{v1,v2}
  \end{fmfgraph*}}
+
\parbox{25mm}{
  \begin{fmfgraph*}(25,10)
    \fmfleft{i1,i2}
    \fmfright{o1,o2}
    \fmf{fermion}{i2,v1}
    \fmf{plain}{v1,v2,v3}
    \fmf{fermion}{v3,o2}
    \fmf{fermion}{i1,v4,o1}
    \fmf{photon,tension=0}{v2,v4}
    \fmf{photon,left,tension=.8}{v1,v3}
  \end{fmfgraph*}}
+
\parbox{25mm}{
  \begin{fmfgraph*}(25,10)
    \fmfleft{i1,i2}
    \fmfright{o1,o2}
    \fmf{fermion}{i1,v1,o1}
    \fmf{fermion}{i2,v2}
    \fmf{plain}{v2,v3,v4,v5,v6}
    \fmf{fermion}{v6,o2}
    \fmf{photon,tension=0}{v1,v4}
    \fmf{photon,left,tension=.7}{v3,v5}
    \fmf{photon,left,tension=.3}{v2,v6}
  \end{fmfgraph*}}
\nonumber\\[2ex]
&+\text{(other loop corrections around the upper vertex)} \nonumber\\
&+\text{(other diagrams, rearranged analogously)} \nonumber\\[2ex]
= &
\parbox{25mm}{
  \begin{fmfgraph*}(25,10)
    \fmfleft{i1,i2}
    \fmfright{o1,o2}
    \fmfblob{10}{v2}
    \fmf{fermion}{i1,v1,o1}
    \fmf{fermion}{i2,v2,o2}
    \fmf{photon,tension=0}{v1,v2}
  \end{fmfgraph*}}
+\left(\begin{array}{c}\text{other diagrams with one upper vertex}\\\text{replaced by an effective vertex}\end{array} \right)
\end{align}
\end{fmffile}

\noindent We can repeat this procedure of rearranging and resumming to write \eqref{eq:eeperturbationseries} entirely in terms of the effective quantities \eqref{eq:eeeffvertex}-\eqref{eq:eeeffelectron}:

\begin{fmffile}{eerenorm}
\begin{eqnarray}
i\scatmat &=& 
\parbox{25mm}{
  \begin{fmfgraph*}(25,15)
    \fmfleft{i1,i2}
    \fmfright{o1,o2}
    \fmfblob{10}{b1,b2}
    \fmf{fermion}{i1,b1,o1}
    \fmf{fermion}{i2,b2,o2}
    \fmf{phantom,tension=0}{b1,b2}
    \fmffreeze
    \fmfblob{10}{b3}
    \fmf{photon}{b1,b3,b2}
  \end{fmfgraph*}}
+\hskip-2mm
\parbox{32mm}{
  \begin{fmfgraph*}(40,15)
    \fmfleft{i1,i2}
    \fmfright{o1,o2}
    \fmfblob{10}{bv1,bv2,bv3,bv4,bf1,bf2}
    \fmf{fermion}{i1,bv1,bf1,bv2,o1}
    \fmf{fermion}{i2,bv3,bf2,bv4,o2}
    \fmffreeze
    \fmfblob{10}{bp1,bp2}
    \fmf{photon}{bv1,bp1,bv3}
    \fmf{photon}{bv2,bp2,bv4}
  \end{fmfgraph*}}
\hskip6mm+\hskip-3mm
\parbox{32mm}{
  \begin{fmfgraph*}(40,15)
    \fmfleft{i1,i2}
    \fmfright{o1,o2}
    \fmfblob{10}{bv1,bv2,bv3,bv4,bf1,bf2}
    \fmf{fermion}{i1,bv1,bf1,bv2,o1}
    \fmf{fermion}{i2,bv3,bf2,bv4,o2}
    \fmffreeze
    \fmfblob{10}{bp1,bp2}
    \fmf{photon}{bv1,bp1}
    \fmf{photon,tension=1.25}{bp1,bv4}
    \fmf{photon,tension=1.25}{bv2,bp2}
    \fmf{photon}{bp2,bv3}
  \end{fmfgraph*}}
\nonumber\\[3ex]
&&+\quad(\text{higher-order diagrams})\nonumber\\[3ex]
&&+\parbox{25mm}{
  \begin{fmfgraph*}(25,15)
    \fmfleft{i1,i2}
    \fmfright{o1,o2}
    \fmfblob{10}{b1,b2}
    \fmf{phantom}{i1,b1,o1}
    \fmf{phantom}{i2,b2,o2}
    \fmf{phantom,tension=0}{b1,b2}
    \fmffreeze
    \fmfblob{10}{b3}
    \fmf{fermion}{i1,b1}
    \fmf{plain}{b1,v}
    \fmf{fermion}{v,o2}
    \fmf{fermion}{i2,b2}
    \fmf{plain}{b2,v}
    \fmf{fermion}{v,o1}
    \fmf{photon}{b1,b3}
    \fmf{photon}{b2,b3}
  \end{fmfgraph*}}
+ \left(\begin{array}{c}\text{higher order diagrams with}\\ \text{final state electrons interchanged}\end{array}\right)
\end{eqnarray}
\end{fmffile}

\noindent Hence we can rewrite the scattering amplitude \eqref{eq:eeperturbationseries} in terms of effective vertices and propagators only.
Now suppose that we have calculated the effective vertex \eqref{eq:eeeffvertex} and propagators \eqref{eq:eeeffphoton}, \eqref{eq:eeeffelectron} up to some loop order: these correspond to coupling constants and masses that depend on the energy scale $\mu$.
If we calculate the scattering amplitude to one-loop order using these running couplings, we have effectively calculated the scattering amplitude in terms of the bare couplings up to some higher loop order: many loop contributions have been absorbed into the effective vertex and propagators.
Hence we can afford to calculate the scattering amplitude up to much lower loop order than before, provided we use the running couplings at the right scale $\mu$ instead of the bare couplings.

\section{Effective Field Theories}\label{s:EFTs}
When we study physics at different energy scales, it is convenient to switch between different descriptions of Nature.
This requires us to use effective field theories (EFTs).
In this section, we review the basic idea behind EFTs and their connection to the Wilsonian Renormalisation Group.
This section has been based on \cite{eft}.

\subsection{Renormalisability and effective field theories}
As mentioned in section \ref{s:renormalisationprocedure}, not all quantum field theories are renormalisable.
If the Lagrangian contains operators with dimensions that are too large, the theory has no predictive power since it has an infinite number of free parameters.
Hence, the prevalent view on renormalisation used to be that a sensible theory describing Nature could only contain renormalisable interactions.
However, this view has changed since Wilson's work on the Renormalisation Group \cite{wilsonkogut,wilson}.
But how can we accept a non-renormalisable theory?
To answer this question, let us review the calculation in section \ref{s:renormalisationprocedure}.

A momentum cutoff $\Lambda$ was introduced as a mathematical trick to tame the divergent integral \eqref{eq:selfenergyintegral}.
The philosophy behind it was to get rid of it by taking $\Lambda\rightarrow\infty$ in the end; this also used to be the prevalent interpretation of renormalisation \cite{delamotte}.\footnote{Renormalisation also has its applications in statistical mechanics, for example in studying phase transitions. In that case, the atomic length scale provides a natural cutoff. However, for quantum field theory no physically meaningful energy cutoff was believed to exist.}
However, by taking this limit we tacitly assume that the theory is valid up to arbitrarily large momenta.
In the modern view on the Renormalisation Group, $\Lambda$ is considered as a scale that does have a physical meaning: it is the scale at which new physics becomes relevant.
For example, the Standard Model does not include gravity.
Since gravitational effects become comparable in size to those of the other interactions at the Planck scale \mbox{$M_\text{Pl}=2.4\cdot 10^{18}$ GeV,} new physics is expected to arise at this scale.
Hence it acts as a natural momentum cutoff:\footnote{Of course it is possible that more than only gravity is missing in the Standard Model. In that case, a lower scale would act as the cutoff for the SM.} above $M_\text{Pl}$, the Standard Model cannot be valid anymore and should be replaced by a more fundamental theory.
This is the motivation for using effective field theories.

\subsection{Basic idea of EFTs}
Suppose we had a `theory of everything': a theory describing all fundamental dynamics of the basic constituents of Nature and unifying different kinds of interactions.
Although this theory could in principle describe all physical phenomena, it would be unnecessarily cumbersome to describe Nature at all physical scales.
For example, the laws of chemistry arise from the electromagnetic interaction, yet it would be unwise to start a quantitative analysis from Quantum Electrodynamics.
Instead, when we wish to analyse a particular physical system, we need to isolate its most relevant ingredients from the rest in order to obtain a simple description without having to understand every detail.

In order to do so, we have to make an appropriate choice of variables that captures the most important physics of the system.
Physics problems usually involve widely separated energy scales, which allow us to study low-energy dynamics without needing to know the details of the high-energy interactions.
The basic idea is to identify the parameters that are large (small) compared to the relevant energy scale and put them to infinity (zero).
Eventually we can improve this approximation by taking into account the corrections of the high energy physics in the form of small perturbations.

\emph{Effective field theories} are the theoretical tool to describe low-energy physics, where `low' means low with respect to some energy scale $\Lambda$.
An EFT only takes into account states with mass $m\ll\Lambda$; heavier excitations with $m\gg\Lambda$ are integrated out from the action.
The information about the heavy states is then contained in the couplings of the low-energy theory: we get non-renormalisable interactions among the light states, organised as an expansion in powers of energy/$\Lambda$.

\subsection{Basic ingredients of EFTs}\label{s:EFTbasic}
An effective field theory is characterised by some effective Lagrangian:
\begin{equation}
\lag = \sum_ic_iO_i
\end{equation}
where the $O_i$ are operators constructed from the light fields and the $c_i$ are couplings containing information on any heavy degrees of freedom.
Since the Lagrangian has dimension 4, dimensional analysis yields:
\begin{equation}
[O_i] \equiv d_i \qquad\Rightarrow\qquad c_i \sim \frac{1}{\Lambda^{d_i-4}}
\end{equation}
where $\Lambda$ is some characteristic heavy scale of the system.
At low energies, the behaviour of these operators is determined by their dimension:
\begin{itemize}
\item Operators with $d_i<4$ are called \emph{relevant}, since they give rise to effects that become large at low energies.
\item Operators with $d_i>4$ are called \emph{irrelevant}: at energy scales $E$ their effects are suppressed by powers of $E/\Lambda$, making them small at low energies.
These are non-renormalisable operators that contain information about the underlying dynamics at higher scales.
\item Operators with $d_i=4$ are called \emph{marginal}, because they are equally important at all energy scales; quantum effects could modify their scaling behaviour to either relevancy or irrelevancy.
\end{itemize}
If we consider physics at an energy scale $E$ that is much lower than the scale of any heavier states (mass $M$), then the effects from irrelevant operators are always suppressed by powers of $E/M$.
This explains why we can include non-renormalisable interactions in an EFT without spoiling its predictive power: at low energies, their effects can either be neglected or incorporated as perturbations in powers of $E/M$.
At high energies, it is more appropriate to use a different EFT.
Thus at sufficiently low energies, an EFT automatically contains only renormalisable (i.e.\ relevant and marginal) operators.

\subsection{Matching}\label{s:matching}
Suppose we have two effective field theories: one that includes a heavy particle and one where its effects are included in the form of higher-dimensional operators, suppressed by inverse powers of the heavy particle mass $M$.
How can we relate the two theories?
The crucial observation is that physics around the mass scale $M$ should not depend on our choice of theory: both EFTs should yield the same physical predictions.
This is called the \textbf{matching condition}: \emph{at the threshold $\mu=M$, the two EFTs should give rise to the same S-matrix elements for light particle scattering.}
In practice, one should match all one-light-particle-irreducible diagrams\footnote{Those diagrams that cannot be disconnected by cutting a single light particle line.} with only external light particles.
This leads to relations between the parameters of the high-energy EFT (the one we use above threshold) and those of the low-energy EFT (the one we use below threshold).
In other words, the matching conditions encode the effects of the heavy field into the low-energy EFT parameters.

\begin{fmffile}{fermi}
\begin{figure}[t]
\begin{minipage}{.44\textwidth}
  \begin{center}
  \begin{fmfgraph*}(30,20)
    \fmfstraight
    \fmfleft{z1,iup,idown,down,z2,z3,z4}
      \fmflabel{$d$}{down}
      \fmflabel{$u$}{iup}
      \fmflabel{$d$}{idown}
    \fmfright{oup,odown,up,x2,x3,x4,x5}
    \fmfcurved
    \fmfright{y1,y2,electron,neutrino}
      \fmflabel{$u$}{up}
      \fmflabel{$u$}{oup}
      \fmflabel{$d$}{odown}
      \fmflabel{$e^-$}{electron}
      \fmflabel{$\bar{\nu}_e$}{neutrino}
    \fmf{fermion}{down,v1,up}\fmffreeze
    \fmf{fermion}{neutrino,v2,electron}
    \fmf{boson,label=$W^-$,label.side=left}{v1,v2}
    \fmf{fermion}{iup,a,oup}
    \fmf{fermion}{idown,b,odown}
    \fmfforce{(0,.3h)}{left}
    \fmfv{label=$n\Bigg\{$,label.angle=180,label.dist=13}{left}
    \fmfforce{(w,.11h)}{right}
    \fmfv{label=$\Bigg\}p$,label.angle=0,label.dist=13}{right}
  \end{fmfgraph*}
  \vskip5mm (a)
  \end{center}
\end{minipage}
\begin{minipage}{.11\textwidth}
$$\underrightarrow{q^2\ll M_W^2}$$
\end{minipage}
\begin{minipage}{.44\textwidth}
  \begin{center}
  \begin{fmfgraph*}(30,20)
    \fmfleft{neutron}
      \fmflabel{$n$}{neutron}
    \fmfright{proton,electron,neutrino}
      \fmflabel{$p$}{proton}
      \fmflabel{$e^-$}{electron}
      \fmflabel{$\bar{\nu}_e$}{neutrino}
    \fmf{fermion,tension=5}{neutron,v,electron}
    \fmf{fermion}{neutrino,v,proton}
  \end{fmfgraph*}
  \vskip5mm (b)
  \end{center}
\end{minipage}
\caption{Tree level diagram for beta decay of a neutron (a) in the Standard Model and (b) in the Fermi theory of weak interactions. In the Standard Model, this decay proceeds through the exchange of a $W$ boson. If the momentum transfer $q$ of the $W$ boson is much smaller than its mass $M_W$, the $W$ boson propagator reduces to a contact interaction. In that case, the Fermi 4-vertex provides an effective description of this decay.}\label{f:betadecay}
\end{figure}
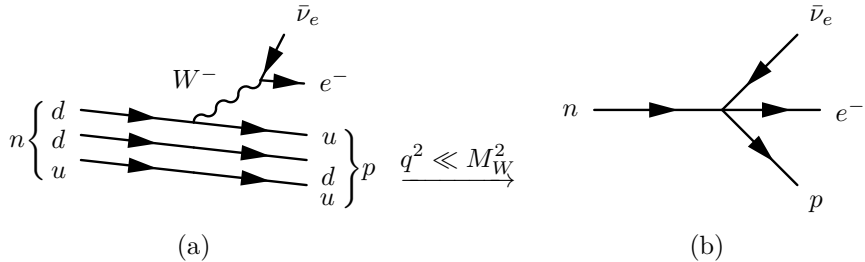

As an example, consider the beta decay of a neutron ($n\rightarrow p\, e\, \bar{\nu}_e$).
In the Standard Model, this decay is mediated by a $W$ boson (figure \ref{f:betadecay}a), which has a propagator:
\begin{equation}
\parbox{25mm}{\begin{fmfgraph*}(15,10)
  \fmfleft{in}
    \fmflabel{$\mu$}{in}
  \fmfright{out}
    \fmflabel{$\nu$}{out}
  \fmf{boson,label=$q$}{in,out}
\end{fmfgraph*}}
= \frac{-ig_{\mu\nu}}{q^2-M_W^2}
\end{equation}
\end{fmffile}
If the momentum transfer $q$ by the $W$ boson is much smaller than its mass, this propagator reduces to a contact interaction:
\begin{equation}
\frac{-ig_{\mu\nu}}{q^2-M_W^2} \qquad\underrightarrow{q^2\ll M_W^2}\qquad \frac{ig_{\mu\nu}}{M_W^2} + \ord{\frac{q^2}{M_W^4}}
\end{equation}
At energies well below the $W$ mass, there is not enough energy available to produce a physical $W$ boson.
Hence we might as well switch to an effective field theory that does not include the $W$ field.
Integrating out the $W$ field from the action, we are left with the Fermi 4-vertex (figure \ref{f:betadecay}b) and higher-order interactions.
Matching the two EFTs at $\mu=M_W$ yields the well-known formula for the Fermi coupling constant:
\begin{equation}
G_F = \frac{\sqrt{2}}{8}\frac{g_2^2}{M_W^2}
\end{equation}
where $g_2$ is the weak coupling constant.
Note that although the $W$ field is not included in the low-energy EFT, its `fingerprints' (namely its coupling constant $g_2$ and mass $M_W$) are still present in the low-energy coupling $G_F$.
Also note that the irrelevant operator corresponding to the Fermi 4-vertex is indeed suppressed by powers of the $W$ mass, as mentioned in section \ref{s:EFTbasic}.
Hence at low energies, the effects of this interaction are highly suppressed; this is why it has been called the `weak interaction'.

\subsection{Travelling along the EFT chain}
\begin{figure}[t]
\begin{center}
\setlength{\unitlength}{1.4pt}
\begin{picture}(200,100)(-120,-50)
  \put(-100,-50){\vector(0,1){100}}
  \multiput(-97,0)(10,0){8}{\line(1,0){5}}
  \multiput(20,0)(10,0){7}{\line(1,0){5}}
  \put(0,50){\vector(0,-1){45}}
  \put(0,-5){\vector(0,-1){45}}
  \put(-107,45){$\mu$}
  \put(-128,-1){$\mu=M_\Phi$}
  \put(-80,30){high-energy EFT}
  \put(-80,20){$\lag(\phi_i)+\lag(\phi_i,\Phi)$}
  \put(-80,-25){low-energy EFT}
  \put(-80,-35){$\lag(\phi_i)+\delta\lag(\phi_i)$}
  \put(30,25){fields $\phi_i,\Phi$}
  \put(30,-30){fields $\phi_i$}
  \put(-15,-1){matching}
  \put(2,25){RG}
  \put(2,-30){RG}
\end{picture}
\setlength{\unitlength}{1mm}
\end{center}
\caption{Schematic display of the procedure for evolving from high to low energies. A high-energy EFT, which contains light fields $\phi_i$ and a heavy field $\Phi$ with mass $M_\Phi$, is evolved down using the Renormalisation Group equations. At the scale $\mu=M_\Phi$ we should switch to a low-energy EFT that includes only the light fields $\phi_i$. The matching conditions yield the masses and couplings of this EFT at this scale. Then we continue to evolve down the theory, now using the RG equations of the low-energy EFT.}\label{f:evolvingEFT}
\end{figure}
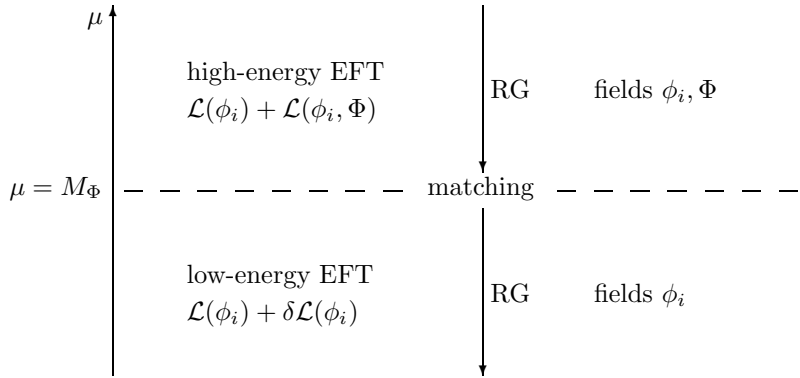
Now that we have reviewed the Renormalisation Group and effective field theories, we know how to evolve a theory from a high energy scale to a low one (see also figure \ref{f:evolvingEFT}).
Suppose that we have an EFT describing physics at some (high) energy scale $\mu$.
The Lagrangian contains a field $\Phi$ with the largest mass $M_\Phi$ and a set of lighter fields $\phi_i$:
\begin{equation}
\lag_\text{high} = \lag(\phi_i) + \lag(\phi_i,\Phi)
\end{equation}
where the first part contains only the light fields and the second part contains the heavy field and its interactions with the light fields.
If we want to describe physics at a lower energy scale, we have to evolve down the running parameters using the Renormalisation Group equations of this EFT.
We can continue to do so until we reach the threshold $\mu=M_\Phi$.
There, we \emph{integrate out} the heavy field $\Phi$ from the action, i.e.\ we switch to a different EFT containing only the light fields $\phi_i$:
\begin{equation}
\lag_\text{low} = \lag(\phi_i) + \delta\lag(\phi_i)
\end{equation}
Note that $\lag(\phi_i)$ contains the same operators in both theories, but with different couplings and masses due to the matching conditions.
The second part $\delta\lag(\phi_i)$ encodes the information on the heavy field $\Phi$.
It contains operators constructed with the light fields $\phi_i$ only, including new higher-order interactions that are suppressed by appropriate powers of $1/M_\Phi$.
By matching the two EFTs at $\mu=M_\Phi$, we fix the values of the running parameters of the low-energy theory.
From there, we can continue to evolve the theory down using the RG equations of the low-energy EFT.

Whenever we reach a new particle threshold, we should integrate out the corresponding field and match the two EFTs.
Thus in the framework of effective field theories, physics is described by a \emph{chain of EFTs}.
Each one has different particle content, and all theories match at the corresponding particle thresholds.
Each theory below a threshold is considered as the low-energy EFT of the theory above the threshold, which is considered as a more fundamental theory.
The ultimate goal of physics then becomes to find the most fundamental theory of Nature, although strictly speaking we can never know whether we have found it (if there is a most fundamental one at all!).

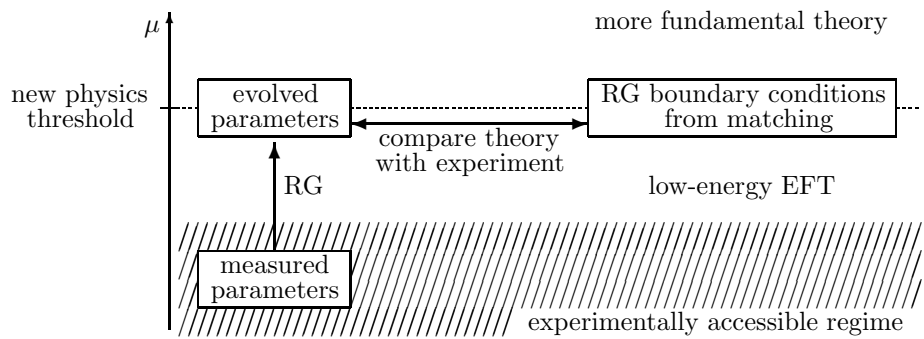
\begin{figure}[t]
\begin{center}
\setlength{\unitlength}{1.2pt}
\begin{picture}(290,100)(-50,-70)
  \put(0,-70){\vector(0,1){100}}
  \put(-8,24){$\mu$}
  \put(-3,0){\line(1,0){6}}
  \multiput(0,0)(2,0){5}{\line(1,0){1}}
  \multiput(57,0)(2,0){38}{\line(1,0){1}}
  \multiput(229,0)(2,0){4}{\line(1,0){1}}
  \put(-50,2){new physics}
  \put(-45,-6){threshold}
  \put(20,2){evolved}
  \put(13,-6){parameters}
  \put(9,-9){\line(1,0){48}}
  \put(9,-9){\line(0,1){18}}
  \put(57,9){\line(-1,0){48}}
  \put(57,9){\line(0,-1){18}}
  \put(66,-12){compare theory}
  \put(65,-20){with experiment}
  \thicklines
  \put(57,-5){\vector(1,0){75}}
  \put(132,-5){\vector(-1,0){75}}
  \thinlines
  \put(136,2){RG boundary conditions}
  \put(156,-6){from matching}
  \put(132,-9){\line(1,0){97}}
  \put(132,-9){\line(0,1){18}}
  \put(229,9){\line(-1,0){97}}
  \put(229,9){\line(0,-1){18}}
  \put(134,25){more fundamental theory}
  \put(151,-27){low-energy EFT}
  \put(113,-70){experimentally accessible regime}
  \multiput(57,-63)(3,0){58}{\line(1,3){9}}
  \put(231,-63){\line(1,3){6}}
  \put(234,-63){\line(1,3){3}}
  \multiput(6,-72)(3,0){34}{\line(1,3){3}}
  \multiput(3,-72)(0,9){3}{\line(1,3){6}}
  \put(3,-45){\line(1,3){3}}
  \multiput(12,-36)(3,0){17}{\line(-1,-3){3}}
  \put(63,-36){\line(-1,-3){6}}
  \put(16,-52){measured}
  \put(13,-60){parameters}
  \put(9,-63){\line(1,0){48}}
  \put(9,-63){\line(0,1){18}}
  \put(57,-45){\line(-1,0){48}}
  \put(57,-45){\line(0,-1){18}}
  \thicklines
  \put(33,-45){\vector(0,1){34}}
  \thinlines
  \put(36,-27){RG}
\end{picture}
\setlength{\unitlength}{1mm}
\end{center}
\caption{Scheme for studying physics at scales beyond experimental access. The running couplings are measured at a scale where the low-energy EFT is applicable. Using the RG equations, they are evolved towards the threshold where new fields presumably enter the theory. Then they can be compared with the matching conditions predicted by the more fundamental theory.}\label{f:EFTtravelling}
\end{figure}
From this point of view, the Standard Model is only a low-energy effective field theory of Nature.
The shortcomings of the SM (see section~\ref{s:SMshortcomings}) hint at the existence of a more fundamental theory.
Even if that more fundamental EFT is appropriate only at energies beyond experimental access, the idea of a chain of EFTs certainly helps us study that more fundamental theory: we could measure the running parameters at a low scale $\mu$ and then evolve them upwards.
At the threshold of the more fundamental theory, the matching conditions act as \emph{boundary conditions} for the Renormalisation Group.
Hence by comparing our evolved masses and couplings with the predicted matching conditions, we can get information on the high-energy theory (figure \ref{f:EFTtravelling}).

\section{Supersymmetry}\label{s:susy}
In this section we will give a short introduction to supersymmetry.
First we will review why it is necessary to extend the Standard Model at all.
Then we will see which shortcomings of the SM are addressed by supersymmetry.
Subsequently, we will summarise what supersymmetry is and how to construct a supersymmetric theory.
Finally, we will consider the Minimal Supersymmetric Standard Model, the most widely used model in the literature.
For an extensive introduction to supersymmetry, see e.g.\ \cite{martin,aitchison}.

\subsection{Going beyond the Standard Model}\label{s:SMshortcomings}
The Standard Model works extremely well in explaining all experimental data collected so far in particle colliders.
So one may wonder why we would want to extend the SM at all: why fix something that is not broken?
The answer is that the SM has some imperfections that strongly hint at physics beyond the Standard Model.
Below we will list some of these imperfections; more can be found in \cite{SMshortcomings}.

\paragraph{Hierarchy problem}
The fact that the Higgs mass should be much smaller than the Planck scale poses a theoretical problem.
Consider the Higgs potential:
\begin{equation}
V = \mu_H^2\phi^\dagger\phi + \frac{\lambda}{4}(\phi^\dagger\phi)^2
\end{equation}
where $\lambda>0$ and $\mu_H^2<0$.
Recall that the minimum of $V$ lies at $|\phi|=\sqrt{-2\mu_H^2/\lambda}\equiv v/\sqrt{2}$.
From measurements of the properties of the weak interactions, we know that $v\approx 246$ GeV.
It sets the scale of all masses in the Standard Model: at tree level, the Higgs mass is given by $M_H=v\sqrt{\lambda/2}$, the $W^\pm$ mass is given by $M_W=g_2v/2\sim 80$ GeV, where $g_2$ is the weak coupling constant, and so on.

However, when we go beyond tree level, the Higgs mass receives loop corrections from every particle it couples to.
Consider for example the contribution from a fermion $f$ with mass $m_f$.
After regularisation, this contribution is (see appendix \ref{a:feynmanrules} for the relevant Feynman rules):
\begin{fmffile}{higgscorrection}
\begin{align}
\parbox{25mm}{
\begin{fmfgraph*}(25,15)
  \fmfleft{in}
  \fmfright{out}
  \fmf{scalar,label=$p$}{in,v1}
  \fmf{scalar,label=$p$}{v2,out}
  \fmf{fermion,right,tension=.5,label=$k$}{v1,v2}
  \fmf{fermion,right,tension=.5,label=$k-p$}{v2,v1}
\end{fmfgraph*}}
&= -|y|^2\int_\Lambda\frac{\d^4k}{(2\pi)^4} \frac{\text{Tr}\left[(\slashed{k}-\slashed{p}+m_f)(\slashed{k}+m_f)\right]}{\left((k-p)^2-m_f^2+i\epsilon\right)(k^2-m_f^2+i\epsilon)} \nonumber\\
&\sim -|y|^2\Lambda^2 + (\text{at most logarithmically diverging terms}) \label{eq:higgscorrection1}
\end{align}
where $y$ is the coupling constant of a Yukawa interaction.
The last step was done by counting powers of $k$ analogous to the argument in section \ref{s:renormalisationprocedure}.
In the Standard Model, there are no diagrams that cancel this quadratic divergence, since the Yukawa couplings are unrelated to the other interactions.
Hence, this divergence needs to be removed by renormalising $\mu_H^2$.
But as we have seen, the value of $\mu_H$ is phenomenologically fixed to at most a few hundred GeV.\footnote{That is, unless $\lambda\gg 1$. But in that case, perturbation theory breaks down anyway. Also, the Standard Model would predict cross sections that cross unitarity bounds. Hence, we assume that $\lambda$ lies in the perturbative regime.}
Its natural value, however, seems to be close to the cutoff $\Lambda$ (this is the problem we referred to in section \ref{s:naturalness}: the Higgs mass is not protected by any symmetry).
In order to arrive at a few hundred GeV after including loop corrections seems to require that the unrenormalised parameter $\mu_H^2$ has a value close to $\Lambda^2$; after including loop corrections some remarkably precise cancellation should produce the phenomenologically correct value.
To see what this implies, suppose the Standard Model is valid up to the Planck scale.
Then $\Lambda\sim 10^{18}$ GeV, and taking the value of $M_H^2$ from $\ord{(10^{18}\text{ GeV})^2}$ to $\ord{(10^2\text{ GeV})^2}$ would require us to choose the value of $\mu_H^2$ precisely up to the $32^\text{nd}$ decimal!
This is called \emph{fine-tuning} and is considered as very unnatural.
Since loop corrections seem to invalidate the hierarchy $M_H\ll M_\text{Pl}$, this problem is called the \emph{hierarchy problem}.

\paragraph{Gravity}
The Standard Model does not describe gravity.
It turns out that General Relativity (GR) is incompatible with quantum field theory: when we try to quantise GR, we get a non-renormalisable theory.
As argued in section \ref{s:EFTbasic}, this means that we cannot combine the SM and gravity and call it a complete theory.
Instead, we would need an underlying theory that leads to non-renormalisable gravity terms at low energies.

\paragraph{Dark matter}
When we look at rotation curves of galaxies, we get different results than we would expect from the visible mass distribution of stars.
This discrepancy can be explained by assuming the existence of dark matter, which interacts only gravitationally.
Many independent observations suggest that about 23\% of the total energy density of the universe is made of this dark matter \cite{darkmatter}.
The Standard Model provides no candidate particle for dark matter.

\subsection{Why supersymmetry?}\label{s:whysusy}
Supersymmetry, a symmetry between bosons and fermions, is a highly developed extension of the Standard Model.
As we will see, it requires the existence of at least twice as many particles as we have in the SM.
At this price, it can solve many problems of the SM: supersymmetry\ldots

\paragraph{\ldots protects the Higgs mass by a symmetry.}
We have seen that the hierarchy problem is caused by quadratically divergent contributions to the Higgs mass.
If supersymmetry is realised in Nature, for each diagram like \eqref{eq:higgscorrection1} there is a corresponding contribution from a scalar $S$ (see appendix \ref{a:feynmanrules} for the relevant Feynman rules):
\vskip3mm
\begin{equation}
\parbox{25mm}{
\begin{fmfgraph*}(25,15)
  \fmfleft{in}
  \fmfright{out}
  \fmf{scalar,label=$p$}{in,v}
  \fmf{scalar,label=$p$}{v,out}
  \fmf{scalar,label=$k$}{v,v}
\end{fmfgraph*}}
= \frac{\lambda_S}{2}\int_\Lambda\frac{\d^4k}{(2\pi)^4} \frac{1}{k^2-m_S^2+i\epsilon} \sim \lambda_S\Lambda^2 \label{eq:higgscorrection2}
\end{equation}
\end{fmffile}
Note that there is a relative sign between \eqref{eq:higgscorrection1} and \eqref{eq:higgscorrection2} because of the fermion loop.
It turns out that if $\lambda_S=|y|^2$ (as is required by supersymmetry), all quadratic divergences will cancel neatly.
Then there will only be logarithmically divergent corrections to $M_H^2$ left, which are of the form:
\begin{equation}
\delta M_H^2 \sim \lambda_S(m_S^2-m_f^2)\log{\Lambda}
\end{equation}
These corrections can be of order $M_H^2$ itself, provided each fermion-boson pair in this cancellation is approximately degenerate in mass.
In that case the fine-tuning problem has evaporated.
Hence supersymmetry could solve the hierarchy problem: it protects the Higgs mass by virtue of a symmetry.

\paragraph{\ldots would connect the Standard Model and gravity.}
As we will see in section \ref{s:whatissusy}, the generators $Q,Q^\dagger$ of supersymmetry are spinors that satisfy:
\begin{equation}
\{Q,Q^\dagger\} \sim P_\mu \label{eq:susyalgebra}
\end{equation}
Here $P_\mu$ is the 4-momentum operator, which is the generator of translations.
If we impose invariance of the theory under \emph{local} supersymmetry transformations, \eqref{eq:susyalgebra} tells us that the theory is automatically invariant under local coordinate shifts.
Then the theory is locally invariant under the full Poincar\'e algebra, which is the underlying principle of General Relativity.
Thus local supersymmetry is inherently related to gravity!
The theory that results from imposing local supersymmetry invariance is therefore called supergravity (SUGRA).

\paragraph{\ldots provides a candidate for cold dark matter.}
In many of the phenomenologically viable supersymmetric theories (those with $R$-parity, see section \ref{s:mssm}), it turns out that the lightest supersymmetric particle (LSP) is stable.
If the LSP is electrically neutral, it interacts only weakly with ordinary matter and becomes an attractive candidate for dark matter.

\begin{figure}[t]
\begin{center}
  \includegraphics[width=.5\textwidth]{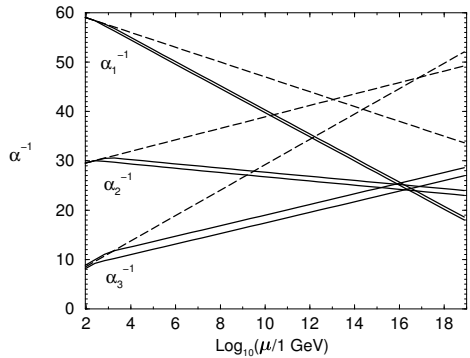}
\end{center}
\caption{Renormalisation Group evolution of the inverse gauge couplings squared $\alpha_a^{-1}(\mu) = 4\pi g_a^{-2}(\mu)$ for $a=1,2,3$ in the Standard Model (dashed lines) and Minimal Supersymmetric Standard Model (solid lines). For the latter, the supersymmetric particle thresholds are varied between 250 GeV and 1 TeV. Figure taken from \cite{martin}.}\label{f:gaugeunification}
\end{figure}

\paragraph{\ldots encourages unification.}
At one-loop order, the inverse gauge couplings squared $\alpha_a^{-1}(\mu)=4\pi g_a^{-2}(\mu)$ for $a=1,2,3$, with $g_1=g'\sqrt{5/3}$, run linearly with $\log{\mu}$ (see figure \ref{f:gaugeunification}).
They do not meet convincingly in the Standard Model, but they do in the Minimal Supersymmetric Standard Model (MSSM).
Thus supersymmetry encourages ideas of a Grand Unified Theory (GUT).
More details will be worked out in section~\ref{s:RGIs}.

\paragraph{\ldots might explain electroweak symmetry breaking.}
As we have seen in section \ref{s:renormalisationgroup}, all parameters in the Lagrangian are scale-dependent.
In the MSSM, if we start with $\mu_H^2$ of order $v^2$ at a scale of $\ord{10^{16}\text{ GeV}}$, RG evolution takes it to a negative value of the correct order of magnitude at a scale of $\ord{100\text{ GeV}}$.
However, in the SM the Higgs potential is introduced rather \emph{ad hoc}.
Hence supersymmetry might explain electroweak symmetry breaking.

\paragraph{\ldots completes the list of possible spacetime symmetries.}
Supersymmetry is a spacetime symmetry.\footnote{This can be seen for instance from \eqref{eq:susyalgebra}: two subsequent supersymmetry transformations are equivalent to a spacetime translation. Hence a supersymmetry transformation is like the `square root' of a spacetime translation, in the same sense that the Dirac operator is the square root of the Klein-Gordon operator.}
The form for such symmetries is restricted by the Haag-Lopuszanski-Sohnius extension of the Coleman-Mandula theorem \cite{CMHLS}.
According to the latter, there are no possible conserved operators with non-trivial Lorentz transformation character other than $M_{\mu\nu}$, the generators of Lorentz transformations, and $P_\mu$.
However, it turns out that their argument does not exclude charges that transform under Lorentz transformations as \emph{spinors}.
Supersymmetry completes the list of possible spacetime symmetries, and one may wonder why Nature would have chosen to incorporate all possible symmetries except one.

\paragraph{\ldots stabilises other theories.}
The phenomenologically most attractive versions of string theory incorporate supersymmetry.
Traces of supersymmetry might still be present in a low-energy effective field theory emerging from string theory.

\subsection{What is supersymmetry?}\label{s:whatissusy}
Supersymmetry is a symmetry relating bosons and fermions.
Its generator $Q$ transforms a bosonic state into a fermionic state, and vice versa:
\begin{equation}
Q|\text{boson}\rangle = |\text{fermion}\rangle \qquad Q|\text{fermion}\rangle = |\text{boson}\rangle
\end{equation}
Since bosons have integer spin and fermions have half-integer spin, the above equation tells us that $Q$ is a spinor.
Hence we get an algebra of anticommutation relations:
\begin{IEEEeqnarray}{rCl}
\{Q_a,Q_b^\dagger\} &=& (\sigma^\mu)_{ab}P_\mu \IEEEyessubnumber\label{eq:susyalgebra1}\\
\{Q_a,Q_b\} &=& \{Q_a^\dagger,Q_b^\dagger\} = 0 \IEEEyessubnumber\label{eq:susyalgebra2}\\
\lbrack P^\mu,Q_a\rbrack &=& \lbrack P^\mu,Q_a^\dagger\rbrack = 0 \IEEEyessubnumber\label{eq:susyalgebra3}
\end{IEEEeqnarray}
where $\sigma^\mu\equiv (1,\sigma_1,\sigma_2,\sigma_3)$ contains the $2\times2$ Pauli spin matrices.
The single-particle states of a supersymmetric theory fall into irreducible representations of the supersymmetry algebra, called \emph{supermultiplets}.
These contain an equal number of bosonic and fermionic degrees of freedom.
The boson and fermion in a supermultiplet are called \emph{superpartners} of each other.
There are two important types of supermultiplets:\footnote{There is a third possibility of a gravitino-graviton supermultiplet with respectively spin $3/2$ and spin $2$, but this is not included in a minimal supersymmetric extension of the Standard Model. Furthermore, there are theories with multiple copies of the generators $Q,Q^\dagger$. However, such theories have no phenomenological prospects, since they do not allow for chiral fermions or parity violation. Hence we will stick to one copy of the generators, i.e.\ $N=1$ supersymmetry.}
\begin{itemize}
\item \emph{Chiral supermultiplets} consist of a two-component Weyl spinor and a complex scalar field.
\item \emph{Gauge supermultiplets} consist of a spin-1 gauge boson field and a spin-$1/2$ Majorana spinor (called the \emph{gaugino} field), which equals its own charge conjugate.
\end{itemize}

Equation \eqref{eq:susyalgebra3} tells us that the supersymmetry generators commute with the mass-squared operator $P^2$; hence members of a supermultiplet have the same mass.
The generators also commute with the generators of gauge transformations, so superpartners have the same gauge quantum numbers (e.g.\ electric charge, colour).
These observations have implications for the particle content of any supersymmetric extension of the Standard Model.
Consider for example the Higgs boson: it only fits into a chiral supermultiplet and should be paired with a spin-$1/2$ fermion.
The quarks are not an option: they have colour whereas the Higgs boson does not.
The leptons are not an option either: the Higgs boson does not carry lepton number.
Using similar arguments for the other Standard Model particles, we come to the conclusion that no two SM particles fit into the same supermultiplet.
Hence supersymmetry requires the existence of at least twice as many particles as we know in the SM: one fermionic (bosonic) `twin brother' for every Standard Model boson (fermion).
These new particles are called `supersymmetric particles' or `sparticles' for short.
Spin-0 sparticles are named by adding the prefix `s' (for `scalar') to its Standard Model partner's name: quarks are paired with \emph{squarks}, electrons with \emph{selectrons} and so on.
Spin-$1/2$ sparticles are given the suffix `ino', resulting in \emph{gluinos}, \emph{winos} and \emph{higgsinos}.

\subsection{How to construct a supersymmetric Lagrangian}\label{s:constructingsusy}
In this section we review how a supersymmetric Lagrangian is constructed and what its ingredients are.
We will only list results and the concepts behind them; for a detailed derivation, see e.g.\ \cite{martin,aitchison}.

A supersymmetric Lagrangian is built from the principles of renormalisability, invariance under a supersymmetry transformation and gauge invariance.
We can find the most general supersymmetric Lagrangian using the following steps:
\paragraph{Specify the supermultiplet content}
As mentioned in section \ref{s:whatissusy}, we only consider chiral and gauge supermultiplets.
Each chiral supermultiplet consists of a complex scalar field $\phi_i$ and a left-handed Weyl spinor $\psi_i$.
Right-handed Weyl spinors should first be conjugated in order to bring them in left-handed form.
In section \ref{s:whatissusy} we mentioned that a supermultiplet has an equal number of bosonic and fermionic degrees of freedom.
However, when the fields are off-shell there is a mismatch between the superpartners: $\phi_i$ has two degrees of freedom, whereas $\psi_i$ has four degrees of freedom off-shell.
Hence an \emph{auxiliary} complex scalar field $F_i$ is introduced to account for this discrepancy.
It will have no kinetic term, so that its equation of motion is $F_i=0$ and the degrees of freedom still match on-shell.
The auxiliary field $F_i$ is only a bookkeeping device to ensure that the supersymmetry algebra closes; after we have introduced interactions it will be eliminated from the Lagrangian by replacing it by its equation of motion.
The Lagrangian for a set of free chiral supermultiplets is:
\begin{equation}
\lag_\text{chiral,free} = \left(\partial^\mu\phi^{*i}\right)\left(\partial_\mu\phi_i\right) + i\psi^{\dagger i}\bar{\sigma}^\mu\partial_\mu\psi_i + F^{*i}F_i
\end{equation}
where $\bar{\sigma}^\mu\equiv(1,-\sigma_1,-\sigma_2,-\sigma_3)$.
	Note that the auxiliary field $F_i$ has dimension~2.
Each gauge supermultiplet consists of a massless gauge boson field $A_\mu^a$ and a Majorana spinor (gaugino field) $\lambda^a$, where the index $a$ runs over the adjoint representation of the gauge group.\footnote{$a=1,\ldots,8$ for $SU(3)_C$ colour; $a=1,2,3$ for $SU(2)_L$ weak isospin; $a=1$ for $U(1)_Y$ weak hypercharge.}
Here the field degrees of freedom do not match off-shell either: $\lambda^a$ has four degrees of freedom, whereas $A_\mu^a$ has only three.
Hence we need a real bosonic auxiliary field $D^a$ to make the supersymmetry algebra close.
The gauge and supersymmetry invariant Lagrangian for a gauge supermultiplet is:
\begin{equation}
\lag_\text{gauge} = -\frac14F_{\mu\nu}^aF^{\mu\nu a} + i\lambda^{\dagger a}\bar{\sigma}^\mu D_\mu\lambda^a + \frac12D^aD^a
\end{equation}
where
\begin{align}
F_{\mu\nu}^a &= \partial_\mu A_\nu^a - \partial_\nu A_\mu^a + gf^{abc}A_\mu^bA_\nu^c \\
D_\mu\lambda^a &= \partial_\mu\lambda^a + gf^{abc}A_\mu^b\lambda^c
\end{align}
are the Yang-Mills field strength and the covariant derivative of the gaugino respectively.\footnote{Note that we now have two different $D$'s: the auxiliary field $D^a$ and the covariant derivative $D_\mu$. Unfortunately this notation is widely used, but their different indices should keep confusion to a minimum.}
Here $g$ is the gauge coupling and $f^{abc}$ are the totally antisymmetric structure constants of the gauge group, i.e.\ the generators $T^a$ of the gauge group satisfy $[T^a,T^b]=if^{abc}T^c$.
Note that the auxiliary field $D^a$ has dimension~2.

\paragraph{Introduce interactions within the chiral supermultiplets}
The next step towards a general supersymmetric Lagrangian is to introduce interactions between the superpartners.
We only have to do this for the chiral supermultiplets, since the interactions within the gauge supermultiplets have already been included by imposing gauge invariance.
The possible interaction terms are determined by renormalisability by power counting (i.e.\ they must have dimension $d\leq4$, see section \ref{s:renormalisationprocedure}) and supersymmetry invariance.
It turns out that the possible interactions can be written as functional derivatives of a \emph{superpotential} $W$, which is an analytic function of the scalar fields $\phi_i$:
\begin{equation}
W = L^i\phi_i + \frac12 M^{ij}\phi_i\phi_j + \frac16 y^{ijk}\phi_i\phi_j\phi_k \label{eq:chiralsuperpotential}
\end{equation}
Here $M^{ij}$ is a symmetric mass matrix and $y^{ijk}$ is a Yukawa coupling.
The $L^i$ term is only allowed when $\phi_i$ is a gauge singlet and hence does not occur in the Minimal Supersymmetric Standard Model.
The interaction part of the Lagrangian becomes:
\begin{equation}
\lag_\text{chiral,int} = \left(-\frac12W^{ij}\psi_i\cdot\psi_j+W^iF_i\right) + \text{h.c.}
\end{equation}
where we have used the spinor product $\psi_i\cdot\psi_j\equiv \psi_i^Ti\sigma_2\psi_j = \psi_j\cdot\psi_i$ and defined
\begin{equation}
W^{ij} \equiv \frac{\delta^2}{\delta\phi_i\delta\phi_j}W \quad \text{and} \quad W^i \equiv \frac{\delta W}{\delta\phi_i}
\end{equation}
The auxiliary fields can now be eliminated from the Lagrangian by their equation of motion $F_i=-W_i^*$, $F^{*i}=-W^i$; the full Lagrangian of the chiral supermultiplets then becomes:
\begin{equation}
\lag_\text{chiral} = \left(\partial^\mu\phi^{*i}\right)\left(\partial_\mu\phi_i\right) + i\psi^{\dagger i}\bar{\sigma}^\mu\partial_\mu\psi_i - \frac12\left(W^{ij}\psi_i\cdot\psi_j+ \text{h.c.}\right) - W^iW_i^*
\end{equation}

\paragraph{Make the Lagrangian gauge-invariant}
Let us assume that the spinors $\psi_i$ and scalar fields $\phi_i$ form multiplets $\psi=(\psi_1,\ldots)$, $\phi=(\phi_1,\ldots)$ in the fundamental representation of the considered gauge group.
The last step is to make the combined Lagrangian $\lag_\text{chiral}+\lag_\text{gauge}$ gauge-invariant.
As in section \ref{s:gaugetheories}, this is done by replacing the ordinary derivatives with covariant derivatives.
This couples the vector bosons in the gauge supermultiplets to the scalars and fermions in the chiral supermultiplets.
In order to maintain supersymmetry invariance, we should also couple the chiral supermultiplet to the gaugino $\lambda^a$ and the auxiliary field $D^a$.
In the end, we can eliminate the auxiliary fields $D^a$ from the Lagrangian by their equation of motion.
Then the full Lagrangian for a renormalisable supersymmetric theory is:
\begin{align}
\lag =& \left(D^\mu\phi^{*i}\right)\left(D_\mu\phi_i\right) + i\psi^{\dagger i}\bar{\sigma}^\mu D_\mu\psi_i - \frac14F_{\mu\nu}^aF^{\mu\nu, a} + i\lambda^{\dagger a}\bar{\sigma}^\mu D_\mu\lambda^a \nonumber\\
&- \frac12\left(W^{ij}\psi_i\cdot\psi_j+ \text{h.c.}\right) - W^iW_i^* - \frac12g^2(\phi^*T^a\phi)^2 \nonumber\\
&- \sqrt{2}g\left((\phi^*T^a\psi)\cdot\lambda^a + \text{h.c.}\right)
\end{align}
To summarise, a supersymmetric Lagrangian is constructed from the principles of renormalisability, invariance under a supersymmetry transformation and gauge invariance.
The Lagrangian is completely determined by specifying the gauge group, field content and superpotential.
The latter is an analytic function of the scalar fields; it can be thought of as a short code for the (non-gauge) interactions in the theory.

\subsection{The Minimal Supersymmetric Standard Model}\label{s:mssm}
\begin{table}[p]
\begin{center}
{\renewcommand{\arraystretch}{1.5}
\begin{tabular}{|c|c|c|c|c|}
\hline
\multirow{2}{3cm}{\begin{center}Name\end{center}}	&	\multirow{2}{1.5cm}{\begin{center}Symbol\end{center}}	&	\multirow{2}{1.5cm}{\begin{center}Spin 0\end{center}}	&	\multirow{2}{1.5cm}{\begin{center}Spin 1/2\end{center}}	&	Gauge group	\\
	&	&	&	& representation	\\\hline\hline
\multirow{3}{3cm}{\begin{center}squarks \& quarks (3 generations)\end{center}}	&	$Q$	&	$\widetilde{Q}=(\widetilde{u}_L \; \widetilde{d}_L)$	&	$(u_L \; d_L)$	&	$(\mathbf{3},\mathbf{2},\frac16)$	\\
		&	$\bar{u}$	&	$\widetilde{u}_R^*$	&	$u_R^c$	&	$(\mathbf{\bar{3}},\mathbf{1},-\frac23)$	\\
		&	$\bar{d}$	&	$\widetilde{d}_R^*$	&	$d_R^c$	&	$(\mathbf{\bar{3}},\mathbf{1},\frac13)$	\\\hline
sleptons \& leptons	&	$L$	&	$\widetilde{L}=(\widetilde{\nu} \; \widetilde{e}_L)$	&	$(\nu \; e_L)$	&	$(\mathbf{1},\mathbf{2},-\frac12)$	\\
(3 generations)		&	$\bar{e}$	&	$\widetilde{e}_R^*$	&	$e_R^c$	&	$(\mathbf{1},\mathbf{1},1)$	\\\hline
\multirow{2}{3cm}{\begin{center}Higgs \& higgsinos\end{center}}	&	$H_u$	&	$(H_u^+ \; H_u^0)$	&	$(\widetilde{H}_u^+ \; \widetilde{H}_u^0)$	&	$(\mathbf{1},\mathbf{2},\frac12)$ \\
	&	$H_d$	&	$(H_d^0 \; H_d^-)$	&	$(\widetilde{H}_d^0 \; \widetilde{H}_d^-)$	&	$(\mathbf{1},\mathbf{2},-\frac12)$ \\\hline
\end{tabular}}
\caption{Chiral supermultiplet content of the Minimal Supersymmetric Standard Model and the corresponding representations of the gauge group $SU(3)_C\times SU(2)_L\times U(1)_Y$. Note that we need an additional Higgs doublet compared to the Standard Model.}\label{t:mssmchiralcontent}
\end{center}
\end{table}
\begin{table}[p]
\begin{center}
{\renewcommand{\arraystretch}{1.5}
\begin{tabular}{|c|c|c|c|}
\hline
\multirow{2}{3cm}{\begin{center}Names\end{center}}	&	\multirow{2}{1.5cm}{\begin{center}Spin 1/2\end{center}}	&	\multirow{2}{1.5cm}{\begin{center}Spin 1\end{center}}	&	Gauge group	\\
	&	&	&	representation	\\\hline\hline
gluino \& gluon		&	$\widetilde{g}$	&	$g$	&	$(\mathbf{8},\mathbf{1},0)$	\\\hline
winos \& $W$ bosons	&	$\widetilde{W}^1 \quad \widetilde{W}^2 \quad \widetilde{W}^3$	&	$W^1 \quad W^2 \quad W^3$	&	$(\mathbf{1},\mathbf{3},0)$	\\\hline
bino \& $B$ boson	&	$\widetilde{B}^0$	&	$B^0$	&	$(\mathbf{1},\mathbf{1},0)$	\\\hline
\end{tabular}}
\caption{Gauge supermultiplet content of the Minimal Supersymmetric Standard Model and the corresponding representations of the gauge group $SU(3)_C\times SU(2)_L\times U(1)_Y$.}\label{t:mssmgaugecontent}
\end{center}
\end{table}
The Minimal Supersymmetric Standard Model (MSSM) is defined to have minimal particle content.
As argued in section \ref{s:whatissusy}, this means that each Standard Model particle has a supersymmetric partner; these are denoted by adding a tilde to the symbol of the corresponding SM particle.
The MSSM particle content is listed in tables \ref{t:mssmchiralcontent}-\ref{t:mssmgaugecontent}.
Note that we need two Higgs doublets rather than one: as we have seen in section \ref{s:higgsmechanism}, up-type fermions acquire mass from the charge conjugate of the Higgs doublet after spontaneous symmetry breaking.
However, the charge conjugate of the Higgs field cannot appear in the superpotential since the latter should be an analytic function of the Higgs field.
Hence the structure of supersymmetric theories requires the existence of an additional Higgs doublet.
The additional Higgs doublet is also needed to prevent gauge anomalies \cite{martin}.

The MSSM has the same gauge group $SU(3)_C\times SU(2)_L\times U(1)_Y$ as the Standard Model.
So the only thing we still need to specify is the MSSM superpotential.
This is given by:
\begin{equation}
W_\text{MSSM} = \widetilde{u}_R^\dagger\mathbf{y_u}\widetilde{Q}\times H_u - \widetilde{d}_R^\dagger\mathbf{y_d}\widetilde{Q}\times H_d - \widetilde{e}_R^\dagger\mathbf{y_e}\widetilde{L}\times H_d + \mu H_u\times H_d \label{eq:mssmsuperpotential}
\end{equation}
where we defined $\widetilde{Q}\times H_u \equiv \widetilde{Q}^Ti\sigma_2H_u = -H_u\times\widetilde{Q}$ in order to combine two doublets into a singlet without using conjugation.
The dimensionless Yukawa coupling parameters $\mathbf{y_u},\mathbf{y_d},\mathbf{y_e}$ are $3\times3$ matrices in family space.
The $\mu$ term is the supersymmetric version of the Higgs boson mass in the Standard Model.
All gauge, family and spinor indices have been suppressed in \eqref{eq:mssmsuperpotential}.

This superpotential is minimal in the sense that it is sufficient to produce a phenomenologically viable model.
In principle we could add other terms to \eqref{eq:mssmsuperpotential}, but these are not included because they violate either baryon number (B) or lepton number (L).
Such interactions would enable proton decay, which is strongly constrained by experiment.
In order to exclude such terms, the MSSM is defined to preserve a new quantum number called \emph{$R$-parity}:\footnote{One may wonder why we do not simply postulate B and L conservation. The answer is that B and L are known to be violated by non-perturbative electroweak effects \cite{BLviolation}, although those effects are negligible for experiments at ordinary energies.}
\begin{equation}
P_R \equiv (-1)^{3(B-L)+2s}
\end{equation}
where $s$ is the spin of the particle.
Note that $P_R=+1$ for all Standard Model particles and $P_R=-1$ for all their superpartners.
A phenomenologically interesting consequence of $R$-parity conservation is that every interaction vertex in the theory contains an even number of sparticles.
This implies that the lightest supersymmetric particle (LSP) is absolutely stable.
If the LSP is electrically neutral, it can make an attractive dark matter candidate, as we already mentioned in section \ref{s:whysusy}.

As a final remark, note that equation \eqref{eq:mssmsuperpotential} has introduced only one new parameter, namely $\mu$, compared to the Standard Model.
However, we get a lot more parameters if supersymmetry is broken.
This will be discussed in the next section.

\subsection{Charginos, neutralinos \& Higgs bosons}\label{s:charginosneutralinos}
Recall from section \ref{s:higgsmechanism} that the interaction eigenstates do not always correspond to the mass eigenstates.
In the Standard Model, the $W^{1,2,3}$ and $B$ bosons are interaction eigenstates of the electroweak gauge bosons.
After electroweak symmetry breaking, the mass eigenstates are mixtures of them: there are massive weak interaction bosons $W^\pm$, $Z^0$ and a massless photon $\gamma$.
Three of the four degrees of freedom of the Higgs doublet have been eaten by the gauge bosons, so only one physical Higgs boson remains.

In the MSSM, the higgsinos and electroweak gauginos also mix with each other because of electroweak symmetry breaking.
The neutral higgsinos $\widetilde{H}_u^0$, $\widetilde{H}_d^0$ and the neutral gauginos $\widetilde{B}$, $\widetilde{W}^3$ combine into four mass eigenstates called \emph{neutralinos}.
These are denoted by $\widetilde{N}_i$, $i=1,2,3,4$.
The charged higgsinos $\widetilde{H}_u^+$, $\widetilde{H}_d^-$ and winos $\widetilde{W}^\pm \equiv \frac{1}{\sqrt{2}}(\widetilde{W}^1\mp i\widetilde{W}^2)$ combine into two mass eigenstates with charge $+1$ and two with charge $-1$.
These are called \emph{charginos} and are denoted by $\widetilde{C}_i^\pm$, $i=1,2$.

Furthermore, we have two complex Higgs doublets with a total of eight degrees of freedom.
Still three of them become the longitudinal polarisation states of the massive $W^\pm$, $Z^0$ bosons.
This leaves us with five physical Higgs bosons: two CP-even neutral scalars $h^0$ and $H^0$ (where $h^0$ is defined to be the lighter one), one CP-odd neutral scalar $A^0$ and two charged scalars $H^+$, $H^-$ with charge $+1$ and $-1$ respectively.

\section{Supersymmetry Breaking}\label{s:susybreaking}

Supersymmetry predicts the existence of a superpartner for each Standard Model particle.
However, we have not discovered any of these sparticles as yet.
Hence supersymmetry cannot be an exact symmetry of Nature: otherwise each sparticle would have the same mass as its Standard Model partner and we would have discovered them already.
If supersymmetry exists, it must be broken somehow.

The literature offers a plethora of possible ways to break supersymmetry.
In this section, we will discuss the categories they can be divided into.
In the first part we will consider to what extent broken supersymmetry can still solve the problems of the SM and which constraints this puts on a realistic supersymmetric theory.
We will also examine the possible terms of a supersymmetry breaking Lagrangian and how these are constrained.
Then we will see how an explicit model for supersymmetry breaking can be constructed and subsequently we will discuss several proposals for breaking mechanisms.\footnote{An introduction to most of these mechanisms can be found in \cite{sparticles}; for more extensive treatments see the literature cited in the relevant sections.}
Finally, we will list some common problems that a realistic supersymmetry breaking model should address.

\subsection{Constraints on broken supersymmetry}\label{s:brokensusyconstraints}
\paragraph{Soft breaking}
The main motivation for a supersymmetric extension of the Standard Model is the elimination of the hierarchy problem.
A nice result from unbroken supersymmetry is the fact that quadratically divergent loop corrections to the Higgs mass vanish to all orders in perturbation theory.
How can we guarantee that this happens even in broken supersymmetry?
As we have seen in section \ref{s:whysusy}, a necessary condition for the cancellation of quadratic terms in $\Lambda$ is $\lambda_S=|y|^2$.
In order to maintain the solution to the hierarchy problem, we need to prevent the couplings $\lambda_S,y$ from acquiring loop corrections from the supersymmetry breaking interactions.
This requires us to consider \emph{soft supersymmetry breaking}.
In this case, the Lagrangian can be written as:
\begin{equation}
\lag = \lag_\text{SUSY} + \lag_\text{soft}
\end{equation}
where $\lag_\text{SUSY}$ is supersymmetry invariant and contains all of the gauge and Yukawa interactions.
The extra term $\lag_\text{soft}$ violates supersymmetry but contains only masses and couplings with \emph{positive} dimension, i.e.\ there are no dimensionless couplings.
This implies that the effects of the terms in $\lag_\text{soft}$ are suppressed at high energies.
The most general soft supersymmetry breaking Lagrangian in the MSSM, compatible with gauge invariance and $R$-parity conservation, is:
\begin{align}
\lag_\text{soft}^\text{MSSM} = &-\frac12\left(M_3\widetilde{g}\cdot\widetilde{g} + M_2\widetilde{W}\cdot\widetilde{W} + M_1\widetilde{B}\cdot\widetilde{B} + \text{h.c.}\right) \nonumber\\
&-\left(\widetilde{u}_R^\dagger\mathbf{a_u}\widetilde{Q}\times H_u - \widetilde{d}_R^\dagger\mathbf{a_d}\widetilde{Q}\times H_d - \widetilde{e}_R^\dagger\mathbf{a_e}\widetilde{L}\times H_d + \text{h.c.}\right) \nonumber\\
&-\widetilde{Q}^\dagger\mathbf{m_Q^2}\widetilde{Q} - \widetilde{L}^\dagger\mathbf{m_L^2}\widetilde{L} - \widetilde{u}_R^\dagger\mathbf{m_{\bar{u}}^2}\widetilde{u}_R - \widetilde{d}_R^\dagger\mathbf{m_{\bar{d}}^2}\widetilde{d}_R - \widetilde{e}_R^\dagger\mathbf{m_{\bar{e}}^2}\widetilde{e}_R \nonumber\\
&-m_{H_u}^2H_u^\dagger H_u - m_{H_d}^2H_d^\dagger H_d - (bH_u\times H_d + \text{h.c.}) \label{eq:softlagrangian}
\end{align}
Here $M_3,M_2,M_1$ are complex gluino, wino and bino mass terms; $\mathbf{a_u}, \mathbf{a_d}, \mathbf{a_e}$ are complex $3\times3$ matrices in family space, similar to the Yukawa couplings in \eqref{eq:mssmsuperpotential}; $\mathbf{m_Q^2}, \mathbf{m_{\bar{u}}^2}, \mathbf{m_{\bar{d}}^2},$ $ \mathbf{m_L^2}, \mathbf{m_{\bar{e}}^2}$ are Hermitian $3\times3$ mass matrices in family space and $m_{H_u}^2$, $m_{H_d}^2$ are real Higgs masses.
The parameter $b$ is a complex parameter, which often appears in the literature as $B\mu$.

\paragraph{TeV scale supersymmetry}
Let $m_\text{soft}$ denote the largest mass scale associated with the soft parameters.
Then all soft non-supersymmetric corrections to the Higgs mass should vanish in the limit $m_\text{soft}\rightarrow0$ (this is why we consider only terms with couplings that have positive dimension).
Dimensional analysis then tells us that these corrections cannot be proportional to $\Lambda^2$.
They also cannot be proportional to $m_\text{soft}\Lambda$, because  in general the loop momentum integrals diverge either quadratically or logarithmically, not linearly.
Hence the corrections must be of the form:
\begin{equation}
\delta M_H^2 \sim m_\text{soft}^2\log{\left(\frac{\Lambda}{m_\text{soft}}\right)}
\end{equation}
Taking $\Lambda\sim M_\text{Pl}$, one finds that $m_\text{soft}$ should not be much larger than 1 TeV.
Hence, if supersymmetry is realised in Nature, we should be able to find sparticles at the Large Hadron Collider.

\paragraph{Spontaneous breaking}
The soft supersymmetry breaking terms add a lot of new parameters to the Lagrangian.
Although supersymmetry itself adds only one new parameter $\mu$ with respect to the SM, supersymmetry \emph{breaking} introduces 97 new masses, mixing angles and phases \cite{manyparameters}.
Thus supersymmetry breaking has added lots of arbitrariness to the theory!

\begin{figure}[t]
\begin{minipage}{.49\textwidth}
  \begin{center}
  \begin{fmfgraph*}(50,20)
    \fmfforce{(0,0)}{mu}
      \fmflabel{$\mu^-$}{mu}
    \fmfforce{(w,0)}{e}
      \fmflabel{$e^-$}{e}
    \fmf{plain,tension=2}{mu,v1}
    \fmf{gaugino,label=$\tilde{B}$,label.side=left}{v1,v2}
    \fmf{plain,tension=2}{v2,e}
    \fmffreeze
    \fmf{dashes_crossed,left,tension=0,label=$\tilde{\mu}_R\qquad\tilde{e}_R$,label.dist=1pt}{v1,v2}
    \fmfforce{(.67w,.45h)}{v3}
    \fmfforce{(w,h)}{gamma}
      \fmflabel{$\gamma$}{gamma}
    \fmf{photon}{v3,gamma}
  \end{fmfgraph*}
  \vskip5mm (a)
  \end{center}
\end{minipage}
\begin{minipage}{.49\textwidth}
  \begin{center}
  \begin{fmfgraph*}(50,20)
    \fmfleft{d,sbar}
      \fmflabel{$d$}{d}
      \fmflabel{$\bar{s}$}{sbar}
    \fmfright{s,dbar}
      \fmflabel{$s$}{s}
      \fmflabel{$\bar{d}$}{dbar}
    \fmf{plain,tension=2}{d,vl1}
    \fmf{dashes_crossed,label.side=left,label=$\tilde{d}_R\qquad\tilde{s}_R$,label.dist=3pt}{vl1,vr1}
    \fmf{plain,tension=2}{vr1,s}
    \fmf{plain,tension=2}{sbar,vl2}
    \fmf{dashes_crossed,label.side=left,label=$\tilde{s}^*_R\qquad\tilde{d}^*_R$,label.dist=3pt}{vl2,vr2}
    \fmf{plain,tension=2}{vr2,dbar}
    \fmf{gaugino,tension=0,label.side=left,label=$\tilde{g}$}{vl1,vl2}
    \fmf{gaugino,tension=0,label.side=right,label=$\tilde{g}$}{vr1,vr2}
  \end{fmfgraph*}
  \vskip5mm (b)
  \end{center}
\end{minipage}
\caption{Some of the diagrams that contribute to flavour changing neutral current (FCNC) processes. Diagram (a) corresponds to a contribution to the process $\mu^-\rightarrow e^-\gamma$, coming from the off-diagonal elements of $\mathbf{m_{\bar{e}}}^2$. Diagram (b) corresponds to a contribution to $K^0\leftrightarrow \bar{K}^0$ mixing, coming from the off-diagonal elements of $\mathbf{m_{\bar{d}}}^2$.}\label{f:flavourmixing}
\end{figure}
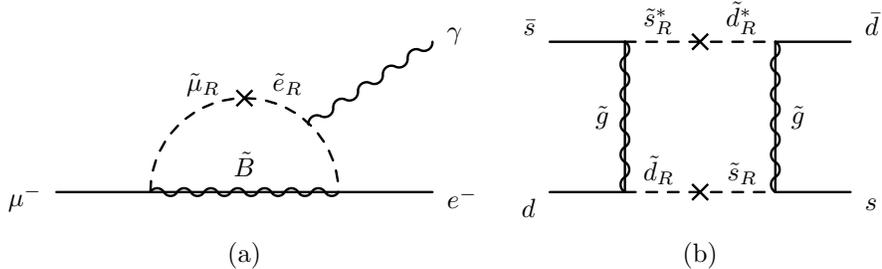

Fortunately, experiments suggest that the soft supersymmetry breaking Lagrangian cannot be completely arbitrary.
Many of the MSSM soft terms would imply flavour-changing neutral current (FCNC) processes (i.e.\ `flavour mixing') which are severely restricted by experiment.
For example, if $\mathbf{m_{\bar{e}}^2}$ were not diagonal in the basis $(\tilde{e}_R,\tilde{\mu}_R,\tilde{\tau}_R)$, the off-diagonal elements would contribute to slepton mixing and individual lepton numbers would not be conserved.
A particularly strong limit on such processes comes from the experimental bound on the process $\mu\rightarrow e\gamma$ (figure \ref{f:flavourmixing}a).
If the right-handed slepton squared-mass matrix were random with all entries of comparable size, then the prediction for the branching ratio of $\mu\rightarrow e\gamma$ would be too large.
\begin{fmffile}{flavourmixing}
\fmfcmd{%new commands for sparticle lines
vardef cross_bar (expr p, len, ang) =
 ((-len/2,0)--(len/2,0))
 rotated (ang + angle direction length(p)/2 of p)
 shifted point length(p)/2 of p
enddef;
style_def dashes_crossed expr p=
 save dpp;
 numeric dpp;
 dpp = ceiling (pixlen (p, 10) / dash_len) / length p;
 for k=0 upto dpp*length(p) - 1:
  cdraw point k/dpp of p ..
   point (k+.5)/dpp of p;
 endfor
 ccutdraw cross_bar (p, 3mm, 45);
 ccutdraw cross_bar (p, 3mm, -45);
enddef;
 style_def gaugino expr p =
 cdraw (wiggly p);
 cdraw p;
enddef;}
\end{fmffile}

Similarly, important experimental constraints on the squark squared-mass matrices come from the neutral kaon system (figure \ref{f:flavourmixing}b).
If the MSSM soft Lagrangian contained terms that mix down squarks and strange squarks, limits on CP-violation due to neutral kaon mixing would be violated.
Many more similar FCNC processes and new sources of CP-violation severely restrict the form of the MSSM soft terms.

These (and many more) processes can be suppressed by assuming so-called universality conditions.
For example, we could avoid all such processes if we assume that all mass matrices in \eqref{eq:softlagrangian} are proportional to the unit matrix and the trilinear scalar couplings are proportional to the Yukawa couplings:
\begin{IEEEeqnarray}{rClrClrClrClrCl}\label{eq:universalityrelations}
\mathbf{m_Q^2} &=& \msq\mathbf{1}\;,\; & \mathbf{m_{\bar{u}}^2} &=& \msu\mathbf{1}\;,\; & \mathbf{m_{\bar{d}}^2} &=& \msd\mathbf{1}\;,\; & \mathbf{m_L^2} &=& \msl\mathbf{1}\;,\; & \mathbf{m_{\bar{e}}^2} &=& \mse\mathbf{1} \qquad \IEEEyessubnumber\\
&&& \mathbf{a_u} &=& A_{u0}\mathbf{y_u}\;,\; & \mathbf{a_d} &=& A_{d0}\mathbf{y_d}\;,\; & \mathbf{a_e} &=& A_{e0}\mathbf{y_e} &&& \IEEEyessubnumber
\end{IEEEeqnarray}
Such an assumption seems strange: first we add lots of arbitrariness to the Lagrangian and then we deal with it by arbitrarily assuming relations between the parameters.
However, from the effective field theoretical point of view, this may make sense after all.
Suppose supersymmetry is exact at very high energies, in a more fundamental theory than the MSSM, but is broken spontaneously at some high energy scale.
Then the soft terms in \eqref{eq:softlagrangian} arise as effective interactions, similar to the way the Fermi 4-vertex arises from the exchange of a $W$ boson (see section \ref{s:matching}).
Relations such as \eqref{eq:universalityrelations} may arise as matching conditions  at the threshold where we switch from the more fundamental theory to the MSSM.

Thus the desire for a theory that naturally explains supersymmetry breaking forces us to consider spontaneously broken supersymmetry, rather than just add explicitly supersymmetry violating terms to the Lagrangian.
There are many different models in the literature that try to explain how supersymmetry is broken at high energies.
Some of them will be discussed in sections \ref{s:SUGRA}-\ref{s:MM}.

\subsection{How to break supersymmetry}
We are looking for a way to break supersymmetry spontaneously: the underlying theory that results in the MSSM should have a Lagrangian that preserves supersymmetry but a vacuum state that breaks it.
Where do we start looking?
An important clue comes from an inspection of the vacuum energy.
Consider the Hamiltonian operator $H$; using \eqref{eq:susyalgebra1} we can write it in terms of the supersymmetry generators as follows:
\begin{equation}
H = P^0 = \frac12\left(Q_1Q_1^\dagger+Q_1^\dagger Q_1 + Q_2Q_2^\dagger+Q_2^\dagger Q_2\right)
\end{equation}
If supersymmetry is unbroken in the vacuum state, then $Q_a|0\rangle=Q_b^\dagger|0\rangle=0$.
It follows that $H|0\rangle=0$ so that the vacuum state has zero energy.
However, if supersymmetry is broken spontaneously, then $Q_a|0\rangle\neq0$ and $Q_b^\dagger|0\rangle\neq0$ for some $a,b$, so that:
\begin{align}
\langle0|H|0\rangle &= \langle0|\frac12\left(Q_1Q_1^\dagger+Q_1^\dagger Q_1 + Q_2Q_2^\dagger+Q_2^\dagger Q_2\right)|0\rangle \nonumber \\
&= \frac12\left(\|Q_1^\dagger|0\rangle\|^2 + \|Q_1|0\rangle\|^2 + \|Q_2^\dagger|0\rangle\|^2 + \|Q_2|0\rangle\|^2\right) >0
\end{align}
Hence a theory with spontaneously broken supersymmetry necessarily has positive vacuum energy!
If spacetime-dependent effects and fermion condensates can be neglected, then $\langle0|H|0\rangle=\langle0|V|0\rangle$, where $V$ is the scalar potential:
\begin{equation}
V(\phi_i,\phi_i^*) = F^{*i}F_i + \frac12D^aD^a
\end{equation}
where $F_i,D^a$ are the auxiliary fields we encountered in section \ref{s:constructingsusy}.
If any state exists in which all auxiliary fields vanish, this state will have zero energy and will be a supersymmetry preserving vacuum state.
Thus one way of guaranteeing that supersymmetry is broken spontaneously is to construct a model in which the equations $F_i=0$ and $D^a=0$ cannot be satisfied simultaneously for any values of the fields.
Below, we will briefly consider two ways of achieving this.

Note that we can get TeV scale supersymmetry even if supersymmetry breaking occurs at a very high energy scale.
Recall that both $F$ and $D$ have mass dimension 2, so $m_\text{soft}$ will be proportional to $\langle F\rangle/M$ or $\langle D\rangle/M$, where $M$ is the energy scale where the auxiliary fields obtain a VEV.
Thus even if $M$ is large, $m_\text{soft}$ can still be of order 1 TeV.

\paragraph{Fayet-Iliopoulos ($D$-term) supersymmetry breaking}
Fayet-Iliopoulos supersymmetry breaking \cite{FayetIliopoulos} is a mechanism that breaks supersymmetry through a non-zero $D$-term VEV.
The idea is to include a $U(1)$ factor in the gauge group and introduce a term linear in the corresponding auxiliary field of the gauge supermultiplet:\footnote{For non-Abelian gauge groups, such a term would not be gauge invariant.}
\begin{equation}
\lag_\text{Fayet-Iliopoulos} = -\kappa D
\end{equation}
where $\kappa$ is a constant.
Then the equation of motion for $D$ becomes:
\begin{equation}
D = \kappa - g\sum_iq_i|\phi_i|^2
\end{equation}
where $q_i$ are the charges of the scalar fields $\phi_i$ under the $U(1)$ gauge group.
If all scalar fields $\phi_i$ that are charged under this $U(1)$ have non-zero superpotential masses, the scalar potential will have the form:
\begin{equation}
V = \sum_i|m_i|^2|\phi_i|^2 + \frac12(\kappa-g\sum_iq_i|\phi_i|^2)^2 \label{eq:Dtermscalarpotential}
\end{equation}
which cannot vanish.
Hence supersymmetry must be broken.

In the MSSM, a Fayet-Iliopoulos term for $U(1)_Y$ cannot work: the squarks and sleptons do not have superpotential mass terms.
If a $D$-term were present, some of the sfermions would just get non-zero VEVs, breaking colour and/or electric charge conservation but not supersymmetry.
This is because without the mass term in \eqref{eq:Dtermscalarpotential}, the scalar potential has a minimum at $V=0$, $D=0$, $\langle\phi_i\rangle\neq0$ for at least one $i$; only with a mass term we can ensure $\langle D\rangle\neq0$.
Instead, one could add a $U(1)$ gauge symmetry that is broken spontaneously at a high energy scale.
However, if this is the dominant source for supersymmetry breaking, it turns out to be difficult to give appropriate masses to all of the MSSM particles: some of the gaugino masses would become so light that we would already have observed them.

\paragraph{O'Raifeartaigh ($F$-term) supersymmetry breaking}
O'Raifeartaigh supersymmetry breaking \cite{ORaifeartaigh} is a mechanism that breaks supersymmetry through a non-zero $F$-term VEV.
The idea is to pick a set of chiral supermultiplets $(\phi_i,\psi_i,F_i)$ and a superpotential $W$ in such a way that the equations $F_i=-W_i^*=0$ have no simultaneous solution.
Consider for example a theory with three chiral supermultiplets and a superpotential:
\begin{equation}
W = -k\phi_1 + m\phi_2\phi_3 + \frac{y}{2}\phi_1\phi_3^2 \label{eq:Ftermpotential}
\end{equation}
The linear term is allowed if $\phi_1$ is a gauge singlet.
Choosing $k,m,y$ to be real and positive by a phase rotation of the fields, the scalar potential following from \eqref{eq:Ftermpotential} is:
\begin{IEEEeqnarray}{rCl}
V &=& |F_1|^2 + |F_2|^2 + |F_3|^2 \IEEEyessubnumber\\
F_1 &=& k-\frac y2\phi_3^{*2} \qquad F_2=-m\phi_3^* \qquad F_3=-m\phi_2^*-y\phi_1^*\phi_3^* \IEEEyessubnumber
\end{IEEEeqnarray}
Clearly $F_1$ and $F_2$ cannot vanish simultaneously if $k\neq0$, so supersymmetry must be broken indeed.
Note that within the MSSM, there is no candidate gauge singlet unless we allow for a right-handed neutrino and its corresponding scalar partner, the sneutrino.

If supersymmetry is found in particle colliders, finding the ultimate source of supersymmetry breaking will be one of the most important issues.
However, to understand collider phenomenology we do not need to know about the underlying dynamics causing supersymmetry breaking: these are only relevant at some high energy scale beyond experimental access.
In the effective field theory that we call the MSSM, we can simply assume some $F$-term has acquired a VEV, and our only concern is the nature of the couplings of this VEV to the MSSM fields.
We will discuss this below.

\subsubsection{Radiative supersymmetry breaking}\label{s:radiativesusybreaking}
Regardless of the way it has happened, spontaneous supersymmetry breaking requires us to extend the MSSM: a $D$-term VEV for $U(1)_Y$ breaks gauge symmetries and does not lead to an acceptable spectrum, and there is no candidate gauge singlet whose $F$-term could develop a VEV.
Therefore we need to find the effects that are responsible for spontaneous supersymmetry breaking, and how the breakdown is `communicated' to the MSSM particles.
It is very difficult to do this working only with renormalisable interactions at tree-level: it turns out to be difficult to give correct masses to all MSSM particles,\footnote{That is, not so light that we should have discovered them already.} and there is no reason why patterns like \eqref{eq:universalityrelations} would emerge.

\begin{figure}[t]
\begin{center}
\includegraphics[width=\textwidth]{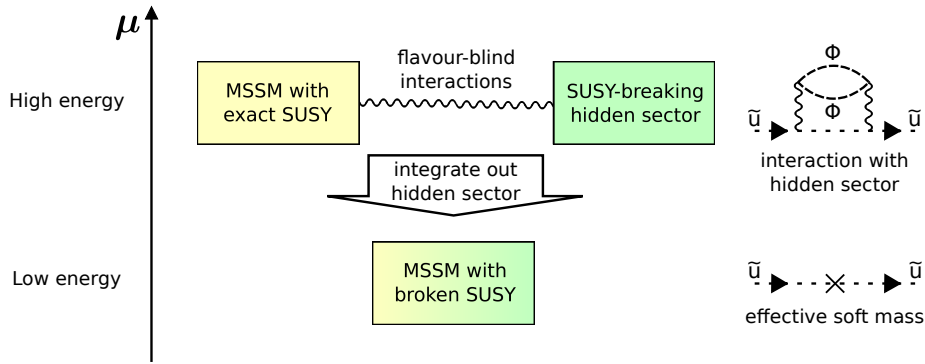}
\caption{Schematic structure of radiative supersymmetry breaking. At very high energies, supersymmetry is exact in the MSSM sector. Supersymmetry is broken by an $F$-term VEV of a field in the \emph{hidden sector}, which consists of very heavy fields. The two sectors are coupled only indirectly through flavour-blind interactions. At low energies, the emerging EFT is the MSSM with soft supersymmetry breaking terms. For example, in the more fundamental theory, the up-squark couples indirectly to the supersymmetry-breaking hidden sector field $\Phi$. When $\Phi$ is integrated out from the action, a soft mass term for the up-squark arises.}\label{f:radiativesusybreaking}
\end{center}
\end{figure}

For these reasons, the MSSM soft terms are expected to arise radiatively, rather than from tree-level renormalisable couplings to the supersymmetry breaking VEVs.
In radiative supersymmetry breaking models, supersymmetry is broken in a \emph{hidden sector}: this sector consists of fields that have no direct couplings to the MSSM fields, which are in the \emph{visible sector}.
The former is called `hidden' because hidden sector fields do not couple directly to the MSSM fields, so we cannot detect them in particle colliders.
The two sectors only interact indirectly; the interactions between them are responsible for mediating the supersymmetry breakdown from the hidden sector to the MSSM (figure \ref{f:radiativesusybreaking}).
For example, the up-squark may interact with a supersymmetry breaking hidden sector field $\Phi$ through the two-loop diagram in figure \ref{f:radiativesusybreaking}.
The hidden sector fields are usually very heavy, so at energy scales accessible by particle colliders we should integrate out the hidden sector fields from the theory.
The resulting interaction becomes a soft mass term for the up-squark: only the MSSM with broken supersymmetry is left.
If the mediating interactions are flavour-blind, then the soft terms of the MSSM will automatically satisfy conditions like \eqref{eq:universalityrelations}.

\subsection{Supergravity}\label{s:SUGRA}
As we already mentioned in section \ref{s:whysusy}, supergravity (SUGRA) \cite{sugra} is the theory that results from imposing local supersymmetry invariance.
Recall from section \ref{s:gaugetheories} that once we promote a global gauge symmetry (with bosonic generators, i.e.\ satisfying commutation relations) to a local one, we have to introduce a bosonic field with predetermined gauge transformation properties.
Similarly, by promoting supersymmetry (which has \emph{fermionic} generators, satisfying \emph{anti}commutation relations) to a local symmetry, we have to introduce a \emph{fermionic} field $\Psi_\mu$ with spin-$3/2$.
This is the \emph{gravitino}, the superpartner of the spin-2 graviton.
It turns out that when chiral and gauge supermultiplets are introduced, all terms in the Lagrangian can be described in terms of three functions of the complex scalar fields:
\begin{itemize}
\item The superpotential $W(\phi_i)$, which we have already encountered in section \ref{s:constructingsusy}. It is an analytic function of the complex scalar fields treated as complex variables and determines the non-gauge interactions of the theory.
\item The \emph{K\"ahler potential} $\mathcal{G}$, which is a real function of both $\phi_i$ and $\phi^{*i}$. Its matrix of second derivatives (called the \emph{K\"ahler metric}) determines the form of the kinetic energy terms of the chiral superfields in the theory.
\item The \emph{gauge kinetic function} $f_{ab}$, which is also an analytic function of the $\phi_i$ and has gauge group indices $a,b$. This function determines the coefficients of the kinetic terms for the gauge superfields.
\end{itemize}
In general, the Lagrangian is nonrenormalisable; there is as yet no renormalisable quantum field theory of gravity.
However, the nonrenormalisable operators are suppressed by inverse powers of $M_\text{Pl}$, so that their effects at low energies are small (see section \ref{s:EFTbasic}).

The spontaneous breakdown of supersymmetry occurs in a hidden sector where the auxiliary component of some superfield gets a VEV.
According to Goldstone's theorem, the spontaneous breaking of a global symmetry yields a massless particle with the same quantum numbers as the broken symmetry generator.
Since the broken generator $Q$ is fermionic, the massless particle is a massless neutral Weyl fermion, called the \emph{goldstino}.
The goldstino then becomes the longitudinal component of the gravitino, which becomes massive.\footnote{Because of the similarities with the Higgs mechanism, where the electroweak gauge bosons `eat' the goldstone bosons and become massive, this mechanism is called the \emph{super-Higgs} mechanism.}
It turns out that when we consider the effects of the supersymmetry breaking VEV, the gravitino mass $m_{3/2}$ sets the scale of all the soft terms.
Moreover, the scalar masses are universal at the scale where supersymmetry becomes broken.

\subsubsection*{Minimal supergravity}
The most widely used model of supersymmetry breaking is minimal supergravity (mSUGRA) \cite{msugra}.
Despite the name, mSUGRA is not a supergravity model, but rather the low-energy EFT resulting from a minimal locally supersymmetric model.
In the underlying model, one uses the simplest possible Ansatz for the superpotential (i.e.\ the sum of two separate superpotentials for the observable sector and the hidden sector) and the K\"ahler potential (which is given a similar additively separated form).
This leads to universal soft supersymmetry breaking parameters in the scalar sector.
Gauge coupling unification in the MSSM suggests an additional simple Ansatz for the gauge kinetic function, which leads to universal gaugino masses.
Then the soft terms at the GUT scale $M_\text{GUT}=2\cdot 10^{16}$ GeV are:
\begin{IEEEeqnarray}{rCl}\label{eq:msugraboundaryconditions}
m_{ij}^2 &=& m_0^2 \,\delta_{ij} \IEEEyessubnumber\\
M_a &=& M_{1/2} \qquad (a=1,2,3) \IEEEyessubnumber\\
A_{ijk} &=& A_0 \IEEEyessubnumber
\end{IEEEeqnarray}
Here $m_{ij}^2$ are the scalar squared masses, $M_a$ are the gaugino masses and $A_{ijk}$ are trilinear couplings defined by $a_{ijk}=y_{ijk}A_{ijk}$ (no summation), cf.\ equation \eqref{eq:universalityrelations}.
From the supergravity point of view, the parameters $m_0,M_{1/2},A_0$ depend on the hidden sector fields and are all proportional to $m_{3/2}$ (for example, one has the relation $m_0=m_{3/2}$).
However, from the perspective of the low-energy EFT that we call mSUGRA, they are simply regarded as model parameters.
The MSSM is assumed to be valid up to the GUT scale, where the relations \eqref{eq:msugraboundaryconditions} serve as RG boundary conditions.
In addition, the soft Higgs mixing term $B=b/\mu$ has the GUT scale value $B_0=A_0-m_{3/2}$.

As an aside, there is a model similar to mSUGRA: it is called constrained MSSM (CMSSM, see e.g.\ \cite{CMSSM}).
It has the same boundary conditions as mSUGRA and these two models are often confused in the literature.
However, mSUGRA arises from a supergravity model whereas the CMSSM does not: the boundary conditions \eqref{eq:msugraboundaryconditions} are simply postulated.
Also, in the CMSSM there is no relation between the model parameters and $m_{3/2}$, and the relation $B_0=A_0-m_{3/2}$ does not hold either.
\vskip3ex
\noindent mSUGRA has the following attractive features:
\begin{itemize}
\item Supergravity models provide a natural framework for supersymmetry breaking: a locally supersymmetric Lagrangian automatically contains terms that can mediate supersymmetry breaking.
\item mSUGRA has great predictive power, since it has only five parameters. Apart from $m_0$, $M_{1/2}$, $A_0$ there are two additional parameters: $\tan{\beta} \equiv \langle H_u^0\rangle/\langle H_d^0\rangle$ and the sign of $\mu$.
\end{itemize}
The disadvantages of general supergravity models are:
\begin{itemize}
\item They must necessarily appeal to Planck scale physics, which is still poorly understood.
\item Though gravity is flavour blind, the supergravity invariance of the Lagrangian cannot prevent the occurrence of (Planck scale suppressed) flavour mixing operators that correspond to tree level interactions between hidden sector fields and observable sector fields. In order to suppress sparticle induced FCNC processes, one must resort to additional generation symmetries.
\end{itemize}

\subsection{Anomaly mediated supersymmetry breaking}\label{s:AMSB}
In some models of supergravity, the visible and hidden sectors are physically separated by extra dimensions \cite{AMSB1,AMSB2}.
In these `braneworld' scenarios, often inspired by string theory, our four-dimensional world is embedded in a higher-dimensional bulk that has additional spatial dimensions, which are curled up.

\begin{figure}[t]
\begin{fmffile}{amsb}
\begin{minipage}{\textwidth}
\begin{center}
  \begin{fmfgraph*}(100,40)
    \fmfforce{(0,0)}{lbrane1}
    \fmfforce{(.2w,.05h)}{lbrane2}
    \fmfforce{(.2w,.9h)}{lbrane3}
    \fmfforce{(0,.85h)}{lbrane4}
      \fmf{plain}{lbrane1,lbrane2,lbrane3,lbrane4,lbrane1}
    \fmfforce{(.17w,.95h)}{lbrane}
      \fmfv{label=3-brane}{lbrane}
    \fmfforce{(.8w,0)}{rbrane1}
    \fmfforce{(w,.05h)}{rbrane2}
    \fmfforce{(w,.9h)}{rbrane3}
    \fmfforce{(.8w,.85h)}{rbrane4}
      \fmf{plain}{rbrane1,rbrane2,rbrane3,rbrane4,rbrane1}
    \fmfforce{(.82w,.95h)}{rbrane}
      \fmfv{label=3-brane}{rbrane}
    \fmfforce{(.17w,.5h)}{mssm}
      \fmfv{label=MSSM}{mssm}
    \fmfforce{(.2w,.42h)}{superfields}
      \fmfv{label=superfields}{superfields}
    \fmfforce{(.82w,.5h)}{hidden}
      \fmfv{label=hidden}{hidden}
    \fmfforce{(.83w,.42h)}{sector}
      \fmfv{label=sector}{sector}
    \fmfforce{(.3w,.5h)}{bulk}
      \fmfv{label=higher-dimensional bulk,label.angle=0}{bulk}
  \end{fmfgraph*}
\end{center}
\end{minipage}
\end{fmffile}
\caption{Schematic picture of Anomaly Mediated Supersymmetry Breaking (AMSB). One assumes that there exist additional spatial dimensions, which are curled up. The visible sector is confined to a three-brane; the hidden sector is confined to another, parallel three-brane. These branes are separated by a higher-dimensional bulk. Only gravity (and maybe new heavy fields) propagates in the bulk. The size of the MSSM soft terms is determined by the anomalous violation of a rescaling symmetry called \emph{superconformal invariance}.}\label{f:AMSB}
\end{figure}
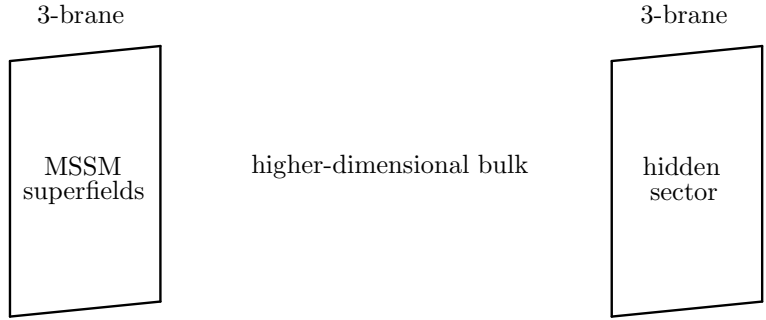

The general idea is that the MSSM fields and the hidden sector fields are confined to parallel, distinct \emph{three-branes} (i.e.\ spacelike hypersurfaces), separated by a distance $r$.
Only the gravity supermultiplet (and possibly new heavy fields) resides in the bulk.
In this scenario every flavour violating term that plagues supergravity, caused by tree level couplings with a bulk field of mass $m$, is suppressed by a factor $e^{-mr}$.
Provided that $r$ is large enough, the flavour violating effects are exponentially suppressed without requiring any fine-tuning.
This class of models is called Anomaly Mediated Supersymmetry Breaking (AMSB), because the size of the soft supersymmetry breaking terms is determined by the loop induced superconformal (Weyl) anomaly \cite{Weylanomaly} (figure \ref{f:AMSB}).
Local superconformal invariance is a rescaling symmetry that is violated at the quantum level.

Anomaly mediated terms are always present in supergravity, but are loop-suppressed with respect to the gravitino mass and hence subleading order contributions to the soft masses.
AMSB is the scenario where there are no supergravity contributions at tree level, so that the anomaly mediated terms become the dominant ones.
At the GUT scale, the soft terms have the following values:
\begin{IEEEeqnarray}{rCl}\label{eq:AMSBboundaryconditions} 
M_a &=& \frac{1}{16\pi^2}b_ag_a^2m_{3/2} \qquad (a=1,2,3)\IEEEyessubnumber\\
m_{ij}^2 &=& \frac12 \dot{\gamma}_{ij}m_{3/2}^2 \IEEEyessubnumber\\
A_{ijk} &=& -\sum_m\left(y_{mjk}\gamma_{im} + y_{imk}\gamma_{jm} + y_{ijm}\gamma_{km}\right)m_{3/2} \IEEEyessubnumber
\end{IEEEeqnarray}
where $b_a = (\frac{33}{5},1,-3)$, $g_a$ are the gauge couplings with $g_1=g'\sqrt{5/3}$, $y_{ijk}$ are the Yukawa couplings, $\gamma_{ij}$ are anomalous dimensions (see appendix \ref{a:anomalousdimensions} for their definition and explicit expressions) and $\dot{\gamma}\equiv d\gamma/d\ln{\mu}$.
\vskip3ex
\noindent Anomaly mediation has the following attractive properties:
\begin{itemize}
\item AMSB naturally conserves flavour, so that the MSSM soft terms introduce no new FCNC amplitudes. In contrast, supergravity requires additional flavour symmetries to accomplish flavour conservation.
\item A CP-preserving supersymmetry breaking sector could be natural in this framework.
\end{itemize}
AMSB also has serious problems:
\begin{itemize}
\item Pure anomaly mediation leads to tachyonic sleptons, i.e.\ their squared soft masses become negative. This would cause them to acquire non-zero VEVs and break electric charge conservation.
\end{itemize}

\subsubsection*{Minimal anomaly mediation}
The minimal AMSB (mAMSB) model uses a phenomenological approach to tackle the tachyonic slepton problem: a universal, non-anomaly-mediated contribution $m_0^2$ is added to the soft scalar masses at the GUT scale.
The origin of these terms may be for example additional fields in the bulk, but in mAMSB $m_0$ is simply considered as a parameter of the model.

\subsection{General gauge mediation}\label{s:GGM}
\begin{figure}[t]
    \begin{minipage}{\textwidth}
      \begin{center}
      \begin{fmffile}{gmsb}
      \fmfcmd{%new commands for sparticle lines
      vardef cross_bar (expr p, len, ang) =
       ((-len/2,0)--(len/2,0))
       rotated (ang + angle direction length(p)/2 of p)
       shifted point length(p)/2 of p
      enddef;
      style_def dashes_crossed expr p=
       save dpp;
       numeric dpp;
       dpp = ceiling (pixlen (p, 10) / dash_len) / length p;
       for k=0 upto dpp*length(p) - 1:
        cdraw point k/dpp of p ..
         point (k+.5)/dpp of p;
       endfor
       ccutdraw cross_bar (p, 3mm, 45);
       ccutdraw cross_bar (p, 3mm, -45);
      enddef;}
      \begin{fmfgraph*}(70,20)
        \fmfleft{i1,i2}
        \fmfright{o1,o2}
        \fmf{dashes}{i1,v1,v2,o1}
        \fmffreeze
        \fmflabel{$\tilde{\text{u}}$}{i1}
        \fmflabel{$\tilde{\text{u}}$}{o1}
        \fmf{phantom}{i2,w1,w2,o2}
        \fmf{wiggly,tension=.5,label=$B$,label.side=left}{v1,w1}
        \fmf{wiggly,tension=.5,label=$B$}{v2,w2}
        \fmffreeze
        \fmf{dashes_crossed,width=2,left=.5,label=$\langle F_X\rangle$}{w1,w2,w1}
        \fmfforce{(.39w,.65h)}{messenger}
          \fmfv{label=$\phi_M$,label.angle=0}{messenger}
      \end{fmfgraph*}
      \end{fmffile}
      \end{center}
    \end{minipage}
  \caption{Contribution to the soft squared mass of the up squark in models of Gauge Mediated Supersymmetry Breaking (GMSB). The auxiliary component of the spurion field $X$ obtains a supersymmetry breaking VEV $\langle F_X\rangle$. The up squark only couples indirectly to this VEV: the scalar component $\phi_M$ of a messenger supermultiplet couples at tree-level to the spurion and through the MSSM gauge fields (in this diagram the $B$ boson) with the MSSM. When the messengers are integrated out from the action, this diagram contributes to the soft up squark mass.}\label{f:GMSB}
\end{figure}
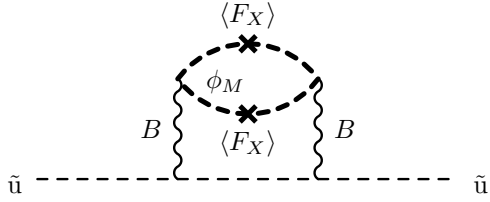

Several models of Gauge Mediated Supersymmetry Breaking (GMSB) have been proposed in the literature (see \cite{GiudiceGMSB} for a review).
Many of these models include a field $X$, called the \emph{spurion}, that acquires a supersymmetry breaking VEV, and a set of weakly coupled fields that are charged under the MSSM.
The latter are called \emph{messenger} fields since they communicate supersymmetry breaking to the MSSM fields: they interact at tree-level with the spurion and through the MSSM gauge fields with the MSSM (see figure \ref{f:GMSB}).

Recently, the framework of General Gauge Mediation (GGM) \cite{GGM1,GGM2} has been proposed to unify all earlier descriptions of GMSB.
It describes the effects of an arbitrary hidden sector on the MSSM.
It starts from the following definition of gauge mediation: \emph{In the limit of vanishing gauge couplings, the theory decouples into the MSSM and a separate, supersymmetry breaking hidden sector}.
For example, the setup described above fits into this definition by taking the messenger and spurion fields as the hidden sector.

In the GGM framework, all MSSM soft terms can be described in terms of a small number of correlation functions involving hidden sector currents.
Essentially, the GGM framework parametrises the effects of the hidden sector on the MSSM.
By constructing the effective Lagrangian, the following soft mass formulae are found:\footnote{For future convenience, a factor $M$ (the messenger scale) has been absorbed into the definition of the $B_a$, cf.\ \cite{Carena2}.}
\begin{IEEEeqnarray}{rCl}
M_a &=& g_a^2 B_a \qquad (a=1,2,3) \IEEEyessubnumber\\
m_{i}^2 &=& g_1^2 Y_i\zeta + \sum_{a=1}^3 g_a^4 C_a(i) A_a \IEEEyessubnumber
\end{IEEEeqnarray}
Here $B_a,\zeta,A_a$ are expressions involving the hidden sector current correlation functions; $Y_i$ is the hypercharge of the scalar field $\Phi_i$ and $C_a(i)$ is the quadratic Casimir (see appendix \ref{a:anomalousdimensions}) of the representation of $\Phi_i$ under the gauge group labeled by $a$.
The term containing $\zeta$ comes from an effective Fayet-Illiopoulos term.
Usually a $\mathbf{Z}_2$ symmetry of the hidden sector is assumed in order to forbid such a term, since it would lead to tachyonic sleptons.
The above conditions are the matching conditions at the scale where we integrate out the hidden sector.
The seven numbers $\zeta,A_a,B_a$ contain information on the hidden sector, but are regarded as parameters of the low energy EFT that we call the MSSM.

The GGM framework does not allow for additional interactions that could generate $\mu$ and $b$ radiatively; that would require interactions between the MSSM and the hidden sector that remain in the limit of vanishing gauge couplings.
The framework would have to be extended to allow for such couplings.
To parameterise the effects of such an extension, additional contributions $\delta_u,\delta_d$ to $m_{H_u}^2,m_{H_d}^2$ are often added.
\vskip3ex
\noindent GGM has the following attractive features:
\begin{itemize}
\item Soft masses are automatically flavour universal, since the gauge interactions are flavour blind.
Thus when the hidden sector is integrated out, this will lead to the same soft masses for each generation\footnote{It is assumed that the messenger scale is well below the Planck scale, so that gravity contributions to the soft masses are small enough not to reintroduce the flavour problem.} (note however that slepton and squark soft masses need not be the same).
\item Since gravity is not incorporated in GGM, the model can be solved using only field-theoretical tools, without facing our present difficulties in treating quantum gravity.
\end{itemize}

The problems of GGM are:
\begin{itemize}
\item The absence of new CP-violating phases is not automatic.
\item The Higgs $\mu$ and $b$ parameters are not generated in pure gauge mediation, so typically additional interactions are assumed to be present.
\item In many gauge mediation models, gaugino masses turn out to be very small compared to the scalar masses. In that case we cannot have both scalar and gaugino masses of $\ord{1}$ TeV at the same time. This may reintroduce the hierarchy problem, see also section \ref{s:brokensusyconstraints}.
\item Gravity is still absent from the theory. From the theoretician's point of view, this may be a disadvantage since GGM only delays the complete unification of all forces.
\end{itemize}

\subsubsection*{Minimal gauge mediation}
Minimal gauge mediation (MGM) is a GGM model that is restricted to a subset of the GGM parameter space, defined by the following constraints:
\begin{IEEEeqnarray}{rCl}
A_1 = A_2 &=& A_3 \equiv A \IEEEyessubnumber\\
B_1 = B_2 &=& B_3 \equiv B \IEEEyessubnumber\\
A &=& 2B^2 \IEEEyessubnumber
\end{IEEEeqnarray}
The Fayet-Iliopoulos term corresponding to $\zeta$ is taken to be zero.
Additional non-gauge contributions $\delta_u$, $\delta_d$ are added to the soft Higgs masses.
Then the expressions for the soft masses become:
\begin{IEEEeqnarray}{rCl}\label{eq:MGMspectrum}
M_a &=& g_a^2B_a \qquad (a=1,2,3) \IEEEyessubnumber \label{eq:MGMgauginos}\\
m_i^2 &=& 2B^2\sum_{a=1}^3 g_a^4C_a(i) \IEEEyessubnumber \label{eq:MGMsfermions}\\
\mhu &=& 2B^2\sum_{a=1}^3 g_a^4C_a(H_u) + \delta_u \IEEEyessubnumber \\
\mhd &=& 2B^2\sum_{a=1}^3 g_a^4C_a(H_d) + \delta_d \IEEEyessubnumber
\end{IEEEeqnarray}
where $m_i^2$ denotes only the squark and slepton masses.

\subsection{Mirage mediation}\label{s:MM}
Rather than restricting oneself to one of the three known mechanisms for radiative supersymmetry breaking (gravity, anomaly or gauge mediation), one could solve the problems of particular models by choosing two (or more) mechanisms and combining the best of both worlds.
For example, one might tackle the tachyonic slepton problem of anomaly mediation by combining it with gauge mediation (see e.g.\ \cite{gauge+anomalymediation}).

Mirage mediation \cite{mirage} is one such scenario in which gravity mediated and anomaly mediated soft terms have comparable contributions.
In this scenario, the gravity mediated terms are suppressed by a relative factor $\log\left(M_\text{pl}/m_{3/2}\right)$, which is numerically of the order of a loop factor.
This results in mirage unification: the gaugino and scalar masses unify at a scale far below the scale where the soft masses are generated.
This \emph{mirage messenger scale} does not correspond to any physical threshold, hence the name.

This class of phenomenological models are based on a class of string models with stabilised moduli, called the KKLT construction.
It solves the tachyonic slepton problem that arises in pure anomaly mediation and has reduced low energy fine-tuning \cite{miragefinetuning}.

\subsubsection*{Deflected mirage mediation}
Deflected mirage mediation \cite{deflectedmirage1,deflectedmirage2} is a scenario in which gravity mediated, gauge mediated and anomaly mediated soft terms all have comparable contributions.
The name refers to the fact that the gauge mediation contribution introduces threshold effects from the messenger fields that deflect the renormalisation group trajectories, so that the gaugino mass unification scale is deflected.
This also solves the tachyonic slepton problem that occurs in anomaly mediation.

The pattern of soft masses depends on the ratios of the contributions from the three different mediation mechanisms.
These ratios are taken to be parameters of the model.
To find general, model-independent gaugino mass formulae, three contributions to the soft masses are assumed.
The contribution from Planck-suppressed operators (i.e.\ gravity mediation) is assumed to be universal in form with a mass scale $M_0$.
The anomaly mediation contribution is proportional to some mass scale $M_g$, which is \emph{a priori} different.
It is assumed that these contributions arise at some high energy scale $\mu_\text{UV}$, where supersymmetry breaking is transmitted from some hidden sector to the observable sector.\footnote{It is common to take $\mu_\text{UV}=M_\text{GUT}$, but in string constructions it might be a different (possibly higher) scale where the supergravity approximation for the effective Lagrangian becomes valid.}
Finally, a gauge mediation contribution is assumed to be proportional to a third mass scale $\Lambda_\text{mess}$.
The messenger sector is assumed to come in complete GUT representations in order to preserve gauge coupling unification.
In particular $N_m$ copies of $\mathbf{5},\bar{\mathbf{5}}$ representations under $SU(5)$ are assumed to exist, which contribute to the soft gaugino masses at an energy scale $\mu_\text{mess}<\mu_\text{UV}$.

At the high scale $\mu_\text{UV}$, the gaugino mass boundary conditions have the simple form:
\begin{equation}
M_a\left(\mu_\text{UV}\right) = M_0 + g_a^2\left(\mu_\text{UV}\right)\frac{b_a'}{16\pi^2}m_{3/2} \qquad (a=1,2,3)
\end{equation}
where $m_{3/2}$ is the gravitino mass and $b_a'=b_a+N_m$ are the $\beta$-function coefficients for the MSSM gauge groups above the messenger scale.
Later on we will need the boundary conditions for the RG equations of the MSSM, so we are interested in the boundary conditions at $\mu_\text{mess}$.
Evolving the soft masses to $\mu_\text{mess}$ using one-loop RG equations, we get the following boundary conditions \cite{DMMgauginomasses}:\footnote{Here it is understood that we consider the soft masses at an infinitesimal value below the messenger scale, so that we are in the region where the MSSM RG equations are valid.}
\begin{equation}
M_a = g_a^2\frac{b_a'}{16\pi^2}m_{3/2} + M_0\left[1-g_a^2\frac{b_a'}{8\pi^2}\log\left(\frac{\mu_\text{UV}}{\mu_\text{mess}}\right)\right] + \Delta M_a \qquad (a=1,2,3)
\end{equation}
where
\begin{equation}
\Delta M_a = -N_m\frac{g_a^2}{16\pi^2}\left(\Lambda_\text{mess}+m_{3/2}\right)
\end{equation}
is a threshold contribution that arises when we integrate out the messenger fields.
The expressions for the soft sfermion masses are more complicated and not particularly enlightening.

\subsection{Puzzles \& problems}
As mentioned at the beginning of this chapter, the literature offers a plethora of models that try to explain supersymmetry breaking.
We have seen that three mechanisms for mediating spontaneous supersymmetry breaking underlie these models.
A specific model may restrict to one of these mechanisms or combine several of them.
Given a choice of mechanism(s), specific models may differ from each other in the specification of hidden sector superfields, potentials, additional symmetries, topology of additional spacetime dimensions and so on.
Apart from the constraints we mentioned in section \ref{s:brokensusyconstraints}, there are several other puzzles (often concerning naturalness) that any realistic model should address.
Below some of these problems are listed.

\subsubsection*{$\mu$ problem}
The $\mu$ problem is a problem concerned with the naturalness of the MSSM parameters.
Consider the conditions for electroweak symmetry breaking:\footnote{These can be derived by demanding that (a) the Higgs scalar potential of the MSSM is bounded from below and allows non-zero VEVs for $H_u,H_d$ and (b) the VEVs $v_u\equiv\langle H_u^0\rangle,v_d\equiv\langle H_d^0\rangle$ are compatible with the condition $v_u^2+v_d^2=2M_Z^2/(g'^2+g_2^2)$. For a derivation see e.g.\ \cite{martin} section 7.1.}
\begin{IEEEeqnarray}{rCl}\label{eq:EWSBconditions}
\sin{(2\beta)} &=& \frac{2b}{m_{H_u}^2+m_{H_d}^2+2|\mu|^2} \IEEEyessubnumber\\
M_Z^2 &=& \frac{|m_{H_d}^2-m_{H_u}^2|}{\sqrt{1-\sin^2(2\beta)}} - m_{H_u}^2-m_{H_d}^2-2|\mu|^2 \IEEEyessubnumber\label{eq:EWSB2}
\end{IEEEeqnarray}
where $\beta$ is defined by $\tan{\beta}\equiv \langle H_u^0\rangle/\langle H_d^0\rangle$ and $M_Z$ is the $Z$ boson mass.
In order to avoid fine-tuning, all input parameters ought to be within an order of magnitude or two of $M_Z^2$.
However, $\mu$ is a supersymmetry respecting parameter appearing in the superpotential, whereas $b,m_{H_u}^2,m_{H_d}^2$ are supersymmetry breaking parameters.
This leads to a naturalness puzzle: why are these parameters with different physical origins of the same order of magnitude?
And why are they so much smaller than the natural cutoff scale $M_\text{Pl}$?

Several different solutions to the $\mu$ problem have been proposed, which all work in roughly the same way.
The $\mu$ parameter is assumed to be absent at tree level before spontaneous supersymmetry breaking.
Then it arises from the VEV(s) of some new field(s), which are in turn determined by minimising a potential that depends on soft supersymmetry breaking terms.
This way, the value of the effective parameter $\mu$ is no longer conceptually distinct from the mechanism of supersymmetry breaking.
Hence if we can explain why $m_\text{soft}\ll M_\text{Pl}$, we can also understand why $\mu$ is of the same order.

\subsubsection*{Little hierarchy problem}
The little hierarchy problem is a `smaller' version of the hierarchy problem.
It appears as a tension between two roles played by the top squark (see e.g.\ \cite{littlehierarchy1,littlehierarchy2,littlehierarchy3}):
\begin{itemize}
\item On the one hand, the top squark must be relatively heavy. The reason is that at tree-level the mass $m_{h^0}^2$ of the lightest Higgs boson is bounded by the $Z$ boson mass $M_Z$. However, this is in conflict with the current experimental lower limit on the Higgs mass. Therefore there must be significant quantum corrections to $m_{h^0}^2$. Loops involving the top squark generate such corrections, but then the top squark must be relatively heavy.
\item On the other hand, the top squark must be not too heavy. Loops involving the top squark also give quantum corrections to the soft mass squared $m_{H_u}^2$. If the top squark is too heavy, these corrections become too large and must be cancelled to within a few percent against the $\mu$ term in \eqref{eq:EWSB2}.
\end{itemize}
An attempt to balance these two effects results in fine-tuning.

\section{How to probe the high scale}\label{s:probinghighscales}
If supersymmetric particles are discovered at the LHC, one of the main puzzles will be to determine how supersymmetry has been broken.
In the last section we have seen that in common models, supersymmetry is broken spontaneously at a very high energy scale, which is the threshold where new physics enters the theory.
This scale may be beyond experimental access in the foreseeable future.
However, the Renormalisation Group provides a useful tool to probe physics at this scale (see also figure \ref{f:EFTtravelling}).
If we measure all running parameters in the energy regime where the MSSM\footnote{Of course the EFT corresponding to Nature need not be the MSSM; it might be some other supersymmetric theory. The arguments made in the rest of this thesis rest on the assumptions that (a) we have discovered supersymmetry and (b) we have determined the appropriate EFT with its particle content, interactions and RG equations. For definiteness and simplicity, we assume this EFT to be the MSSM. The analysis in the rest of this thesis may be applied to extensions of the MSSM as well; for relevant RG equations and RG invariants of some extensions, see \cite{Demir}.} is an appropriate EFT, we can evolve them towards the new physics threshold using the known RG equations.
Then we can compare the parameter values with the matching equations predicted by the supersymmetry breaking models.

The RG equations of the MSSM (Appendix \ref{a:mssmrges}) are coupled differential equations that cannot be solved analytically.
Therefore any analysis that involves RG evolution has to be done numerically.
Several approaches to the above scheme have been studied in the literature.
First we will discuss the two most widely studied methods, which are known as the \emph{top-down} and the \emph{bottom-up} method.
Then we will examine another approach that has gained some interest recently.
This method uses Renormalisation Group invariants, which are the main subject of this thesis.
We will see what they are, how they can be found and how they can be used to study high scale physics.

\subsection{Top-down method}
The top-down method (see e.g.\ \cite{topdown1,topdown2,topdown3}) is the most widely studied method in the literature.
It is called this way because a top-down study is started from the high scale (e.g.\ the GUT scale) and the theory is evolved down to the collider scale ($\ord{1\text{ TeV}}$).
One starts by choosing a model with few parameters; the most popular model is mSUGRA because it has only five parameters.
Then one procedes as follows:
\begin{itemize}
\item Pick a point in the parameter space of the model and translate this into values of the running parameters at the high scale (masses, gauge and Yukawa couplings, etc.).
\item Evolve the running parameters down to the collider scale using the Renormalisation Group equations.
\item Using the resulting parameter values, perform a detector simulation to calculate relevant branching ratios and cross sections.
\item Compare the results to experimental data and extract constraints on the parameter space of the supersymmetry breaking model.
\end{itemize}
The top-down method is suitable for making general predictions of supersymmetric phenomenology.
It helps us to recognise supersymmetry when it appears in the LHC: the top-down method is used to find collider signatures that are characteristic for supersymmetry.
However, for the purpose of determining the supersymmetry breaking mechanism, this method has some serious limitations:
\begin{itemize}
\item The supersymmetry breaking scenario must be assumed beforehand. With the top-down method one determines the regions in parameter space that are consistent with the data. If only small portions of the total parameter space seem phenomenologically viable, one might conclude that the model is neither likely to be correct nor natural. However, it seems unlikely that we can strictly exclude models this way.
\item Scanning the entire parameter space is very time consuming. In order to scan the parameter space, one would want to use a reasonably fine grid and check each point separately. But the parameter space is usually too big to perform a full detector simulation for each point. For general predictions of supersymmetric phenomenology, one usually resorts to using a limited set of benchmark points, because many points in parameter space have a very similar phenomenology (see e.g.\ \cite{Irene}). But for the purpose of excluding a certain supersymmetry breaking mechanism, this is no satisfying solution.
\item Fitting the numerical results to the experimental data becomes much more difficult as the number of model parameters is increased. Therefore one always limits oneself to a model with few parameters. But there is no reason to think that Nature would restrict itself to (say) only five parameters in the next EFT beyond the MSSM.
\end{itemize}

\subsection{Bottom-up method}
The bottom-up method (see e.g.\ \cite{bottomup1,bottomup2,bottomup3}), as its name suggests, works by evolving the theory upwards from the collider scale to the new physics threshold.
A bottom-up analysis works as follows:
\begin{itemize}
\item Convert experimental data into the running parameters at the collider scale.
\item Using the Renormalisation Group equations, evolve them towards the scale where new physics presumably comes into play.
\item Analyse the structure of the high scale parameters: do they fit the matching conditions predicted by any model?
\end{itemize}
This method seems more suitable for determining the supersymmetry breaking mechanism than the top-down method, because one does not have to assume a breaking mechanism beforehand.
Also, there is no practical need to only consider models with few parameters.
Furthermore, no time consuming scanning of the parameter space is involved.
However, the bottom-up method has its own severe limitations:
\begin{itemize}
\item The running parameters at the collider scale will come with experimental errors. To determine the uncertainty in these parameters at a higher scale, we also have to evolve the error bars. These may become larger under RG flow. This makes it difficult to tell for example whether certain parameters unify or not.
\item We do not know the value of the high scale that should be taken as the new physics threshold. This scale has to be guessed; in practice, the GUT scale is often chosen. But the real scale of new physics may be higher if the GUT scale does not correspond to a threshold (cf.\ mirage mediation, section \ref{s:MM}). Also, there may be an intermediate new physics scale even though no unification occurs there. In both cases, we would extract incorrect boundary conditions for matching with the underlying EFT.
\item Because the RG equations are coupled, all running parameters must be known. If we fail to measure (say) one sparticle mass, the bottom-up method cannot be used.
\end{itemize}

\subsection{Renormalisation Group invariants}\label{s:RGIs}
Recently, a third approach to probing the high scale has gained some interest \cite{Demir,Carena1,Carena2,Jaeckel}.
This method makes clever use of quantities called \emph{Renormalisation Group invariants} (RGIs).
In this section, we will see what RGIs are and illustrate their usefulness with a test for gauge coupling unification.
Then we will list all one-loop RGIs for the MSSM; for a derivation see \cite{Carena1} and appendix~\ref{a:derivingRGIs}.
After that, we will see how RGIs have been used in the literature and consider a new strategy for using them; this is the main subject of this thesis.
The advantages and limitations of the RGI method will be discussed in section \ref{s:discussion}, since they will be easier to identify after we have seen RGIs in action.

\paragraph{RGIs and their use}
As the name suggests, Renormalisation Group invariants are quantities that are invariant under Renormalisation Group flow.
More specifically, they are combinations of running parameters chosen in such a way that they are independent of the renormalisation scale $\mu$.
A well-known example of an RGI is the following combination of the gauge couplings $g_1,g_2,g_3$:
\begin{equation}
I_g \equiv \left(b_2-b_3\right)g_1^{-2} + \left(b_3-b_1\right)g_2^{-2} + \left(b_1-b_2\right)g_3^{-2} \label{eq:gaugecouplingRGI}
\end{equation}
where the gauge couplings satisfy the following Renormalisation Group equations at the one-loop level (see also appendix \ref{a:mssmrges}):
\begin{equation}
16\pi^2\frac{\d g_a}{\d t} = b_ag_a^3 \label{eq:gaugecouplingRGEs}
\end{equation}
Here $t\equiv\ln{\left(\mu/\mu_0\right)}$ is a rescaled version of the energy scale $\mu$, where $\mu_0$ is a reference scale that makes the argument of the logarithm dimensionless; its value is arbitrary since it drops out of the RG equations.
The coefficients $b_a$ are constants depending on the particle content of the model; for the MSSM their values are $b_a=(\frac{33}{5},1,-3)$, whereas for the SM they are $b_a=(\frac{41}{10},-\frac{19}{6},-7)$.
It is easy to verify that $I_g$ is RG invariant after rewriting \eqref{eq:gaugecouplingRGEs} as:
\begin{equation}
16\pi^2\frac{\d}{\d t}g_a^{-2} = -2b_a
\end{equation}
Then it follows immediately that:
\begin{equation}
16\pi^2\frac{\d}{\d t}I_g = \left(b_2-b_3\right)(-2b_1) + \left(b_3-b_1\right)(-2b_2) + \left(b_1-b_2\right)(-2b_3) = 0
\end{equation}
so $I_g$ is indeed independent of the renormalisation scale.
Note that this quantity is not exactly RG invariant, since we used the one-loop RG equations to construct it.
We will come back to this issue in section \ref{s:discussion}.

How could we use RG invariant quantities?
A crucial observation is that if we measure their values at the collider scale, \emph{we will immediately know their values at the threshold of new physics}.
This fact allows us to probe physics beyond the MSSM without having to evolve all parameters!
We only have to find RGIs that enable us to test predictions about the matching conditions at the new physics threshold.

As an example, $I_g$ has been tailor-made to test gauge coupling unification.
Note that if the gauge couplings $g_1,g_2,g_3$ unify, i.e.\ they have the same value $g_U$ at \emph{some} energy scale, then $I_g$ will vanish.
But since $I_g$ is RG invariant, it has to vanish at \emph{every} scale where the RG equations of the MSSM are valid.
Hence, if we measure the gauge couplings at a scale where the MSSM is an appropriate EFT (say, at $\ord{1\text{ TeV}}$), we should insert their values into \eqref{eq:gaugecouplingRGI}.
Then we will immediately see whether the hypothesis of gauge coupling unification is compatible with experiment or not.\footnote{If we do this for the SM, using the measured gauge couplings at $\mu=M_Z$, we find no compatibility with unification. This can also be seen in figure~\ref{f:gaugeunification}.}
Note that we have not specified at which scale the couplings unify; we needed only the hypothesis that they unify \emph{at all}.

To summarise, RGIs provide a tool for probing matching conditions at high energy scales.
They allow us to circumvent the need to evolve the running parameters numerically; we do not even need to know exactly at which energy scale new physics arises.
The only task that rests us is to find suitable RG invariants for testing predictions from supersymmetry breaking models.
A logical first step in this direction is to find all RGIs of the MSSM.

\paragraph{One-loop RGIs for the MSSM}
It should be noted that if we have a set of RGIs, then any function of those RGIs will also be RG invariant.
Therefore in order to find all RGIs, one should look for a maximal set of \emph{independent} RGIs, i.e.\ invariants that cannot be expressed in terms of each other.\footnote{It is tempting to call this a `basis of RGIs', as in \cite{Carena1}. Note however that it is not the same as a basis of a vector space. One should keep in mind that once we have found such a set, we are not restricted to making linear combinations of them, but can also take products, quotients and so on.}
Recently an almost complete list of independent one-loop RGIs for the MSSM was derived in \cite{Carena1} (there exist two more, see below).
They have been derived under the following assumptions:
\begin{itemize}
\item The soft sfermion masses are flavour diagonal and the first and second generation masses are degenerate. This assumption is motivated by low-energy flavour mixing constraints, such as those in section \ref{s:brokensusyconstraints}.
\item Similarly, there are no sources of CP-violation in the soft supersymmetry breaking sector beyond those induced by the Yukawa couplings.
\item First and second generation Yukawa and trilinear couplings are neglected, because they give very small contributions to the evolution of the soft supersymmetry breaking parameters. These contributions are smaller than the two-loop corrections associated with the gauge couplings and third generation Yukawa couplings.
\item The right-handed neutrino (if it exists) effectively decouples from the spectrum and the RG equations.
\end{itemize}
Under these assumptions, there are 12 soft scalar masses left to consider.
We denote them as $\msqa,\msqc,\msua,\msuc,\msda,\msdc,\msla,\mslc,\msea,\msec,\mhu,\mhd$ in accordance with the notation in table \ref{t:mssmchiralcontent}; the subscripts 1 and 3 refer to the first and the third generation respectively.
Furthermore, we have three gauge couplings $g_1,g_2,g_3$ and corresponding gaugino masses $M_1,M_2,M_3$, three third-generation Yukawa couplings $y_t,y_b,y_\tau$, three soft trilinear couplings $A_t$, $A_b$, $A_\tau$ and the parameters $\mu,b$.
Their corresponding $\beta$-functions can be found in appendix \ref{a:mssmrges}.
The one-loop RGIs for the MSSM are listed in table \ref{t:mssmrgis}; they are derived in appendix \ref{a:derivingRGIs}.
Note that apart from those in table \ref{t:mssmrgis}, there exist two additional RGIs; they have been found in \cite{Demir}:
\begin{IEEEeqnarray}{rCl}\label{eq:additionalRGIs}
I_2 &=& \mu\left[\frac{g_2^9g_3^{256/3}}{y_t^{27}y_b^{21}y_\tau^{10}g_1^{73/33}}\right]^{1/61} \label{eq:rgiI2} \\
I_4 &=& B - \frac{27}{61}A_t -\frac{21}{61}A_b -\frac{10}{61}A_\tau -\frac{256}{183}M_3 -\frac{9}{61}M_2 +\frac{73}{2013}M_1 \label{eq:rgiI4}
\end{IEEEeqnarray}
where $B=b/\mu$.
These RGIs have not been listed in table \ref{t:mssmrgis} because they will not be relevant to our analysis.
We will explain why at the end of this section.

Any other (one-loop) RGI we can think of can be written in terms of those in table \ref{t:mssmrgis}.
For example, the RGI in the above example can be written as:
\begin{align}
I_g &= 4g_1^{-2} - \frac{48}{5}g_2^{-2} + \frac{28}{5}g_3^{-3} \nonumber\\
&= \frac{16}{11}I_{g_2} + \frac{28}{11}I_{g_3}
\end{align}

\begin{table}[p]
{\renewcommand{\arraystretch}{3}
\begin{tabular}{|c||c|}
\hline
\textbf{Invariant}	& \textbf{Definition} \\
\hline\hline
$D_{B_{13}}$	& $2\left(\msqa-\msqc\right)-\msua+\msuc-\msda+\msdc$	\\\hline
$D_{L_{13}}$	& $2\left(\msla-\mslc\right)-\msea+\msec$	\\\hline
$D_{\chi_1}$	& $3\left(3\msda-2\left(\msqa-\msla\right)-\msua\right)-\msea$	\\\hline
\multirow{2}{*}{$D_{Y_{13H}}$}	& $\msqa-2\msua+\msda-\msla+\msea$	\\
	& $-\frac{10}{13}\left(\msqc-2\msuc+\msdc-\mslc+\msec+\mhu-\mhd\right)$	\\\hline
$D_Z$		& $3\left(\msdc-\msda\right)+2\left(\mslc-\mhd\right)$	\\\hline
$I_{Y_\alpha}$	& $\frac1{g_1^2}\left(\mhu-\mhd+\sum_{gen}\left(\msq-2\msu+\msd-\msl+\mse\right)\right)$	\\\hline
$I_{B_a}$	& \Large{$\frac{M_a}{g_a^2}$}	\\\hline
$I_{M_1}$	& $M_1^2-\frac{33}8\left(\msda-\msua-\msea\right)$	\\\hline
$I_{M_2}$	& $M_2^2+\frac1{24}\left(9\left(\msda-\msua\right)+16\msla-\msea\right)$	\\\hline
$I_{M_3}$	& $M_3^2-\frac3{16}\left(5\msda+\msua-\msea\right)$	\\\hline
$I_{g_2}$	& \Large{$\frac1{g_1^2}-\frac{33}{5g_2^2}$}	\\\hline
$I_{g_3}$	& \Large{$\frac1{g_1^2}+\frac{11}{5g_3^2}$}	\\\hline
\end{tabular}}
\caption{One-loop Renormalisation Group invariants for the MSSM. The sum in $I_{Y_\alpha}$ runs over the three sfermion generations.}\label{t:mssmrgis}
\end{table}

\paragraph{RGIs in the literature}
As we have seen, RGIs provide a new tool to test predictions about high scale physics, such as gauge coupling unification.
The trick is to find sum rules for high scale physics that can be written in terms of RGIs.
In the literature, several such sum rules can be found.

Consider for example minimal gauge mediation, which has been studied in the context of RGIs in \cite{Carena2}.
If one inserts the spectrum \eqref{eq:MGMspectrum} at the messenger scale into the RGI expressions, one immediately finds that $D_{B_{13}} = D_{L_{13}} = D_{\chi_1} = 0$.
In terms of the model parameters of MGM, the non-vanishing RGIs are:
\begin{IEEEeqnarray}{rCl}
D_{Y_{13H}} &=& -\frac{10}{13}\left(\delta_u-\delta_d\right) \IEEEyessubnumber\label{eq:mgmrgi1}\\
D_Z &=& -2\delta_d \IEEEyessubnumber\label{eq:mgmrgi2}\\
I_{Y_\alpha} &=& g_1^{-2}(M)\left(\delta_u-\delta_d\right) \IEEEyessubnumber\label{eq:mgmrgi3}\\
I_{B_1} &=& B \IEEEyessubnumber\label{eq:mgmrgi4}\\
I_{B_2} &=& B \IEEEyessubnumber\label{eq:mgmrgi5}\\
I_{B_3} &=& B \IEEEyessubnumber\label{eq:mgmrgi6}\\
I_{M_1} &=& \frac{38}{5}g_1^4(M)B^2 \IEEEyessubnumber\label{eq:mgmrgi7}\\
I_{M_2} &=& 2g_2^4(M)B^2 \IEEEyessubnumber\label{eq:mgmrgi8}\\
I_{M_3} &=& -2g_3^4(M)B^2 \IEEEyessubnumber\label{eq:mgmrgi9}\\
I_{g_2} &=& g_1^{-2}(M) -\frac{33}{5}g_2^{-2}(M) \IEEEyessubnumber\label{eq:mgmrgi10}\\
I_{g_3} &=& g_1^{-2}(M) + \frac{11}{5}g_3^{-2}(M) \IEEEyessubnumber\label{eq:mgmrgi11}
\end{IEEEeqnarray}
where $M$ is the messenger scale.
This gives us eleven equations in terms of six unknowns ($\delta_u$, $\delta_d$, $B$, $g_1(M)$, $g_2(M)$, $g_3(M)$).
We can trade each unknown for an equation, i.e.\ for each parameter we use one of the above equations to express it in terms of RGIs only.
Since we have more independent equations than unknowns, we can substitute the resulting six expressions into the remaining five equations to obtain five sum rules in terms of RGIs only.
For example, using equation \eqref{eq:mgmrgi4} we can eliminate the model parameter $B$ from the remaining equations.
Then equations \eqref{eq:mgmrgi7}-\eqref{eq:mgmrgi9} can be used to eliminate the gauge couplings at the messenger scale.
We can get the value of $\delta_d$ from equation \eqref{eq:mgmrgi2}, and then \eqref{eq:mgmrgi1} gives the value of $\delta_u$.
After substituting the resulting six expressions into the five remaining equations, we are left with the following sum rules:
\begin{align}
0 &= I_{Y_\alpha} + \frac{13}{10}D_{Y_{13H}}I_{B_1}\sqrt{\frac{38}{5I_{M_1}}} \tag{\ref{eq:mgmrgi3}}\\
0 &= I_{B_1} - I_{B_2} \tag{\ref{eq:mgmrgi5}}\\
0 &= I_{B_1} - I_{B_3} \tag{\ref{eq:mgmrgi6}}\\
0 &= I_{B_1}\sqrt{\frac{38}{5I_{M_1}}} - \frac{33}{5}I_{B_1}\sqrt{\frac{2}{I_{M_2}}} - I_{g_2} \tag{\ref{eq:mgmrgi10}}\\
0 &= I_{B_1}\sqrt{\frac{38}{5I_{M_1}}} + \frac{11}{5}I_{B_1}\sqrt{\frac{-2}{I_{M_3}}} - I_{g_3} \tag{\ref{eq:mgmrgi11}}
\end{align}
To summarise, we have chosen a specific supersymmetry breaking model and expressed the RGIs in terms of model parameters.
Since we ended up with more equations than unknowns, we could construct eight sum rules in terms of RGIs only: three from vanishing RGIs and five by eliminating the model parameters.
If any of these sum rules are violated, MGM is not consistent with experimental data.

As an aside, we can now see why the RGIs \eqref{eq:additionalRGIs} are not useful.
Suppose we wish to test a specific breaking model.
Let us denote the values of $B$ and $\mu$ at the new physics threshold in this model as $B_\text{thr}$ and $\mu_\text{thr}$ respectively.
Now we apply the above procedure to this model: we express $B_\text{thr}$ and $\mu_\text{thr}$ in terms of RGIs and the other couplings at the high scale; then we can insert these expressions into the remaining equations.
But since $B$ and $\mu$ both appear in only one independent RGI, there are no equations to insert these expressions into!
In the above example, it was possible to combine all RGIs into sum rules because each running parameter appeared in more than one RGI.
Since $B$ and $\mu$ do not, their corresponding RGIs become useless to our analysis.
Hence we will have to restrict ourselves to the RGIs that do not contain these running parameters.
As will be shown in appendix \ref{a:derivingRGIs}, it is not possible to construct RGIs that contain the Yukawa and soft trilinear couplings without using $B$ and $\mu$.
That is why we will only use RGIs constructed out of soft masses and/or gauge couplings, i.e.\ those listed in table \ref{t:mssmrgis}.

\paragraph{Using RGIs effectively}
In studies of RGIs such as \cite{Carena1,Carena2,Jaeckel}, a certain breaking mechanism is usually presupposed.
Then one constructs sum rules that are tailor-made for that breaking mechanism.
For example, the sum rules constructed above all provide a test for consistency of MGM with experimental data.
However, some of these sum rules will also hold for other breaking mechanisms.
It is not always clear to what extent the validity of the sum rules depends on the unique features of the breaking mechanism under study.
For example, in minimal gauge mediation the quantity $M_a/g_a^2$ unifies at the messenger scale; this follows directly from the matching condition \eqref{eq:MGMgauginos}.
However, in mSUGRA this quantity also unifies, but for a different reason: it is the consequence of the assumption of gauge coupling unification and gaugino mass unification at the same energy scale!
Hence the sum rules that test this unification property cannot be used to confirm that either of these specific models corresponds to reality.
They can only provide consistency checks that should be satisfied if any of these models are realised in Nature.

Therefore, we will look for RG invariant sum rules using a different approach.
We will not presume any spectrum specific to a certain breaking mechanism.
Instead, we will search for sum rules that test properties that are common in supersymmetry breaking models (e.g.\ $M_a/g_a^2$ unification).
Then any breaking model, be it an existing one such as those described in section \ref{s:susybreaking} or a new one contrived in the future, can be tested directly if it predicts any of these properties.
For example, if the sum rules for $M_a/g_a^2$ unification are not satisfied by experimental data, then models that predict this property (mSUGRA and MGM, but not necessarily GGM) are falsified.
But also anyone who would concoct a new model that has this property, would have to go back to the drawing board at once.
In the next section, we will look for common properties to test and find sum rules for them.

\section{Results}\label{s:results}
In the previous section, we proposed a new strategy for using RGIs to study the supersymmetry breaking mechanism: we will look for RG invariant sum rules that could test properties of the high scale spectrum.
First we will list the properties that we wish to test and argue why we should test them.
Subsequently, we will construct sum rules to test these properties.
After that we will consider sum rules specific to the breaking mechanisms from section \ref{s:susybreaking}; this will help us evaluate the quality of our sum rules.
We will conclude this section with a discussion of the results and comment on the advantages and limitations of the RGI method.

\subsection{Common properties of matching conditions}
Supersymmetry breaking models, such as those in section \ref{s:susybreaking}, predict relations between the running parameters as a result of matching conditions at the new physics threshold.
These relations mostly involve the unification of certain parameters.
Therefore we will construct sum rules that test the following properties:

\paragraph{Gauge coupling unification}
As can be seen from figure \ref{f:gaugeunification}, the MSSM may be consistent with gauge coupling unification, depending on the values of the sparticle thresholds.
The hypothesis that the gauge couplings unify is often made in supersymmetry breaking models, for example in mSUGRA.
Therefore it will be important to determine whether gauge coupling unification occurs in Nature.

\paragraph{Flavour-universality of high scale sfermion masses}
In many theories, the sfermion masses are assumed to be flavour-universal, i.e.\ the first (and therefore also the second) and third generation masses are equal: $\msqa=\msqc\equiv\msq$, $\msua=\msuc\equiv\msu$, $\msda=\msdc\equiv\msd$, $\msla=\mslc\equiv\msl$ and \mbox{$\msea=\msec\equiv\mse$.}
This hypothesis is motivated by the need to suppress FCNC amplitudes.
Flavour-universality may be a consequence of flavour symmetries (as postulated in mSUGRA) or of the flavour-blindness of the interactions that mediate supersymmetry breaking (as in GGM).
Since this property occurs in many models, we will test flavour-universality of the soft masses.

\paragraph{Scalar mass unification}
Unification of the soft scalar masses, which occurs in mSUGRA, is very predictive: many matching conditions depend on a single parameter $m_0$, which allows us to construct multiple sum rules.
Sometimes non-universality of the soft Higgs masses is assumed, because suppression of FCNC amplitudes does not require them to be universal with squark and slepton masses.
Therefore we will distinguish between two cases: one where the soft Higgs masses have the same value $m_0$ as other soft scalar masses, and another where $\mhu$ and $\mhd$ have additional non-universal contributions $\delta_u$ and $\delta_d$ respectively.

\paragraph{Unification of $\mathbf{M_a/g_a^2}$}
As we mentioned in section \ref{s:RGIs}, the quantity $M_a/g_a^2$ may unify for different reasons.
It could be the consequence of gaugino mass and gauge coupling unification at the same scale (as in mSUGRA) or it may be the result of the gaugino mass matching conditions (as in MGM).
Therefore we will also test this property.

\paragraph{Gaugino mass unification}
In mSUGRA, the gaugino masses are assumed to unify.
Since this model is widely used, we will need to check whether gaugino mass unification occurs in Nature.

\paragraph{Multiple unifications at one scale}
It is possible that several of these unification properties will turn out to be consistent with experimental data.
In that case, we could test whether these unifications occur at the same energy scale.

At first sight, it may seem strange to consider the possibility of two kinds of unifications at different energy scales.
In supersymmetry breaking models such unifications usually occur at a threshold where new physics enters the theory.
Thus even if the MSSM were consistent with two kinds of unifications, the RG trajectories of the running parameters could be deflected from the MSSM trajectories after the first threshold, spoiling the second unification.
However, recall that in mirage mediation (see section \ref{s:MM}) the scale where the soft masses unify is lower than the scale at which the soft masses are generated.
Thus it is possible that the unification scale does not correspond to any physical threshold.
Therefore we will separately check whether multiple unifications occur at the same scale.

\subsection{Sum rules}
In this section we construct sum rules that test the properties described above.
For each one, we determine the values of the RGIs in terms of the unknown parameters.
Then we eliminate the unknown parameters to obtain RG invariant sum rules.

\subsubsection{Gauge coupling unification}
Suppose we have $g_1=g_2=g_3\equiv g$ at some scale $M_\text{gc}$.
If we insert these relations into the RGI expressions, we get:
\begin{IEEEeqnarray}{rCl}
D_{B_{13}} &=& 2\left(\msqa-\msqc\right) - \msua + \msuc - \msda + \msdc \IEEEyessubnumber\\
D_{L_{13}} &=& 2\left(\msla-\mslc\right) - \msea + \msec \IEEEyessubnumber\\
D_{\chi_1} &=& 3\left(3\msda-2\left(\msqa-\msla\right)-\msua\right) - \msea \IEEEyessubnumber\\
D_{Y_{13H}} &=& \msqa-2\msua+\msda-\msla+\msea \nonumber\\
&&-\frac{10}{13}\left(\msqc-2\msuc+\msdc-\mslc+\msec+\mhu-\mhd\right) \IEEEyessubnumber\\
D_Z &=& 3\left(\msdc-\msda\right) + 2\left(\mslc-\mhd\right) \IEEEyessubnumber\\
I_{Y_\alpha} &=& \frac{1}{g^2}\left(\mhu-\mhd + \sum_\text{gen}\left(\msq-2\msu+\msd-\msl+\mse\right)\right) \IEEEyessubnumber\\
I_{B_1} &=& \frac{M_1}{g^2} \IEEEyessubnumber\\
I_{B_2} &=& \frac{M_2}{g^2} \IEEEyessubnumber\\
I_{B_3} &=& \frac{M_3}{g^2} \IEEEyessubnumber\\
I_{M_1} &=& M_1^2 - \frac{33}{8}\left(\msda-\msua-\msea\right) \IEEEyessubnumber\\
I_{M_2} &=& M_2^2 + \frac{1}{24}\left(9\left(\msda-\msua\right) + 16\msla-\msea\right) \IEEEyessubnumber\\
I_{M_3} &=& M_3^2 - \frac{3}{16}\left(5\msda+\msua-\msea\right) \IEEEyessubnumber\\
I_{g_2} &=& -\frac{28}{5}g^{-2} \IEEEyessubnumber\label{eq:gcu1}\\
I_{g_3} &=& \frac{16}{5}g^{-2} \IEEEyessubnumber\label{eq:gcu2}
\end{IEEEeqnarray}
where the soft scalar and gaugino masses are understood to be evaluated at the scale $M_\text{gc}$.
This amounts to 14 equations with 16 unknowns, so at first sight we expect to find no sum rules.
However, equations \eqref{eq:gcu1}-\eqref{eq:gcu2} constitute two equations with only one unknown $g$.
Hence we can make one sum rule:
\begin{equation}
I_{g_2} + \frac74I_{g_3} = 0 \label{eq:gcusumrule}
\end{equation}

\subsubsection{Flavour-universality of high scale sfermion masses}
Suppose the sfermion masses are flavour-universal at some scale $M_\text{fu}$: $\msqa(M_\text{fu})=\msqc(M_\text{fu})\equiv\msq$, $\msua(M_\text{fu})=\msuc(M_\text{fu})\equiv\msu$, $\msda(M_\text{fu})=\msdc(M_\text{fu})\equiv\msd$, $\msla(M_\text{fu})=\mslc(M_\text{fu})\equiv\msl$ and $\msea(M_\text{fu})=\msec(M_\text{fu})\equiv\mse$ .
If we insert this into the RGIs, we get:
\begin{IEEEeqnarray}{rCl}
D_{B_{13}} &=& 0 \IEEEyessubnumber\\
D_{L_{13}} &=& 0 \IEEEyessubnumber\\
D_{\chi_1} &=& 3\left(3\msd-2\left(\msq-\msl\right)-\msu\right) - \mse \IEEEyessubnumber\\
D_{Y_{13H}} &=& \frac{1}{13}\left(3\left(\msq-2\msu+\msd-\msl+\mse\right)+10\left(\mhd-\mhu\right)\right) \IEEEyessubnumber\\
D_Z &=& 2\left(\msl-\mhd\right) \IEEEyessubnumber\\
I_{Y_\alpha} &=& \frac{1}{g_1^2}\left(3\left(\msq-2\msu+\msd-\msl+\mse\right)-\mhd+\mhu\right) \IEEEyessubnumber\\
I_{B_1} &=& \frac{M_1}{g_1^2} \IEEEyessubnumber\\
I_{B_2} &=& \frac{M_2}{g_2^2} \IEEEyessubnumber\\
I_{B_3} &=& \frac{M_3}{g_3^2} \IEEEyessubnumber\\
I_{M_1} &=& M_1^2 - \frac{33}{8}\left(\msd-\msu-\mse\right) \IEEEyessubnumber\\
I_{M_2} &=& M_2^2 + \frac{1}{24}\left(9\left(\msd-\msu\right) + 16\msl-\mse\right) \IEEEyessubnumber\\
I_{M_3} &=& M_3^2 - \frac{3}{16}\left(5\msd+\msu-\mse\right) \IEEEyessubnumber\\
I_{g_2} &=& g_1^{-2} - \frac{33}{5}g_2^{-2} \IEEEyessubnumber\\
I_{g_3} &=& g_1^{-2} + \frac{11}{5}g_3^{-2} \IEEEyessubnumber
\end{IEEEeqnarray}
where the gauge couplings and gaugino masses are understood to be evaluated at $M_\text{fu}$.
The first two equations immediately give us two sum rules:
\begin{align}
D_{B_{13}} &= 0 \label{eq:flavouruni1}\\
D_{L_{13}} &= 0 \label{eq:flavouruni2}
\end{align}
The remaining 12 equations depend on 13 unknowns, hence we cannot construct any other sum rules.

\subsubsection{Scalar mass unification}
Suppose that the scalar masses have a universal value $m_0$ at some scale $M_\text{S}$; we allow for non-universal Higgs masses $\mhu(M_\text{S})=m_0^2+\delta_u$, $\mhd(M_\text{S})=m_0^2+\delta_d$.
Note that this hypothesis implies flavour-universality of the scalar masses, so $D_{B_{13}}$ and $D_{L_{13}}$ vanish.
The remaining RGIs have the following values:
\begin{IEEEeqnarray}{rCl}
D_{\chi_1} &=& 5m_0^2 \IEEEyessubnumber\\
D_{Y_{13H}} &=& -\frac{10}{13}\left(\delta_u-\delta_d\right) \IEEEyessubnumber\\
D_Z &=& -2\delta_d \IEEEyessubnumber\\
I_{Y_\alpha} &=& g_1^{-2}\left(\delta_u-\delta_d\right) \IEEEyessubnumber\\
I_{B_1} &=& \frac{M_1}{g_1^2} \IEEEyessubnumber\\
I_{B_2} &=& \frac{M_2}{g_2^2} \IEEEyessubnumber\\
I_{B_3} &=& \frac{M_3}{g_3^2} \IEEEyessubnumber\\
I_{M_1} &=& M_1^2+\frac{33}{8}m_0^2 \IEEEyessubnumber\\
I_{M_2} &=& M_2^2+\frac{5}{8}m_0^2 \IEEEyessubnumber\\
I_{M_3} &=& M_3^2-\frac{15}{16}m_0^2 \IEEEyessubnumber\\
I_{g_2} &=& g_1^{-2}-\frac{33}{5}g_2^{-2} \IEEEyessubnumber\\
I_{g_3} &=& g_1^{-2}+\frac{11}{5}g_3^{-2} \IEEEyessubnumber
\end{IEEEeqnarray}
where the gauge couplings and gaugino masses are understood to be evaluated at $M_\text{S}$.
This amounts to twelve equations with nine unknowns, hence we can construct three sum rules:
\begin{align}
I_{g_2} &= \left(I_{M_1}-\frac{33}{40}D_{\chi_1}\right)^{-1/2}I_{B_1} - \frac{33}{5}\left(I_{M_2}-\frac{1}{8}D_{\chi_1}\right)^{-1/2}I_{B_2} \label{eq:NUHM1}\\
I_{g_3} &= \left(I_{M_1}-\frac{33}{40}D_{\chi_1}\right)^{-1/2}I_{B_1} + \frac{11}{5}\left(I_{M_3}+\frac{3}{16}D_{\chi_1}\right)^{-1/2}I_{B_3} \label{eq:NUHM2}\\
0 &= I_{Y_\alpha}\sqrt{I_{M_1}-\frac{33}{40}D_{\chi_1}} + \frac{13}{10}I_{B_1}D_{Y_{13H}} \label{eq:NUHM3}
\end{align}
Furthermore, non-universality of the Higgs masses can be tested directly because we can extract $\delta_u$ and $\delta_d$ from the RGIs:
\begin{IEEEeqnarray}{rCl}
\delta_d &=& -\frac12D_Z \neq 0 \label{eq:Higgsnonuniversality1}\\
\delta_u &=& -\frac{13}{10}D_{Y_{13H}} - \frac12D_Z \neq 0 \label{eq:Higgsnonuniversality2}
\end{IEEEeqnarray}

\paragraph{Scalar mass unification with universal Higgs masses}
Now we consider the same scenario, but with $\delta_u=\delta_d=0$.
Apart from the vanishing of $D_{B_{13}}$ and $D_{L_{13}}$ we find:
\begin{IEEEeqnarray}{rCl}
D_{\chi_1} &=& 5m_0^2 \IEEEyessubnumber\\
D_{Y_{13H}} &=& 0 \IEEEyessubnumber\\
D_Z &=& 0 \IEEEyessubnumber\\
I_{Y_\alpha} &=& 0 \IEEEyessubnumber\\
I_{B_1} &=& \frac{M_1}{g_1^2} \IEEEyessubnumber\\
I_{B_2} &=& \frac{M_2}{g_2^2} \IEEEyessubnumber\\
I_{B_3} &=& \frac{M_3}{g_3^2} \IEEEyessubnumber\\
I_{M_1} &=& M_1^2+\frac{33}{8}m_0^2 \IEEEyessubnumber\\
I_{M_2} &=& M_2^2+\frac{5}{8}m_0^2 \IEEEyessubnumber\\
I_{M_3} &=& M_3^2-\frac{15}{16}m_0^2 \IEEEyessubnumber\\
I_{g_2} &=& g_1^{-2}-\frac{33}{5}g_2^{-2} \IEEEyessubnumber\\
I_{g_3} &=& g_1^{-2}+\frac{11}{5}g_3^{-2} \IEEEyessubnumber
\end{IEEEeqnarray}
again with the gaugino masses and gauge couplings evaluated at $M_\text{S}$.
We directly find three new sum rules:
\begin{align}
D_{Y_{13H}} &= 0 \label{eq:UHM1}\\
D_Z &= 0 \label{eq:UHM2}\\
I_{Y_\alpha} &= 0 \label{eq:UHM3}
\end{align}
The remaining nine equations contain seven parameters, which allows us to construct two more sum rules:
\begin{align}
I_{g_2} &= \left(I_{M_1}-\frac{33}{40}D_{\chi_1}\right)^{-1/2}I_{B_1} - \frac{33}{5}\left(I_{M_2}-\frac{1}{8}D_{\chi_1}\right)^{-1/2}I_{B_2} \\
I_{g_3} &= \left(I_{M_1}-\frac{33}{40}D_{\chi_1}\right)^{-1/2}I_{B_1} + \frac{11}{5}\left(I_{M_3}+\frac{3}{16}D_{\chi_1}\right)^{-1/2}I_{B_3}
\end{align}
Note that these two sum rules are the same as in the non-universal case.
This makes sense: since $\delta_u=\delta_d=0$ is a special case of non-universality in the Higgs sector, the sum rules \eqref{eq:NUHM1}-\eqref{eq:NUHM3} will also hold for universal Higgs masses.
However, sum rule \eqref{eq:NUHM3} has become redundant because it is automatically satisfied if \eqref{eq:UHM1} and \eqref{eq:UHM3} hold.

\subsubsection{Unification of $\mathbf{M_a/g_a^2}$}
Suppose that the quantities $M_a/g_a^2$ have a common value $C$ at a scale $M_\text{C}$.
Then the RGIs have the following values:
\begin{IEEEeqnarray}{rCl}
D_{B_{13}} &=& 2\left(\msqa-\msqc\right) - \msua + \msuc - \msda + \msdc \IEEEyessubnumber\\
D_{L_{13}} &=& 2\left(\msla-\mslc\right) - \msea + \msec \IEEEyessubnumber\\
D_{\chi_1} &=& 3\left(3\msda-2\left(\msqa-\msla\right)-\msua\right) - \msea \IEEEyessubnumber\\
D_{Y_{13H}} &=& \msqa-2\msua+\msda-\msla+\msea \nonumber\\
&&-\frac{10}{13}\left(\msqc-2\msuc+\msdc-\mslc+\msec+\mhu-\mhd\right) \IEEEyessubnumber\\
D_Z &=& 3\left(\msdc-\msda\right) + 2\left(\mslc-\mhd\right) \IEEEyessubnumber\\
I_{Y_\alpha} &=& \frac{1}{g_1^2}\left(\mhu-\mhd + \sum_\text{gen}\left(\msq-2\msu+\msd-\msl+\mse\right)\right) \IEEEyessubnumber\\
I_{B_1} &=& C \IEEEyessubnumber\label{eq:Mg1}\\
I_{B_2} &=& C \IEEEyessubnumber\label{eq:Mg2}\\
I_{B_3} &=& C \IEEEyessubnumber\label{eq:Mg3}\\
I_{M_1} &=& C^2g_1^4 - \frac{33}{8}\left(\msda-\msua-\msea\right) \IEEEyessubnumber\\
I_{M_2} &=& C^2g_2^4 + \frac{1}{24}\left(9\left(\msda-\msua\right) + 16\msla-\msea\right) \IEEEyessubnumber\\
I_{M_3} &=& C^2g_3^4 - \frac{3}{16}\left(5\msda+\msua-\msea\right) \IEEEyessubnumber\\
I_{g_2} &=& g_1^{-2} - \frac{33}{5}g_2^{-2} \IEEEyessubnumber\\
I_{g_3} &=& g_1^{-2} + \frac{11}{5}g_3^{-2} \IEEEyessubnumber
\end{IEEEeqnarray}
where the scalar masses and gauge couplings are understood to be evaluated at $M_\text{C}$.
This gives us 14 equations with 16 unknowns, so at first sight we expect no sum rules.
However, equations \eqref{eq:Mg1}-\eqref{eq:Mg3} constitute a subset of three equations depending on one unknown.
This gives us two sum rules:
\begin{align}
I_{B_1} &= I_{B_2} \label{eq:Mgunification1}\\
I_{B_1} &= I_{B_3} \label{eq:Mgunification2}
\end{align}

\subsubsection{Gaugino mass unification}
Suppose we have universal gaugino masses $M_1=M_2=M_3\equiv M_{1/2}$ at some energy scale $M_\text{G}$.
Then the RGIs have the following values:
\begin{IEEEeqnarray}{rCl}
D_{B_{13}} &=& 2\left(\msqa-\msqc\right) - \msua + \msuc - \msda + \msdc \IEEEyessubnumber\\
D_{L_{13}} &=& 2\left(\msla-\mslc\right) - \msea + \msec \IEEEyessubnumber\\
D_{\chi_1} &=& 3\left(3\msda-2\left(\msqa-\msla\right)-\msua\right) - \msea \IEEEyessubnumber\\
D_{Y_{13H}} &=& \msqa-2\msua+\msda-\msla+\msea \nonumber\\
&&-\frac{10}{13}\left(\msqc-2\msuc+\msdc-\mslc+\msec+\mhu-\mhd\right) \IEEEyessubnumber\\
D_Z &=& 3\left(\msdc-\msda\right) + 2\left(\mslc-\mhd\right) \IEEEyessubnumber\\
I_{Y_\alpha} &=& \frac{1}{g_1^2}\left(\mhu-\mhd + \sum_\text{gen}\left(\msq-2\msu+\msd-\msl+\mse\right)\right) \IEEEyessubnumber\\
I_{B_1} &=& \frac{M_{1/2}}{g_1^2} \IEEEyessubnumber\label{eq:Gu1}\\
I_{B_2} &=& \frac{M_{1/2}}{g_2^2} \IEEEyessubnumber\label{eq:Gu2}\\
I_{B_3} &=& \frac{M_{1/2}}{g_3^2} \IEEEyessubnumber\label{eq:Gu3}\\
I_{M_1} &=& M_{1/2}^2 - \frac{33}{8}\left(\msda-\msua-\msea\right) \IEEEyessubnumber\\
I_{M_2} &=& M_{1/2}^2 + \frac{1}{24}\left(9\left(\msda-\msua\right) + 16\msla-\msea\right) \IEEEyessubnumber\\
I_{M_3} &=& M_{1/2}^2 - \frac{3}{16}\left(5\msda+\msua-\msea\right) \IEEEyessubnumber\\
I_{g_2} &=& g_1^{-2} - \frac{33}{5}g_2^{-2} \IEEEyessubnumber\label{eq:Gu4}\\
I_{g_3} &=& g_1^{-2} + \frac{11}{5}g_3^{-2} \IEEEyessubnumber\label{eq:Gu5}
\end{IEEEeqnarray}
where the scalar masses and gauge couplings should be evaluated at $M_\text{G}$.
This adds up to 14 equations with 16 unknowns, so we expect to find no sum rules at first sight.
However, equations \eqref{eq:Gu1}-\eqref{eq:Gu3} and \eqref{eq:Gu4}-\eqref{eq:Gu5} form a subset of five equations with four unknowns.
This allows us to construct one sum rule:
\begin{equation}
\left(I_{B_1}-\frac{33}{5}I_{B_2}\right)I_{g_3} = \left(I_{B_1}+\frac{11}{5}I_{B_3}\right)I_{g_2} \label{eq:GMU}
\end{equation}

\subsubsection{Scalar + gaugino mass unification}
Suppose that experimental data are consistent with both scalar mass unification (with or without universal Higgs masses) and gaugino mass unification.
Again $D_{B_{13}}$ and $D_{L_{13}}$ vanish because scalar mass unification implies flavour-universal sfermion masses.
If the scalar mass unification scale $M_\text{S}$ equals the gaugino mass unification scale $M_\text{G}$, the remaining RGIs have the following values:
\begin{IEEEeqnarray}{rCl}
D_{\chi_1} &=& 5m_0^2 \IEEEyessubnumber\\
D_{Y_{13H}} &=& -\frac{10}{13}\left(\delta_u-\delta_d\right) \IEEEyessubnumber\\
D_Z &=& -2\delta_d \IEEEyessubnumber\\
I_{Y_\alpha} &=& g_1^{-2}\left(\delta_u-\delta_d\right) \IEEEyessubnumber\\
I_{B_1} &=& \frac{M_{1/2}}{g_1^2} \IEEEyessubnumber\\
I_{B_2} &=& \frac{M_{1/2}}{g_2^2} \IEEEyessubnumber\\
I_{B_3} &=& \frac{M_{1/2}}{g_3^2} \IEEEyessubnumber\\
I_{M_1} &=& M_{1/2}^2+\frac{33}{8}m_0^2 \IEEEyessubnumber\\
I_{M_2} &=& M_{1/2}^2+\frac{5}{8}m_0^2 \IEEEyessubnumber\\
I_{M_3} &=& M_{1/2}^2-\frac{15}{16}m_0^2 \IEEEyessubnumber\\
I_{g_2} &=& g_1^{-2}-\frac{33}{5}g_2^{-2} \IEEEyessubnumber\\
I_{g_3} &=& g_1^{-2}+\frac{11}{5}g_3^{-2} \IEEEyessubnumber
\end{IEEEeqnarray}
The gauge couplings are understood to be evaluated at $M_\text{S}=M_\text{G}$ and we assume non-universal Higgs masses for the moment.
Here we have twelve equations with seven unknowns, yielding five sum rules.
Among these are:
\begin{itemize}
\item Equations \eqref{eq:NUHM1}-\eqref{eq:NUHM3} for scalar mass unification.
\item Equation \eqref{eq:GMU} for gaugino mass unification.
\end{itemize}
Hence there is one new sum rule, which we find to be:
\begin{equation}
I_{M_1} - \frac{81}{25}I_{M_2} + \frac{56}{25}I_{M_3} = 0 \label{eq:scalargauginosumrule}
\end{equation}
Note that this result is independent of universality in the Higgs sector: if we had chosen $\delta_u=\delta_d=0$, we would only get the additional sum rules $D_{Y_{13H}} = D_Z = I_{Y_\alpha} = 0$ that test universality of the Higgs masses.

\subsubsection{Gaugino mass + gauge coupling unification}
Suppose that experimental data are consistent with both gaugino mass unification and gauge coupling unification.
If the gaugino mass unification scale $M_\text{G}$ equals the gauge coupling unification scale $M_\text{gc}$, the RGIs have the values:
\begin{IEEEeqnarray}{rCl}
D_{B_{13}} &=& 2\left(\msqa-\msqc\right) - \msua + \msuc - \msda + \msdc \IEEEyessubnumber\\
D_{L_{13}} &=& 2\left(\msla-\mslc\right) - \msea + \msec \IEEEyessubnumber\\
D_{\chi_1} &=& 3\left(3\msda-2\left(\msqa-\msla\right)-\msua\right) - \msea \IEEEyessubnumber\\
D_{Y_{13H}} &=& \msqa-2\msua+\msda-\msla+\msea \nonumber\\
&&-\frac{10}{13}\left(\msqc-2\msuc+\msdc-\mslc+\msec+\mhu-\mhd\right) \IEEEyessubnumber\\
D_Z &=& 3\left(\msdc-\msda\right) + 2\left(\mslc-\mhd\right) \IEEEyessubnumber\\
I_{Y_\alpha} &=& \frac{1}{g^2}\left(\mhu-\mhd + \sum_\text{gen}\left(\msq-2\msu+\msd-\msl+\mse\right)\right) \IEEEyessubnumber\\
I_{B_1} &=& \frac{M_{1/2}}{g^2} \IEEEyessubnumber\label{eq:Ggu1}\\
I_{B_2} &=& \frac{M_{1/2}}{g^2} \IEEEyessubnumber\label{eq:Ggu2}\\
I_{B_3} &=& \frac{M_{1/2}}{g^2} \IEEEyessubnumber\label{eq:Ggu3}\\
I_{M_1} &=& M_{1/2}^2 - \frac{33}{8}\left(\msda-\msua-\msea\right) \IEEEyessubnumber\\
I_{M_2} &=& M_{1/2}^2 + \frac{1}{24}\left(9\left(\msda-\msua\right) + 16\msla-\msea\right) \IEEEyessubnumber\\
I_{M_3} &=& M_{1/2}^2 - \frac{3}{16}\left(5\msda+\msua-\msea\right) \IEEEyessubnumber\\
I_{g_2} &=& -\frac{28}{5}g^{-2} \IEEEyessubnumber\label{eq:Ggu4}\\
I_{g_3} &=& \frac{16}{5}g^{-2} \IEEEyessubnumber\label{eq:Ggu5}
\end{IEEEeqnarray}
Here the scalar masses are understood to be evaluated at $M_\text{G}=M_\text{gc}$.
We have 14 equations with 14 unknowns, yielding no sum rules at first sight.
However, equations \eqref{eq:Ggu1}-\eqref{eq:Ggu3} and \eqref{eq:Ggu4}-\eqref{eq:Ggu5} constitute five equations with two unknowns, giving us three sum rules.
Among them are:
\begin{itemize}
\item Equation \eqref{eq:gcusumrule} for gauge coupling unification.
\item Equations \eqref{eq:Mgunification1} and \eqref{eq:Mgunification2} for $M_a/g_a^2$ unification.
\item Equation \eqref{eq:GMU} for gaugino mass unification. However, this one is redundant since it is automatically satisfied if \eqref{eq:gcusumrule}, \eqref{eq:Mgunification1} and \eqref{eq:Mgunification2} are satisfied.
\end{itemize}
This adds up to three sum rules, hence we can make no new sum rules.
We could have expected this: given the fact that the gaugino masses and gauge couplings unify separately, the hypothesis that they unify at the same scale is equivalent to the hypothesis of $M_a/g_a^2$ unification.

\subsubsection{Scalar mass + gauge coupling unification}
Suppose that experimental data are consistent with both scalar mass unification (for the moment we assume non-universal Higgs masses) and gaugino mass unification.
Again $D_{B_{13}}$ and $D_{L_{13}}$ vanish because scalar mass unification implies flavour-universal sfermion masses.
If the scalar mass unification scale $M_\text{S}$ equals the gauge coupling unification scale $M_\text{gc}$, the RGIs have the values:
\begin{IEEEeqnarray}{rCl}
D_{\chi_1} &=& 5m_0^2 \IEEEyessubnumber\\
D_{Y_{13H}} &=& -\frac{10}{13}\left(\delta_u-\delta_d\right) \IEEEyessubnumber\\
D_Z &=& -2\delta_d \IEEEyessubnumber\\
I_{Y_\alpha} &=& g^{-2}\left(\delta_u-\delta_d\right) \IEEEyessubnumber\\
I_{B_1} &=& \frac{M_1}{g^2} \IEEEyessubnumber\\
I_{B_2} &=& \frac{M_2}{g^2} \IEEEyessubnumber\\
I_{B_3} &=& \frac{M_3}{g^2} \IEEEyessubnumber\\
I_{M_1} &=& M_1^2+\frac{33}{8}m_0^2 \IEEEyessubnumber\\
I_{M_2} &=& M_2^2+\frac{5}{8}m_0^2 \IEEEyessubnumber\\
I_{M_3} &=& M_3^2-\frac{15}{16}m_0^2 \IEEEyessubnumber\\
I_{g_2} &=& -\frac{28}{5}g^{-2} \IEEEyessubnumber\\
I_{g_3} &=& \frac{16}{5}g^{-2} \IEEEyessubnumber
\end{IEEEeqnarray}
Here the gaugino masses are understood to be evaluated at $M_\text{S}=M_\text{gc}$.
This amounts to twelve equations containing seven parameters, so we can construct five sum rules.
These include:
\begin{itemize}
\item Equations \eqref{eq:NUHM1}-\eqref{eq:NUHM3} for scalar mass unification.
\item Equation \eqref{eq:gcusumrule} for gauge coupling unification.
\end{itemize}
We find one new sum rule:
\begin{equation}
I_{Y_\alpha} = \frac{13}{56}I_{g_2}D_{Y_{13H}} \label{eq:scalargcsumrule}
\end{equation}
In the case of universal Higgs masses, we have $I_{Y_\alpha}=D_{Y_{13H}}=0$ and the above sum rule becomes redundant.

\subsubsection{Scalar mass + gaugino mass + gauge coupling unification}
Now suppose experimental data are consistent with universal scalar masses, universal gaugino masses and gauge coupling unification.
Again \mbox{$D_{B_{13}} = D_{L_{13}} = 0$} because scalar mass unification implies flavour-universality.
If the scalar mass unification scale $M_\text{S}$, the gaugino mass unification scale $M_\text{G}$ and the gauge coupling unification scale $M_\text{gc}$ are all equal, the RGIs have the values:
\begin{IEEEeqnarray}{rCl}
D_{\chi_1} &=& 5m_0^2 \IEEEyessubnumber\\
D_{Y_{13H}} &=& -\frac{10}{13}\left(\delta_u-\delta_d\right) \IEEEyessubnumber\\
D_Z &=& -2\delta_d \IEEEyessubnumber\\
I_{Y_\alpha} &=& g^{-2}\left(\delta_u-\delta_d\right) \IEEEyessubnumber\\
I_{B_1} &=& \frac{M_{1/2}}{g^2} \IEEEyessubnumber\\
I_{B_2} &=& \frac{M_{1/2}}{g^2} \IEEEyessubnumber\\
I_{B_3} &=& \frac{M_{1/2}}{g^2} \IEEEyessubnumber\\
I_{M_1} &=& M_{1/2}^2+\frac{33}{8}m_0^2 \IEEEyessubnumber\\
I_{M_2} &=& M_{1/2}^2+\frac{5}{8}m_0^2 \IEEEyessubnumber\\
I_{M_3} &=& M_{1/2}^2-\frac{15}{16}m_0^2 \IEEEyessubnumber\\
I_{g_2} &=& -\frac{28}{5}g^{-2} \IEEEyessubnumber\\
I_{g_3} &=& \frac{16}{5}g^{-2} \IEEEyessubnumber
\end{IEEEeqnarray}
Here we assume non-universal Higgs masses for the moment.
We have twelve equations with only five parameters, so we can construct seven sum rules.
These include:
\begin{itemize}
\item Equation \eqref{eq:gcusumrule} for gauge coupling unification.
\item Equations \eqref{eq:Mgunification1}-\eqref{eq:Mgunification2} for $M_a/g_a^2$ unification.
\item Equation \eqref{eq:GMU} for gaugino mass unification. However, this one is redundant since it is automatically satisfied if \eqref{eq:gcusumrule}, \eqref{eq:Mgunification1} and \eqref{eq:Mgunification2} are satisfied.
\item Equations \eqref{eq:NUHM1}-\eqref{eq:NUHM3} for scalar mass unification.
\item Equation \eqref{eq:scalargcsumrule} for equality of $M_\text{S}$ and $M_\text{gc}$.
\item Equation \eqref{eq:scalargauginosumrule} for equality of $M_\text{S}$ and $M_\text{G}$. However, this one has become redundant: by combining equations \eqref{eq:NUHM1}, \eqref{eq:NUHM2}, \eqref{eq:NUHM3}, \eqref{eq:Mgunification1}, \eqref{eq:Mgunification2} and \eqref{eq:scalargcsumrule} one can retrieve equation \eqref{eq:scalargauginosumrule}. We could have expected this, because if we have consistency with scalar mass, gaugino mass and gauge coupling unification and have established both $M_\text{G}=M_\text{gc}$ and $M_\text{gc}=M_\text{S}$, then it follows automatically that $M_\text{G}=M_\text{S}$.
\end{itemize}
This adds up to seven independent sum rules, so there are no new ones.

\subsubsection{Sum rules summary}
\begin{figure}[p]
\begin{center}
  \includegraphics[width=\textwidth]{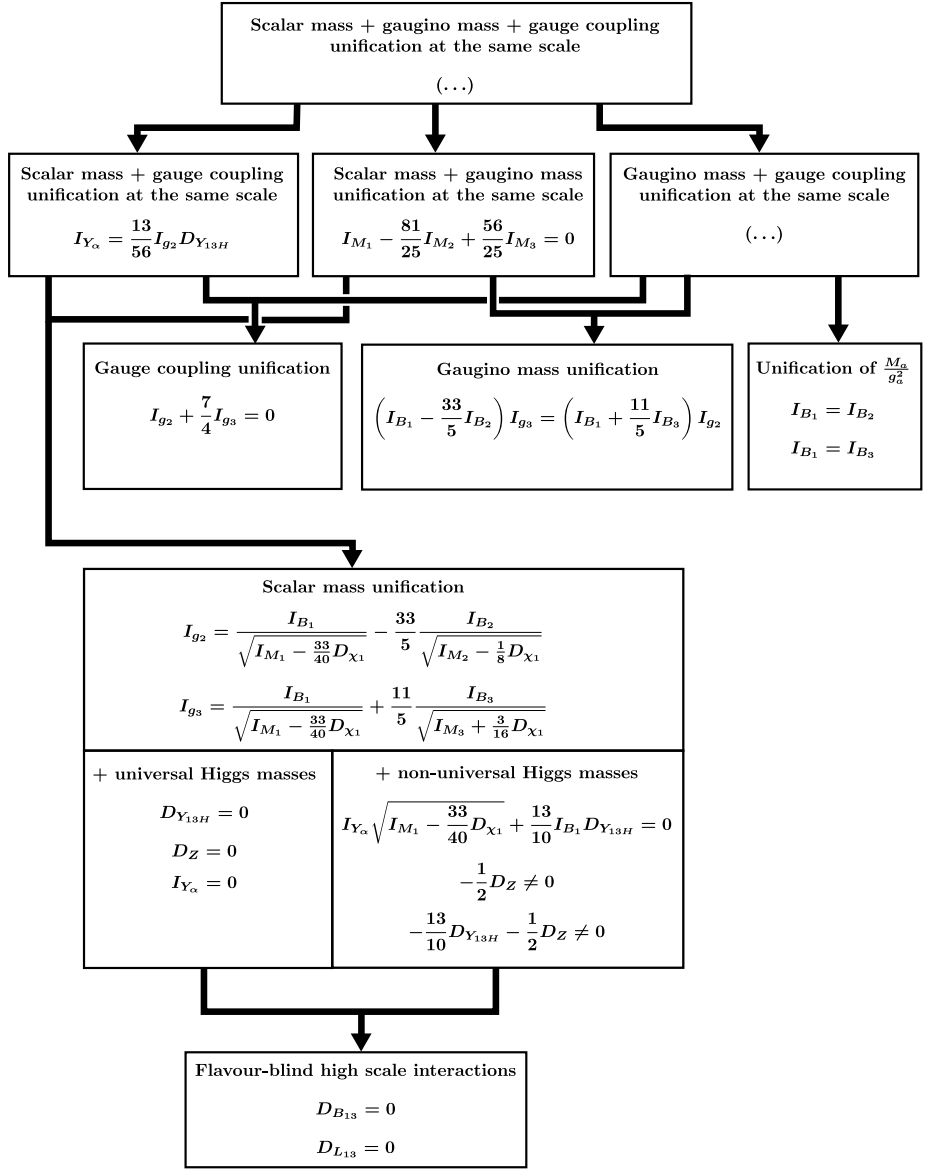}
\end{center}
\caption{Scheme for testing hypotheses about the spectrum at the new physics threshold. For a given hypothesis, the arrows point towards its underlying hypotheses. To test a specific one, check whether the corresponding sum rules are satisfied. Then follow the arrows downwards all the way to the bottom and for each sum rule along the way, check whether it is satisfied.}\label{f:flowchart}
\end{figure}

All hypotheses discussed above and their corresponding sum rules have been summarised in figure \ref{f:flowchart}.
Related hypotheses have been connected: if one starts at a given hypothesis, one should follow the arrows downwards to arrive at the underlying hypotheses.
When we have determined the values of the RGIs from experimental data, we can test whether the listed hypotheses are consistent with the data.
One should proceed as follows: to test a hypothesis, check the validity of the sum rules in the corresponding box.
Then check the validity of the sum rules in all boxes one encounters by following the arrows all the way down.
If all these sum rules are satisfied, the hypothesis is consistent with the experimental data (as far as our sum rules are concerned).

\subsection{Model-specific sum rules}
Until now we have only considered hypotheses concerned with relations between the running parameters of the MSSM.
These hypotheses do not refer to any model-specific parameters.
However, we can find additional sum rules for certain models because the soft masses are related by only a few parameters.
For example, in MGM the gaugino and sfermion masses are determined by the gauge couplings and a single parameter $B$; see equations \eqref{eq:MGMgauginos}-\eqref{eq:MGMsfermions}.
Furthermore, the question whether the messenger scale equals the gauge coupling unification scale only makes sense if we consider gauge mediation models.
Therefore we consider model-specific sum rules separately in this section.

In the following, we will look for sum rules for GGM and AMSB that do not follow from the hypotheses we have discussed above.
We will not discuss mSUGRA, because for our purposes its spectrum is completely characterised by simultaneous scalar mass, gaugino mass and gauge coupling unification.

\subsubsection{General gauge mediation}
Recall that the RG boundary conditions for GGM (with non-universal Higgs masses and without a Fayet-Iliopoulos term $\zeta$) are:
\begin{IEEEeqnarray}{rCl}
M_a &=& g_a^2B_a \qquad (a=1,2,3) \IEEEyessubnumber\\
m_i^2 &=& \sum_{a=1}^3 g_a^4C_a(i)A_a \IEEEyessubnumber\\
\mhu &=& \sum_{a=1}^3 g_a^4C_a(H_u)A_a + \delta_u \IEEEyessubnumber\\
\mhd &=& \sum_{a=1}^3 g_a^4C_a(H_d)A_a + \delta_d \IEEEyessubnumber
\end{IEEEeqnarray}
Using the Casimir invariants $C_a(i)$ from appendix \ref{a:anomalousdimensions} we get explicit values:
\begin{IEEEeqnarray}{rCl}
M_1 &=& g_1^2B_1 \IEEEyessubnumber\\
M_2 &=& g_2^2B_2 \IEEEyessubnumber\\
M_3 &=& g_3^2B_3 \IEEEyessubnumber\\
\msq &=& \frac{1}{60}g_1^4A_1 + \frac34g_2^4A_2 + \frac43g_3^4A_3 \IEEEyessubnumber\\
\msu &=& \frac{4}{15}g_1^4A_1 + \frac43g_3^4A_3 \IEEEyessubnumber\\
\msd &=& \frac{1}{15}g_1^4A_1 + \frac43g_3^4A_3 \IEEEyessubnumber\\
\msl &=& \frac{3}{20}g_1^4A_1 + \frac34g_2^4A_2 \IEEEyessubnumber\\
\mse &=& \frac35g_1^4A_1 \IEEEyessubnumber\\
\mhu &=& \frac{3}{20}g_1^4A_1 + \frac34g_2^4A_2 + \delta_u \IEEEyessubnumber\\
\mhd &=& \frac{3}{20}g_1^4A_1 + \frac34g_2^4A_2 + \delta_d \IEEEyessubnumber
\end{IEEEeqnarray}
If we insert these expressions into the RGIs, we immediately find \mbox{$D_{B_{13}}=D_{L_{13}}=0$,} as expected from flavour-universality.
We get one new sum rule:
\begin{equation}
D_{\chi_1}=0
\end{equation}
The remaining RGIs have the following values:
\begin{IEEEeqnarray}{rCl}\label{eq:GGMRGIs}
D_{Y_{13H}} &=& -\frac{10}{13}\left(\delta_u-\delta_d\right) \IEEEyessubnumber\\
D_Z &=& -2\delta_d \IEEEyessubnumber\\
I_{Y_\alpha} &=& g_1^{-2}\left(\delta_u-\delta_d\right) \IEEEyessubnumber\\
I_{B_1} &=& B_1 \IEEEyessubnumber\\
I_{B_2} &=& B_2 \IEEEyessubnumber\\
I_{B_3} &=& B_3 \IEEEyessubnumber\\
I_{M_1} &=& g_1^4\left(B_1^2 + \frac{33}{10}A_1\right) \IEEEyessubnumber\\
I_{M_2} &=& g_2^4\left(B_2^2 + \frac12A_2\right) \IEEEyessubnumber\\
I_{M_3} &=& g_3^4\left(B_3^2 - \frac32A_3\right) \IEEEyessubnumber\\
I_{g_2} &=& g_1^{-2} -\frac{33}{5}g_2^{-2} \IEEEyessubnumber\\
I_{g_3} &=& g_1^{-2} + \frac{11}{5}g_3^{-2} \IEEEyessubnumber
\end{IEEEeqnarray}
where the gauge couplings are understood to be evaluated at the messenger scale.
This amounts to eleven equations with eleven unknowns, hence no additional sum rules can be constructed.
Note that we can again verify non-universality in the Higgs sector using equations \eqref{eq:Higgsnonuniversality1}-\eqref{eq:Higgsnonuniversality2}.

\paragraph{Equality of $\mathbf{M_\text{gc}}$ and the messenger scale}
If both gauge coupling unification and GGM are compatible with experimental data, we may ask ourselves if the messenger scale equals the scale $M_\text{gc}$ of gauge coupling unification.
In that case the non-vanishing RGIs have the values:
\begin{IEEEeqnarray}{rCl}
D_{Y_{13H}} &=& -\frac{10}{13}\left(\delta_u-\delta_d\right) \IEEEyessubnumber\\
D_Z &=& -2\delta_d \IEEEyessubnumber\\
I_{Y_\alpha} &=& g^{-2}\left(\delta_u-\delta_d\right) \IEEEyessubnumber\\
I_{B_1} &=& B_1 \IEEEyessubnumber\\
I_{B_2} &=& B_2 \IEEEyessubnumber\\
I_{B_3} &=& B_3 \IEEEyessubnumber\\
I_{M_1} &=& g^4\left(B_1^2 + \frac{33}{10}A_1\right) \IEEEyessubnumber\\
I_{M_2} &=& g^4\left(B_2^2 + \frac12A_2\right) \IEEEyessubnumber\\
I_{M_3} &=& g^4\left(B_3^2 - \frac32A_3\right) \IEEEyessubnumber\\
I_{g_2} &=& -\frac{28}{5}g^{-2} \IEEEyessubnumber\\
I_{g_3} &=& \frac{16}{5}g^{-2} \IEEEyessubnumber
\end{IEEEeqnarray}
This adds up to eleven equations with nine unknowns, so we can make two sum rules.
This includes equation \eqref{eq:gcusumrule} for gauge coupling unification.
Hence, there is only one new sum rule:
\begin{equation}
I_{Y_\alpha} = \frac{13}{56}I_{g_2}D_{Y_{13H}} \label{eq:messengergcsumrule}
\end{equation}
Note that this sum rule becomes redundant in the case of universal Higgs masses, since then $I_{Y_\alpha}=D_{Y_{13H}}=0$.
Also note that this sum rule equals \eqref{eq:scalargcsumrule}, which tests equality of $M_\text{S}$ and $M_\text{gc}$.

\paragraph{Minimal gauge mediation}
Recall that MGM is a GGM model restricted to a subset of the GGM parameter space defined by $A_a=A$, $B_a=B$, $A=2B^2$.
Inserting this into the RGI values \eqref{eq:GGMRGIs} of GGM, we find the non-vanishing RGIs to be:
\begin{IEEEeqnarray}{rCl}
D_{Y_{13H}} &=& -\frac{10}{13}\left(\delta_u-\delta_d\right) \IEEEyessubnumber\\
D_Z &=& -2\delta_d \IEEEyessubnumber\\
I_{Y_\alpha} &=& g_1^{-2}\left(\delta_u-\delta_d\right) \IEEEyessubnumber\\
I_{B_1} &=& B \IEEEyessubnumber\\
I_{B_2} &=& B \IEEEyessubnumber\\
I_{B_3} &=& B \IEEEyessubnumber\\
I_{M_1} &=& \frac{38}{5}g_1^4B^2 \IEEEyessubnumber\\
I_{M_2} &=& 2g_2^4B^2 \IEEEyessubnumber\\
I_{M_3} &=& -2g_3^4B^2 \IEEEyessubnumber\\
I_{g_2} &=& g_1^{-2} -\frac{33}{5}g_2^{-2} \IEEEyessubnumber\\
I_{g_3} &=& g_1^{-2} + \frac{11}{5}g_3^{-2} \IEEEyessubnumber
\end{IEEEeqnarray}
This amounts to eleven equations with six parameters, so we can construct five sum rules.
These include \eqref{eq:Mgunification1}-\eqref{eq:Mgunification2} for $M_a/g_a^2$ unification.
There are three new sum rules:\footnote{Note that in MGM, we can safely divide by $I_{M_a}$: if one of the $I_{M_a}$ vanished, then $B=0$ and the gaugino masses would vanish at the messenger scale. Their $\beta$-functions, being proportional to the gaugino masses, would vanish as well. Then at one-loop order, gauginos would be massless at all scales (only through two-loop effects the masses will be non-vanishing). In that case we would have observed them already. Thus the $I_{M_a}$ cannot vanish.}
\begin{IEEEeqnarray}{rCl}
0 &=& I_{Y_\alpha} + \frac{13}{10}D_{Y_{13H}}I_{B_1}\sqrt{\frac{38}{5I_{M_1}}} \label{eq:MGMsumrule1}\\
0 &=& I_{B_1}\sqrt{\frac{38}{5I_{M_1}}} - \frac{33}{5}I_{B_1}\sqrt{\frac{2}{I_{M_2}}} - I_{g_2} \\
0 &=& I_{B_1}\sqrt{\frac{38}{5I_{M_1}}} + \frac{11}{5}I_{B_1}\sqrt{\frac{-2}{I_{M_3}}} - I_{g_3}
\end{IEEEeqnarray}
Note that \eqref{eq:MGMsumrule1} becomes redundant in the case of universal Higgs masses, because then $I_{Y_\alpha}=D_{Y_{13H}}=0$.

\subsubsection{Anomaly mediation}
Recall that the RG boundary conditions for AMSB are:
\begin{IEEEeqnarray}{rCl}\label{eq:AMSBspectrum}
M_a &=& \frac{1}{16\pi^2}b_ag_a^2m_{3/2} \qquad (a=1,2,3) \IEEEyessubnumber\\
m_i^2 &=& \frac12 \dot{\gamma}_im_{3/2}^2 \IEEEyessubnumber\label{eq:AMSBspectrum2}
\end{IEEEeqnarray}
where we have used the fact that the anomalous dimensions are diagonal (see appendix \ref{a:anomalousdimensions}).
Inserting \eqref{eq:AMSBspectrum} and the expressions in appendix \ref{a:anomalousdimensions} into the RGIs, we immediately find nine sum rules:
\begin{equation}
D_{B_{13}} = D_{L_{13}} = D_{\chi_1} = D_{Y_{13H}} = D_Z = I_{Y_\alpha} = I_{M_1} = I_{M_2} = I_{M_3} = 0
\end{equation}
Note that $D_{B_{13}}$ and $D_{L_{13}}$ vanish although the sfermion masses are not flavour-universal.
The non-vanishing RGIs have the values:
\begin{IEEEeqnarray}{rCl}\label{eq:AMSBRGI}
I_{B_1} &=& \frac{33}{5}\frac{m_{3/2}}{16\pi^2} \IEEEyessubnumber\label{eq:AMSBRGI1}\\
I_{B_2} &=& \frac{m_{3/2}}{16\pi^2} \IEEEyessubnumber\label{eq:AMSBRGI2}\\
I_{B_3} &=& -3\frac{m_{3/2}}{16\pi^2} \IEEEyessubnumber\label{eq:AMSBRGI3}\\
I_{g_2} &=& g_1^{-2} - \frac{33}{5}g_2^{-2} \IEEEyessubnumber\label{eq:AMSBRGI4}\\
I_{g_3} &=& g_1^{-2} + \frac{11}{5}g_3^{-2} \IEEEyessubnumber\label{eq:AMSBRGI5}
\end{IEEEeqnarray}
where the gauge couplings should be evaluated at the scale of supersymmetry breaking.
This amounts to five equations with four unknowns, but we can do better: equations \eqref{eq:AMSBRGI1}-\eqref{eq:AMSBRGI3} constitute three equations with one unknown.
This yields two additional sum rules:
\begin{align}
0 &= I_{B_1} - \frac{33}{5}I_{B_2} \\
0 &= I_{B_1} + \frac{11}{5}I_{B_3}
\end{align}

\paragraph{Supersymmetry breaking at $\mathbf{M_\text{gc}}$}
If both AMSB and gauge coupling unification turn out to be consistent with experimental data, we may ask ourselves whether supersymmetry breaking occurs at the scale $M_\text{gc}$ of gauge coupling unification.
In that case we should insert $g_a=g$ into \eqref{eq:AMSBRGI}.
But this will only affect equations \eqref{eq:AMSBRGI4} and \eqref{eq:AMSBRGI5}, which we have not used to make the above sum rules.
This amounts to two equations with only one parameter, so we get one additional sum rule.
This must be the sum rule \eqref{eq:gcusumrule} for gauge coupling unification, hence there are no sum rules that test whether supersymmetry breaking occurs at $M_\text{gc}$.

\paragraph{Minimal anomaly mediation}
Recall that in minimal AMSB, a universal additional term $m_0^2$ is added to the soft scalar masses \eqref{eq:AMSBspectrum2}:
\begin{equation}
m_i^2 = \frac12 \dot{\gamma}_im_{3/2}^2 + m_0^2
\end{equation}
If we insert this into the RGI expressions, we immediately find five sum rules:
\begin{equation}
D_{B_{13}} = D_{L_{13}} = D_{Y_{13H}} = D_Z = I_{Y_\alpha} = 0 \\
\end{equation}
The non-vanishing RGIs have the values:
\begin{IEEEeqnarray}{rCl}
D_{\chi_1} &=& 5m_0^2 \IEEEyessubnumber\\
I_{B_1} &=& \frac{33}{5}\frac{m_{3/2}}{16\pi^2} \IEEEyessubnumber\\
I_{B_2} &=& \frac{m_{3/2}}{16\pi^2} \IEEEyessubnumber\\
I_{B_3} &=& -3\frac{m_{3/2}}{16\pi^2} \IEEEyessubnumber\\
I_{M_1} &=& \frac{33}{8}m_0^2 \IEEEyessubnumber\\
I_{M_2} &=& \frac58m_0^2 \IEEEyessubnumber\\
I_{M_3} &=& -\frac{15}{16}m_0^2 \IEEEyessubnumber\\
I_{g_2} &=& g_1^{-2} - \frac{33}{5}g_2^{-2} \IEEEyessubnumber\label{eq:mAMSBRGI1}\\
I_{g_3} &=& g_1^{-2} + \frac{11}{5}g_3^{-2} \IEEEyessubnumber\label{eq:mAMSBRGI2}
\end{IEEEeqnarray}
where again the gauge couplings should be evaluated at the scale of supersymmetry breaking.
This adds up to nine equations with five unknowns, so we expect to find four additional sum rules.
However if we leave out equations \eqref{eq:mAMSBRGI1}-\eqref{eq:mAMSBRGI2}, we are left with seven equations with only two unknowns.
This yields five sum rules:
\begin{align}
0 &= I_{B_1} - \frac{33}{5}I_{B_2} \label{eq:mAMSB1}\\
0 &= I_{B_1} + \frac{11}{5}I_{B_3} \label{eq:mAMSB2}\\
0 &= D_{\chi_1} - \frac{40}{33}I_{M_1} \label{eq:mAMSB3}\\
0 &= D_{\chi_1} - 8I_{M_2} \label{eq:mAMSB4}\\
0 &= D_{\chi_1} + \frac{16}{3}I_{M_3} \label{eq:mAMSB5}
\end{align}
Here equations \eqref{eq:mAMSB1}-\eqref{eq:mAMSB2} also hold for AMSB.
Equations \eqref{eq:mAMSB3}-\eqref{eq:mAMSB5} are automatically satisfied in AMSB because $D_{\chi_1}$ and $I_{M_a}$ vanish.

\subsection{Discussion}\label{s:discussion}
In this section, we have found (a) sum rules that test general properties of the RG boundary conditions and (b) sum rules that test the consistency of specific model spectra.
Comparing both sets of sum rules will help us to determine how good the sum rules are at distinguishing between several properties and model spectra.
In the sum rules we observe the following ambiguities:
\begin{itemize}
\item If the sum rules \eqref{eq:gcusumrule} for gauge coupling unification and \eqref{eq:Mgunification1}-\eqref{eq:Mgunification2} for $M_a/g_a^2$ unification are both satisfied, then the sum rule \eqref{eq:GMU} for gaugino mass unification is automatically satisfied. But gaugino mass unification is implied by gauge coupling unification and $M_a/g_a^2$ unification only if both unifications occur at the same scale! Hence, if \eqref{eq:gcusumrule}, \eqref{eq:Mgunification1} and \eqref{eq:Mgunification2} are satisfied by experimental data, then we cannot determine unambiguously whether the gaugino masses unify. At this point, we should use the bottom-up method to examine the running of the parameters. Then we could see whether the unification scales are the same.
\item Equation \eqref{eq:scalargcsumrule} checks whether scalar masses and gauge couplings unify at the same scale. Equation \eqref{eq:messengergcsumrule} checks whether the gauge couplings unify at the messenger scale in GGM. Yet these sum rules happen to be the same. However, this does not mean we cannot distinguish between these two scenarios. The former scenario also requires that the sum rules \eqref{eq:NUHM1}-\eqref{eq:NUHM3} for scalar mass unification are valid. In the latter scenario, these sum rules are not satisfied. Thus the double role of \eqref{eq:scalargcsumrule} poses no problem.
\item In AMSB and mAMSB, the sum rules \eqref{eq:flavouruni1} and \eqref{eq:flavouruni2} for flavour-universality are satisfied, although the sfermion masses in these models are clearly non-universal. Fortunately, (m)AMSB has a lot more sum rules, which could help discern these models from flavour-universal ones. For example, the vanishing of $D_{Y_{13H}}$, $D_Z$ and $I_{Y_\alpha}$ is typical for (m)AMSB. Equations \eqref{eq:NUHM1}-\eqref{eq:NUHM2} then help us discern (m)AMSB from scalar mass unification with universal Higgs masses. Again, satisfying a single sum rule may be ambiguous, but other sum rules eliminate this ambiguity.
\item Because $D_{Y_{13H}}$ and $I_{Y_\alpha}$ vanish in (m)AMSB, the sum rule \eqref{eq:scalargcsumrule} for simultaneous scalar and gauge coupling unification is automatically satisfied. However, the sum rules for scalar mass unification and gauge coupling unification again help us distinguish between both scenarios.
\item The vanishing of $I_{M_a}$ in AMSB and the sum rules \eqref{eq:mAMSB3}-\eqref{eq:mAMSB5} of mAMSB both imply that the sum rule \eqref{eq:scalargauginosumrule} for simultaneous scalar mass and gaugino mass unification is satisfied. However, the sum rules for scalar mass unification and gaugino mass unification help us distinguish between both scenarios.
\end{itemize}

If we only consider the spectrum properties and breaking mechanisms we discussed in this section, our sum rules work surprisingly well.
Many of the sum rules are not unambiguous by themselves, but in most cases the other sum rules remove the ambiguity.
Only when the data are consistent with both gauge coupling unification and $M_a/g_a^2$ unification, we have to resort to other methods (such as the bottom-up method) to determine whether the gaugino masses also unify (or equivalently, whether both unifications occur at the same scale).

Of course, it is possible that a new supersymmetry breaking model is concocted in the future, and that some of its corresponding sum rules introduce similar ambiguities.
These may or may not be resolved by other sum rules.
Therefore, we should keep in mind that if the sum rules of a model or hypothesis are satisfied, this is not a confirmation that it is correct.
\emph{The true power of our sum rules is their falsifying power: the failure to satisfy just one sum rule implies that the corresponding hypothesis or model is incorrect.}
\vskip3ex
\noindent Now that we have an idea of the quality of the RGI sum rules, we can finally examine the advantages and limitations of the RGI method.

\paragraph{Advantages of RGIs}
\begin{itemize}
\item The RGI method requires less input than the other methods we have discussed. We only need the values of all soft masses and gauge couplings at one scale. These are sufficient to reconstruct the values of the RGIs in table \ref{t:mssmrgis}. In contrast to the bottom-up method, we do not need the values of the Yukawa couplings, soft trilinear couplings and $\mu$, $b$ because we could not use them anyway. Also, the value of the new physics threshold does not have to be known.
\item The RGI method is very simple: it is entirely algebraical and does not require the numerical integration of Renormalisation Group equations. Therefore it avoids the complicated propagation of errors between the collider scale and the new physics threshold. Also, it is not as time-consuming as the top-down method.
\end{itemize}

\paragraph{Limitations of the RGI method}
\begin{itemize}
\item As we mentioned before, RG invariance only holds up to a certain loop level. The RGIs in table \ref{t:mssmrgis} have been determined using the one-loop RG equations. Higher order loop effects will certainly spoil RG invariance. We could of course try to find RGIs for the MSSM at a higher loop order. But already at the two-loop level the RG equations for the MSSM (see e.g.\ \cite{twoloopRGEs}) are too complicated to retain the simplicity of this method, if it is possible to find RGIs at all.

However, the relevant question is to what extent we should worry about this approximate RG invariance. It has been demonstrated in \cite{Carena1} that two-loop contributions to the RGIs are smaller than the expected experimental errors of the one-loop RGIs, even in the optimistic scenario of 1\% experimental uncertainties in the determination of soft masses at the collider scale. Thus for all practical purposes we can safely treat the one-loop RGIs as true invariants.
\item It may seem like the RGI method magically reduces the uncertainties of the running parameters, compared to RG-evolved parameters. However, we have paid a price for this reduction, namely \emph{information}. We can directly see this from table \ref{t:mssmrgis}: we started with 18 running parameters (12 scalar masses, 3 gaugino masses and 3 gauge couplings) and have reduced them to only 14 invariants.

We can easily understand why we have to give up information to gain smaller errors. Consider for example the RG equations for $\msqa$ and $\msqc$ (see appendix \ref{a:mssmrges} for the definitions of $D_Y,X_t,X_b$):
\begin{IEEEeqnarray}{rCl}
16\pi^2\frac{\d \msqa}{\d t} &=& -\frac{2}{15}g_1^2M_1^2 - 6g_2^2M_2^2 - \frac{32}{3}g_3^2M_3^2 + \frac15g_1^2D_Y \IEEEyessubnumber\\
16\pi^2\frac{\d \msqc}{\d t} &=& X_t + X_b -\frac{2}{15}g_1^2M_1^2 - 6g_2^2M_2^2 - \frac{32}{3}g_3^2M_3^2 + \frac15g_1^2D_Y \qquad\quad \IEEEyessubnumber
\end{IEEEeqnarray}
Note that in the RG equations of all soft masses, dependence on the gaugino mass $M_2$ occurs only as terms proportional to $g_2^2M_2^2$.
Hence we can eliminate the $M_2$ dependence by taking suitable linear combinations of MSSM parameters. For example, the RG equation for the quantity $\msqa-\msqc$ (which occurs in $D_{B_{13}}$) does not depend on $M_2$ any more, so its experimental uncertainty will spread less under RG flow. However, in this process we have thrown away information about the value of $\msqa+\msqc$. Thus we have to reduce the number of independent quantities to reduce the spread of uncertainties under RG flow.

This may become a limitation of the RGI method in the following sense. A minimal model such as mSUGRA, with only three parameters that govern the soft masses plus gauge couplings at the GUT scale ($m_0,M_{1/2},g\equiv g_a(M_\text{GUT})$), allows us to construct sum rules because we have more RGIs than mSUGRA has parameters. However, if we have a not-so-minimal model with (say) 15 parameters that determine the high scale spectrum, we do not have enough RGIs to make any sum rules.\footnote{That is, unless a subset of $n$ RGIs accidentally depends on less than $n$ model parameters.} Hence, despite the simplicity of the method, we are still limited to models with few parameters.
\item The applicability of the RGI method to the study of supersymmetry breaking depends crucially on the assumption that the MSSM Renormalisation Group equations are valid all the way up to the scale of supersymmetry breaking. But suppose that in Nature, a new field $\Phi$ (or possibly more than one) enters the theory at a high scale $\mu_\Phi$ that is not the scale of supersymmetry breaking; instead supersymmetry is broken at an even higher scale $\mu_\text{SUSY}$. Then at $\mu_\Phi$ the physical RG trajectories of the running parameters will be deflected from their MSSM trajectories. Thus we might mistakingly see gaugino mass unification where it is absent, or vice versa. Hence, if we want to study supersymmetry breaking directly from RGIs, we have to assume that new physics, if present, does not alter the one-loop RG equations for the MSSM up to the scale of supersymmetry breaking.
\item In order to reconstruct the values of all RGIs, all soft masses and gauge couplings need to be known at one energy scale. This may prove difficult in practice. First of all, due to mixing effects the gauge eigenstates do not always correspond to the mass eigenstates (see section \ref{s:charginosneutralinos}). Reconstructing the soft masses from measured pole masses will introduce additional uncertainties. Furthermore, determining all soft masses and gauge couplings is one thing, but determining all of them \emph{at the same energy scale} may prove difficult. Note however that the bottom-up method also suffers from these complications.
\end{itemize}

\section{Conclusions \& outlook}\label{s:conclusion}
In this master thesis, we have examined a method for determining the mechanism that breaks supersymmetry.
We have argued that important clues are to be found in patterns between the high scale soft supersymmetry breaking parameters.
Also, we have seen that we need the Renormalisation Group to study the high scale spectrum.
We have discussed several methods to do this, and proposed a new strategy to make effective use of Renormalisation Group invariants.

Assuming that the MSSM is an appropriate EFT beyond the Standard Model, we have constructed a set of RG invariant sum rules that test properties that are common in supersymmetry breaking models.
If a certain property is realised in Nature, all corresponding sum rules must be satisfied.
None of these sum rules refer to any parameters that are specific to some supersymmetry breaking mechanism.
Therefore, the sum rules are useful regardless of the way supersymmetry has been broken in Nature.

We have also considered sum rules that are tailor-made for testing specific supersymmetry breaking mechanisms.
We have used these to determine the effectiveness of our sum rules by looking for ambiguities among the sum rules.
It was found that some sum rules do not provide unambiguous checks by themselves; however, in almost all cases the other sum rules lift the ambiguity.
Hence, in light of the currently known supersymmetry breaking mechanisms, our sum rules are surprisingly effective.
When they are not, one may have to resort to other methods to resolve the ambiguity.

It is possible that new breaking mechanisms will be proposed in the future, and that their corresponding sum rules introduce new ambiguities.
Therefore, it should be kept in mind that the main strength of RG invariant sum rules is their falsifying power.
If we are able to determine all soft masses and gauge couplings, RGIs will put severe constraints on any realistic model of supersymmetry breaking.
In conclusion, RGIs provide a simple yet powerful tool to check whether such models are compatible with experiment.

Apart from establishing the validity of the MSSM, the main obstacle to using our RG invariant sum rules is the necessity of knowing all soft masses and gauge couplings at one scale.
Therefore, an important topic for future study will be to determine how well this can be done.
Another important issue is to find out how we can reconstruct the soft mass parameters from the sparticle mass eigenstates.

It is possible that the next EFT beyond the Standard Model is not the MSSM.
It may as well be a non-minimal supersymmetric extension of the SM, or even a non-supersymmetric theory.
Nevertheless, our scheme for probing high scale properties of running parameters may be applied just the same.
In order to perform an analogous study, we need to determine the particle content, interactions and $\beta$-functions of the appropriate EFT.
Then we should find all independent RGIs for this EFT and construct sum rules in the same way that we have done in this work.
Given the large amount of structure in the MSSM $\beta$-functions, i.e.\ the limited number of combinations in which running parameters appear in them, an interesting topic is to determine the form of the (one-loop) $\beta$-functions for a theory more general than the MSSM, and to see what RGIs can be found directly for such a general theory.
We leave these issues to future work.

\appendix

\section{Feynman rules}\label{a:feynmanrules}
This appendix lists the momentum space Feynman rules used in this thesis.
\subsubsection*{$\phi^3$ theory}
\begin{fmffile}{appendixfeynmanrules}
\begin{align}
\text{Scalar propagator:}\qquad\parbox{20mm}{
  \begin{fmfgraph*}(20,15)
    \fmfleft{in}
    \fmfright{out}
    \fmf{scalar,label=$p$}{in,out}
  \end{fmfgraph*}}
&= \frac{i}{p^2-m_0^2+i\epsilon} \\
\text{Trilinear scalar interaction:}\qquad\parbox{20mm}{
  \begin{fmfgraph*}(20,15)
    \fmfleft{in}
    \fmfright{out1,out2}
    \fmf{dashes}{out2,v,out1}
    \fmf{dashes,tension=1.5}{v,in}
  \end{fmfgraph*}}
&= -ig_0 
\end{align}

\subsubsection*{Higgs sector}
\begin{align}
\text{Scalar propagator:}\qquad\parbox{20mm}{
  \begin{fmfgraph*}(20,15)
    \fmfleft{in}
    \fmfright{out}
    \fmf{scalar,label=$p$}{in,out}
  \end{fmfgraph*}}
&= \frac{i}{p^2-m^2+i\epsilon} \\
\text{Fermion propagator:}\qquad\parbox{20mm}{
  \begin{fmfgraph*}(20,15)
    \fmfleft{in}
    \fmfright{out}
    \fmf{fermion,label=$p$}{in,out}
  \end{fmfgraph*}}
&= \frac{i(\gamma^\mu p_\mu + m)}{p^2-m^2+i\epsilon} \\
\text{Yukawa interaction:}\qquad\parbox{20mm}{
  \begin{fmfgraph*}(20,15)
    \fmfleft{in}
    \fmfright{out1,out2}
    \fmf{scalar}{in,v}
    \fmf{fermion}{out1,v,out2}
  \end{fmfgraph*}}
&= -iy \\
\text{Quartic scalar self-interaction:}\qquad\parbox{20mm}{
  \begin{fmfgraph*}(20,15)
    \fmfleft{in1,in2}
    \fmfright{out1,out2}
    \fmf{scalar}{in1,v,in2}
    \fmf{scalar}{out1,v,out2}
  \end{fmfgraph*}}
&= -i\lambda
\end{align}
\end{fmffile}

\section{One-loop self-energy in $\phi^3$ theory}\label{a:oneloopselfenergy}
In order to calculate loop diagrams, one has to deal with integrals over Minkowskian 4-momenta.
Such integrals require a toolbox of tricks to solve them.
This appendix illustrates some of them using the one-loop self-energy in $\phi^3$ theory as an example.

The one-loop self-energy is given by:
\begin{align}
\Sigma_1(p) &\equiv i\times \reusediagram{35}{oneloopselfenergy} \nonumber\\
&= \frac{ig_0^2}{2}\int\frac{\mathrm{d}^4k}{(2\pi)^4}\frac{1}{\left((k-p)^2-m_0^2+i\epsilon\right)\left(k^2-m_0^2+i\epsilon\right)}
\end{align}
The strategy for solving such an integral is as follows: first we squeeze the two denominator factors into the square of a single quadratic polynomial in $k$.
Then we complete the square and shift the integration variable in order to eliminate the linear term from this polynomial.
After that, we turn the integral over the Minkowskian four-momentum into an integral over a \emph{Euclidean} four-momentum.
Finally, we are able to switch to spherical coordinates to drastically simplify the integral.

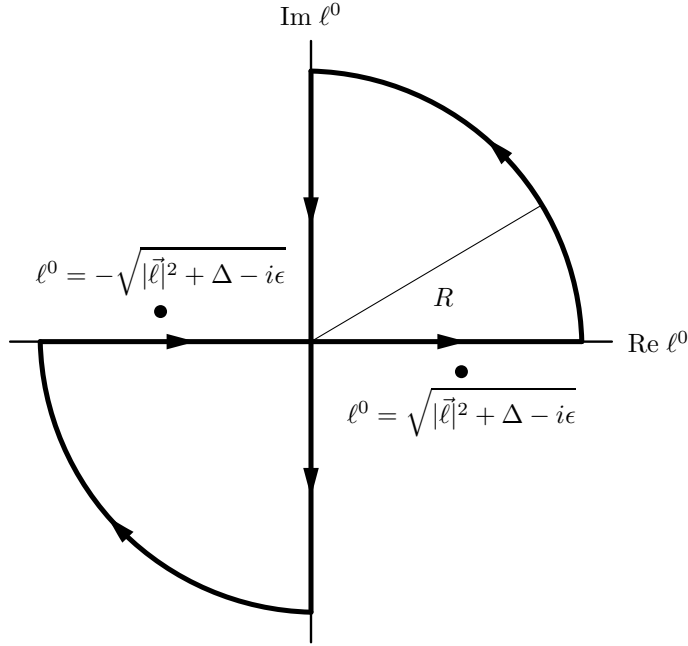
\begin{figure}[t]
\begin{fmffile}{wickrotation}
\begin{minipage}{\textwidth}
\begin{center}
  \begin{fmfgraph*}(80,80)
    \fmfforce{(.5w,h)}{no}
    \fmfforce{(w,.5h)}{eo}
    \fmfforce{(.5w,0)}{so}
    \fmfforce{(0,.5h)}{wo}
    \fmfforce{(.5w,.95h)}{n}
    \fmfforce{(.95w,.5h)}{e}
    \fmfforce{(.5w,.05h)}{s}
    \fmfforce{(.05w,.5h)}{w}
    \fmfforce{(.5w,.5h)}{c}
    \fmfforce{(.75w,.45h)}{pp}
    \fmfforce{(.25w,.55h)}{pm}
    \fmfforce{(.88w,.725h)}{r}
    \fmf{plain}{wo,eo}
    \fmf{plain}{no,so}
    \fmf{plain_arrow,width=2pt}{w,c,e}
    \fmf{plain_arrow,width=2pt}{n,c,s}
    \fmf{plain_arrow,width=2pt,right=.4}{e,n}
    \fmf{plain_arrow,width=2pt,left=.4}{s,w}
    \fmf{plain,label=$R$,width=.5pt}{c,r}
    \fmfdot{pp,pm}
    \fmfv{label=$\text{Re}\;\ell^0$,label.angle=0}{eo}
    \fmfv{label=$\text{Im}\;\ell^0$,label.angle=90}{no}
    \fmfv{label=$\ell^0=\sqrt{|\vec{\ell}|^2+\Delta-i\epsilon}$,label.angle=-90}{pp}
    \fmfv{label=$\ell^0=-\sqrt{|\vec{\ell}|^2+\Delta-i\epsilon}$,label.angle=90}{pm}
  \end{fmfgraph*}
\end{center}
\end{minipage}
\end{fmffile}
\caption{`Figure eight' contour in the complex $\ell^0$-plane. The integrand in \eqref{eq:l0integral} has poles at $\ell^0=\pm\sqrt{|\vec{\ell}|^2+\Delta-i\epsilon}$, which lie outside this contour. It follows from Cauchy's integral theorem that for $R\rightarrow\infty$ the integral from $-\infty$ to $+\infty$ equals the integral from $-i\infty$ to $+i\infty$. In other words, we are allowed to \emph{Wick-rotate} the $\ell^0$ integration contour of \eqref{eq:l0integral} $90^\circ$ counterclockwise.}\label{f:wickrotation}
\end{figure}

For the first step we use the following identity:
\begin{equation}
\frac{1}{A_1A_2\ldots A_n} = \int_0^1\mathrm{d}x_1\ldots\mathrm{d}x_n \;\delta\left(\sum_ix_i-1\right) \frac{(n-1)!}{\left[x_1A_1+x_2A_2+\ldots+x_nA_n\right]^n}
\end{equation}
The variables $x_1,\ldots,x_n$ are called \emph{Feynman parameters}.
Introducing them for $n=2$ and completing the square, we find (after integrating out the delta function):
\begin{align}
\Sigma_1(p) &= \frac{ig_0^2}{2}\int\frac{\mathrm{d}^4k}{(2\pi)^4} \int_0^1\mathrm{d}x \frac{1}{\left[x\left((k-p)^2-m_0^2+i\epsilon\right)+(1-x)(k^2-m_0^2+i\epsilon)\right]^2} \nonumber\\
&= \frac{ig_0^2}{2}\int_0^1\mathrm{d}x\int\frac{\mathrm{d}^4k}{(2\pi)^4} \frac{1}{\left[k^2-2xp\cdot k+xp^2-m_0^2+i\epsilon\right]^2} \nonumber\\
&= \frac{ig_0^2}{2}\int_0^1\mathrm{d}x\int\frac{\mathrm{d}^4k}{(2\pi)^4} \frac{1}{\left[(k-xp)^2-x^2p^2+xp^2-m_0^2+i\epsilon\right]^2}
\end{align}
Now we define $\ell\equiv k-xp$, $\Delta\equiv-x(1-x)p^2+m_0^2$:
\begin{align}
\Sigma_1(p) &= \frac{ig_0^2}{2}\int_0^1\mathrm{d}x \int\frac{\mathrm{d}^4\ell}{(2\pi)^4} \frac{1}{\left[\ell^2-\Delta+i\epsilon\right]^2} \nonumber\\
&= \frac{ig_0^2}{2}\int_0^1\mathrm{d}x \int\frac{\mathrm{d}^3\vec{\ell}}{(2\pi)^3} \int_{-\infty}^\infty\frac{\mathrm{d}\ell^0}{2\pi} \frac{1}{\left[(\ell^0)^2-|\vec{\ell}|^2-\Delta+i\epsilon\right]^2} \label{eq:l0integral}
\end{align}
At this point, let us consider the $\ell^0$-integral more closely.
In the complex $\ell^0$-plane, the integrand has poles at $\ell^0=\pm\sqrt{|\vec{\ell}|^2+\Delta-i\epsilon}$.
Hence according to Cauchy's integral theorem, integration along the `figure eight' contour in figure~\ref{f:wickrotation} yields zero, since it encloses no poles.
As we let the radius of the two quarter-circles become infinitely large, the contribution of the integral along these arcs will go to zero: the arc length increases as $R$ while the integrand decreases as $1/R^4$.
It follows that:
\begin{equation}
\Sigma_1(p) = \frac{ig_0^2}{2}\int_0^1\mathrm{d}x \int\frac{\mathrm{d}^3\vec{\ell}}{(2\pi)^3} \int_{-i\infty}^{i\infty}\frac{\mathrm{d}\ell^0}{2\pi} \frac{1}{\left[(\ell^0)^2-|\vec{\ell}|^2-\Delta+i\epsilon\right]^2}
\end{equation}
This trick is known as \emph{Wick rotation}, since we are effectively rotating the integration contour $90^\circ$ such that it crosses no poles (that is, counterclockwise).
Now we can turn the integral over the Minkowskian four-momentum $\ell$ into an integral over a Euclidean four-momentum $\ell_E$ by substituting $i\ell_E^0\equiv\ell^0$, $\vec{\ell}_E\equiv\vec{\ell}$:
\begin{align}
\Sigma_1(p) &= \frac{ig_0^2}{2}\int_0^1\mathrm{d}x \int\frac{\mathrm{d}^3\vec{\ell}_E}{(2\pi)^3} \int_{-\infty}^{\infty}\frac{i\mathrm{d}\ell_E^0}{2\pi}\frac{1}{\left[(i\ell_E^0)^2-|\vec{\ell}_E|^2-\Delta+i\epsilon\right]^2} \nonumber\\
&= -\frac{g_0^2}{2}\int_0^1\mathrm{d}x\int\frac{\mathrm{d}^4\ell_E}{(2\pi)^4} \frac{1}{\left[\ell_E^2+\Delta-i\epsilon\right]^2}
\end{align}
Now that we have an integral over Euclidean space, we can switch to four-dimensional spherical coordinates:
\begin{equation}
\int\mathrm{d}^4\ell_E \rightarrow \int\mathrm{d}\Omega_4\int_0^\infty\mathrm{d}\ell_E \ell_E^3 = 2\pi^2\int_0^\infty\mathrm{d}\ell_E \ell_E^3
\end{equation}
Then our integral becomes:
\begin{equation}
\Sigma_1(p) = -\frac{g_0^2}{16\pi^2}\int_0^1\mathrm{d}x\int_0^\infty\mathrm{d}\ell_E \frac{\ell_E^3}{\left[\ell_E^2+\Delta-i\epsilon\right]^2}
\end{equation}
At this point we introduce a momentum cutoff $\Lambda$ in the integral:
\begin{align}
\Sigma_{1,\Lambda}(p) &= -\frac{g_0^2}{16\pi^2}\int_0^1\mathrm{d}x\int_0^\Lambda\mathrm{d}\ell_E \frac{\ell_E^3}{\left[\ell_E^2+\Delta-i\epsilon\right]^2} \nonumber\\
&= -\frac{g_0^2}{16\pi^2}\int_0^1\mathrm{d}x\frac12\left.\left(\log{\left(\ell_E^2+\Delta-i\epsilon\right)}+\frac{\Delta-i\epsilon}{\ell_E^2+\Delta-i\epsilon}\right)\right\arrowvert_0^\Lambda \nonumber\\
&= -\frac{g_0^2}{32\pi^2}\int_0^1\mathrm{d}x\left[\log{\left(\frac{\Lambda^2+\Delta-i\epsilon}{\Delta-i\epsilon}\right)} + \frac{\Delta-i\epsilon}{\Lambda^2+\Delta-i\epsilon} - 1\right] \nonumber\\
&= -\frac{g_0^2}{32\pi^2}\int_0^1\mathrm{d}x\left[\log{\left(\frac{\Lambda^2+\Delta-i\epsilon}{\Delta-i\epsilon}\right)} - \frac{\Lambda^2}{\Lambda^2+\Delta-i\epsilon}\right]
\end{align}
The last term is finite for $\Lambda\rightarrow\infty$, but the first term diverges logarithmically in $\Lambda$, as we already observed in section \ref{s:renormalisationprocedure} from power counting.

\section{Renormalisation Group equations of the MSSM}\label{a:mssmrges}
In this appendix we give the Renormalisation Group equations of the MSSM that have been used in this thesis.
They have been taken from \cite{martin} and are one-loop equations that have been simplified by making the following approximations:
\begin{itemize}
\item The soft supersymmetry breaking mass matrices are assumed to be flavour diagonal. The first and second generation masses are assumed to be degenerate.
\item The soft supersymmetry breaking trilinear couplings are taken proportional to the Yukawa couplings: $\mathbf{a_u}=A_u\mathbf{y_u}$, $\mathbf{a_d}=A_d\mathbf{y_d}$, $\mathbf{a_e}=A_e\mathbf{y_e}$.
\item First and second generation Yukawa and soft trilinear couplings are neglected.
\end{itemize}
For general two-loop RG equations, see e.g.\ \cite{twoloopRGEs}.
It is convenient to use the $\beta$-functions, which differ from their corresponding RG equations by a constant:
\begin{equation}
\beta(p) \equiv 16\pi^2\frac{\d p}{\d t}
\end{equation}
Here $p$ is a running parameter and $t\equiv\log{\left(\mu/\mu_0\right)}$, where $\mu$ is the renormalisation scale and $\mu_0$ an (arbitrary) energy scale that makes the argument of the logarithm dimensionless.
Under the above approximations, we are left with the following running parameters:
\begin{IEEEeqnarray*}{ll}
g_a \qquad(a=1,2,3) & \text{Gauge couplings} \\
M_a \qquad(a=1,2,3) & \text{Soft supersymmetry breaking gaugino masses} \\
\msq, \msu, \msd, \msl, \mse \quad & \text{Soft supersymmetry breaking sfermion masses} \\
\mhu, \mhd & \text{Soft supersymmetry breaking Higgs mass parameters} \\
y_t, y_b, y_\tau & \text{Yukawa couplings for the third generation (s)fermions} \\
A_t, A_b, A_\tau & \text{Soft supersymmetry breaking trilinear couplings for the} \\
& \text{third generation sfermions} \\
\mu & \text{Supersymmetry respecting Higgs mixing parameter} \\
B & \text{Soft supersymmetry breaking Higgs mixing parameter}
\end{IEEEeqnarray*}
Here we use the soft Higgs mixing parameter $B=b/\mu$ rather than $b$ because its $\beta$-function is simpler.
For the sfermion masses, we denote the first and third generation with a subscript 1 and 3 respectively.

The following notation is used for parameters that enter the RG equations through common combinations of Dynkin indices and quadratic Casimir invariants:
\begin{equation}
b_a = \left(\frac{33}{5}, 1, -3\right)
\end{equation}
It is also convenient to define the following combination of running parameters, which appears in the RG equations of the sfermion masses:
\begin{align}
D_Y &\equiv \text{Tr}\left(Ym^2\right) \nonumber \\
&= \sum_\text{gen} \left(\msq-2\msu+\msd-\msl+\mse\right) + \mhu-\mhd
\end{align}
Here the trace runs over all chiral multiplets and the sum runs over the three sfermion generations.
Note that $D_Y$ is often called $S$ in the literature.
Furthermore, we define the useful combinations:
\begin{IEEEeqnarray}{lCl}
X_t &=& 2|y_t|^2\left(\mhu + \msqc + \msuc + |A_t|^2\right) \IEEEyessubnumber\\
X_b &=& 2|y_b|^2\left(\mhd + \msqc + \msdc + |A_b|^2\right) \IEEEyessubnumber\\
X_\tau &=& 2|y_\tau|^2\left(\mhd + \mslc + \msec + |A_\tau|^2\right) \IEEEyessubnumber
\end{IEEEeqnarray}
Then the resulting $\beta$-functions for the MSSM are:
\begin{IEEEeqnarray}{lCl}
\beta(g_a) &=& b_ag_a^3 \qquad (a=1,2,3) \IEEEyessubnumber\\
\beta(M_a) &=& 2b_ag_a^2M_a \qquad (a=1,2,3) \IEEEyessubnumber\\
\beta(m_{\tilde{Q}_{1,2}}^2) &=& -\frac{2}{15}g_1^2M_1^2 - 6g_2^2M_2^2 - \frac{32}{3}g_3^2M_3^2 + \frac{1}{5}g_1^2D_Y \IEEEyessubnumber\\
\beta(m_{\tilde{\bar{u}}_{1,2}}^2) &=& -\frac{32}{15}g_1^2M_1^2 - \frac{32}{3}g_3^2M_3^2 - \frac{4}{5}g_1^2D_Y \IEEEyessubnumber\\
\beta(m_{\tilde{\bar{d}}_{1,2}}^2) &=& -\frac{8}{15}g_1^2M_1^2 - \frac{32}{3}g_3^2M_3^2 + \frac{2}{5}g_1^2D_Y \IEEEyessubnumber\\
\beta(m_{\tilde{L}_{1,2}}^2) &=& -\frac{6}{5}g_1^2M_1^2 - 6g_2^2M_2^2 - \frac{3}{5}g_1^2D_Y \IEEEyessubnumber\\
\beta(m_{\tilde{\bar{e}}_{1,2}}^2) &=& -\frac{24}{5}g_1^2M_1^2 + \frac{6}{5}g_1^2D_Y \IEEEyessubnumber\\
\beta(m_{\tilde{Q}_3}^2) &=& X_t + X_b - \frac{2}{15}g_1^2M_1^2 - 6g_2^2M_2^2 - \frac{32}{3}g_3^2M_3^2 + \frac{1}{5}g_1^2D_Y \IEEEyessubnumber\\
\beta(m_{\tilde{\bar{u}}_3}^2) &=& 2X_t - \frac{32}{15}g_1^2M_1^2 - \frac{32}{3}g_3^2M_3^2 - \frac{4}{5}g_1^2D_Y \IEEEyessubnumber\\
\beta(m_{\tilde{\bar{d}}_3}^2) &=& 2X_b - \frac{8}{15}g_1^2M_1^2 - \frac{32}{3}g_3^2M_3^2 + \frac{2}{5}g_1^2D_Y \IEEEyessubnumber\\
\beta(m_{\tilde{L}_3}^2) &=& X_\tau - \frac{6}{5}g_1^2M_1^2 - 6g_2^2M_2^2 - \frac{3}{5}g_1^2D_Y \IEEEyessubnumber\\
\beta(m_{\tilde{\bar{e}}_3}^2) &=& 2X_\tau - \frac{24}{5}g_1^2M_1^2 + \frac{6}{5}g_1^2D_Y \IEEEyessubnumber\\
\beta(m_{H_u}^2) &=& 3X_t - \frac{6}{5}g_1^2M_1^2 - 6g_2^2M_2^2 + \frac{3}{5}g_1^2D_Y \IEEEyessubnumber\\
\beta(m_{H_d}^2) &=& 3X_b + X_\tau - \frac{6}{5}g_1^2M_1^2 - 6g_2^2M_2^2 - \frac{3}{5}g_1^2D_Y \IEEEyessubnumber\\
\beta(y_t) &=& y_t\left[6|y_t|^2 + |y_b|^2 - \frac{13}{15}g_1^2 - 3g_2^2 - \frac{16}{3}g_3^2\right] \IEEEyessubnumber\\
\beta(y_b) &=& y_b\left[6|y_b|^2 + |y_t|^2 + |y_\tau|^2 - \frac{7}{15}g_1^2 - 3g_2^2 - \frac{16}{3}g_3^2\right] \IEEEyessubnumber\\
\beta(y_\tau) &=& y_\tau\left[4|y_\tau|^2 + 3|y_b|^2 - \frac{9}{5}g_1^2 - 3g_2^2\right] \IEEEyessubnumber\\
\beta(\mu) &=& \mu\left[3|y_t|^2 + 3|y_b|^2 + |y_\tau|^2 - \frac{3}{5}g_1^2 - 3g_2^2\right] \IEEEyessubnumber\\
\beta(A_t) &=& 12A_t|y_t|^2 + 2A_b|y_b|^2 + \frac{26}{15}g_1^2M_1 + 6g_2^2M_2 + \frac{32}{3}g_3^2M_3 \IEEEyessubnumber\\
\beta(A_b) &=& 12A_b|y_b|^2 + 2A_t|y_t|^2 + 2A_\tau|y_\tau|^2 \nonumber\\
&&\qquad\qquad\qquad\qquad\quad + \frac{14}{15}g_1^2M_1 + 6g_2^2M_2 + \frac{32}{3}g_3^2M_3 \IEEEyessubnumber\\
\beta(A_\tau) &=& 8A_\tau|y_\tau|^2 + 6A_b|y_b|^2 + \frac{18}{5}g_1^2M_1 + 6g_2^2M_2 \IEEEyessubnumber\\
\beta(B) &=& 6A_t|y_t|^2 + 6A_b|y_b|^2 + 2A_\tau|y_\tau|^2 + \frac{6}{5}g_1^2M_1 + 6g_2^2M_2 \IEEEyessubnumber
\end{IEEEeqnarray}

\section{Anomalous dimensions}\label{a:anomalousdimensions}
This appendix lists the expressions for the anomalous dimensions that have been used in this thesis.
The anomalous dimensions $\gamma^i_j$ at one-loop order are given by \cite{martin}:
\begin{equation}
\gamma^i_j = \frac{1}{16\pi^2}\left[\frac12y^{imn}y_{jmn}^* - 2g_a^2C_a(i)\delta^i_j\right]
\end{equation}
Here the label $i$ is not to be summed over; the labels $i,j,m,n$ denote particles with a specific colour and weak isospin label.
The $y$'s are Yukawa couplings\footnote{Note that these are the Yukawa couplings as defined in \eqref{eq:chiralsuperpotential}; the Yukawa couplings in the MSSM superpotential \eqref{eq:mssmsuperpotential} are symmetrised versions of $y^{ijk}$.} and the $g_a$ are gauge couplings.
$C_a(i)$ is defined as the quadratic Casimir invariant $c_2(\rho)$ of the representation $\rho$ of the field $\Phi_i$ under the gauge group labelled by $a$.\footnote{I.e.\ $a=1$ for $U(1)_Y$, $a=2$ for $SU(2)_L$ and $a=3$ for $SU(3)_C$.}
The quadratic Casimir invariant $c_2(\rho)$ of the representation $\rho$ is defined in terms of the Lie algebra generators $T^b$ as:
\begin{equation}
\sum_b \rho(T^b)^2 = c_2(\rho)I
\end{equation}
where $\rho(T^b)$ is the generator $T^b$ belonging to the representation $\rho$ and $I$ is the identity matrix.
For the MSSM supermultiplets, the explicit values of the $C_a(i)$ are:
\begin{IEEEeqnarray}{rCl}
C_1(i) &=& \frac35Y_i^2 \text{ for each $\Phi_i$ with hypercharge $Y_i$.} \IEEEyessubnumber\\
C_2(i) &=& \begin{cases} \frac34 & \text{ for $\Phi_i=\widetilde{Q},\widetilde{L},H_u,H_d$} \\ 0 & \text{ for $\Phi_i=\widetilde{u}_R,\widetilde{d}_R,\widetilde{e}_R$} \end{cases} \IEEEyessubnumber\\
C_3(i) &=& \begin{cases} \frac43 & \text{ for $\Phi_i=\widetilde{Q},\widetilde{u}_R,\widetilde{d}_R$} \\ 0 & \text{ for $\Phi_i=\widetilde{L},\widetilde{e}_R,H_u,H_d$} \end{cases} \IEEEyessubnumber
\end{IEEEeqnarray}
Again, we assume that only the Yukawa couplings of the third generation are significant.
Then the anomalous dimensions become diagonal matrices, with the following values at one-loop order:
\begin{IEEEeqnarray}{rCl}
16\pi^2\gamma_{H_u} &=& 3|y_t|^2 - \frac32g_2^2 - \frac{3}{10}g_1^2 \IEEEyessubnumber\\
16\pi^2\gamma_{H_d} &=& 3|y_b|^2 + |y_\tau|^2 -\frac32g_2^2 - \frac{3}{10}g_1^2 \IEEEyessubnumber\\
16\pi^2\gamma_{\widetilde{Q}_i} &=& \delta_{i3}\left(|y_t|^2+|y_b|^2\right) - \frac83g_3^2 - \frac32g_2^2 - \frac{1}{30}g_1^2 \IEEEyessubnumber\\
16\pi^2\gamma_{\widetilde{\bar{u}}_i} &=& \delta_{i3}\cdot 2|y_t|^2 - \frac83g_3^2 - \frac{8}{15}g_1^2 \IEEEyessubnumber\\
16\pi^2\gamma_{\widetilde{\bar{d}}_i} &=& \delta_{i3}\cdot 2|y_b|^2 - \frac83g_3^2 - \frac{2}{15}g_1^2 \IEEEyessubnumber\\
16\pi^2\gamma_{\widetilde{L}_i} &=& \delta_{i3}\cdot |y_\tau|^2 - \frac32g_2^2 - \frac{3}{10}g_1^2 \IEEEyessubnumber\\
16\pi^2\gamma_{\widetilde{\bar{e}}_i} &=& \delta_{i3}\cdot 2|y_\tau|^2 - \frac65g_1^2 \IEEEyessubnumber
\end{IEEEeqnarray}

For the RG boundary conditions of anomaly mediation (see equation \eqref{eq:AMSBboundaryconditions}) we need their derivatives with respect to $t=\ln{\left(\mu/\mu_0\right)}$.
These are given by:
\begin{IEEEeqnarray}{rCl}
(16\pi^2)^2\dot{\gamma}_{H_u} &=& 6|y_t|^2B_t - 3g_2^4 - \frac{99}{25}g_1^4 \IEEEyessubnumber\\
(16\pi^2)^2\dot{\gamma}_{H_d} &=& 6|y_b|^2B_b + 2|y_\tau|^2B_\tau - 3g_2^4 - \frac{99}{25}g_1^4 \IEEEyessubnumber\\
(16\pi^2)^2\dot{\gamma}_{\widetilde{Q}_i} &=& \delta_{i3}\left(2|y_t|^2B_t+2|y_b|^2B_b\right) + 16g_3^4 - 3g_2^4 - \frac{11}{25}g_1^4 \IEEEyessubnumber\\
(16\pi^2)^2\dot{\gamma}_{\widetilde{\bar{u}}_i} &=& \delta_{i3}\cdot4|y_t|^2B_t + 16g_3^4 - \frac{176}{25}g_1^4 \IEEEyessubnumber\\
(16\pi^2)^2\dot{\gamma}_{\widetilde{\bar{d}}_i} &=& \delta_{i3}\cdot4|y_b|^2B_b + 16g_3^4 - \frac{44}{25}g_1^4 \IEEEyessubnumber\\
(16\pi^2)^2\dot{\gamma}_{\widetilde{L}_i} &=& \delta_{i3}\cdot2|y_\tau|^2B_\tau - 3g_2^4 - \frac{99}{25}g_1^4 \IEEEyessubnumber\\
(16\pi^2)^2\dot{\gamma}_{\widetilde{\bar{e}}_i} &=& \delta_{i3}\cdot4|y_\tau|^2B_\tau - \frac{396}{25}g_1^4 \IEEEyessubnumber
\end{IEEEeqnarray}
where we have defined the following quantities for convenience:
\begin{IEEEeqnarray}{rCl}
B_t &\equiv& 6|y_t|^2 + |y_b|^2 - \frac{16}{3}g_3^2 - 3g_2^2 - \frac{13}{15}g_1^2 \IEEEyessubnumber\\
B_b &\equiv& 6|y_b|^2 + |y_t|^2 + |y_\tau|^2 - \frac{16}{3}g_3^2 - 3g_2^2 - \frac{7}{15}g_1^2 \IEEEyessubnumber\\
B_\tau &\equiv& 4|y_\tau|^2 + 3|y_b|^2 - 3g_2^2 - \frac95g_1^2 \IEEEyessubnumber
\end{IEEEeqnarray}

\section{Deriving the one-loop RGIs for the MSSM}\label{a:derivingRGIs}
In this appendix we will derive a maximal set of independent RGIs for the MSSM.
First we will determine invariants that contain the running parameters $\mu$ and $B=b/\mu$.
We will see that there is only one independent RGI for each of them, making them useless for our study.
Then we will argue that we are restricted to RGIs containing only soft masses and/or gauge couplings.
We will derive all of them systematically; our approach will be globally the same as in \cite{Carena1}, but using different arguments to show that we do indeed find all RGIs.

Let us consider the parameter $\mu$.
The only $\beta$-function containing $\mu$ is that of $\mu$ itself.
Note that we can write $\beta(\mu)$ more conveniently as:
\begin{equation}
\beta(\log{\mu}) = 3|y_t|^2 + 3|y_b|^2 + |y_\tau|^2 - \frac35g_1^2 - 3g_2^2
\end{equation}
The only other $\beta$-functions containing terms linear in $|y_t|^2,|y_b|^2,|y_\tau|^2$ are those of the logarithms of the Yukawa couplings:
\begin{IEEEeqnarray}{rCl}
\beta(\log{y_t}) &=& 6|y_t|^2 + |y_b|^2 - \frac{13}{15}g_1^2 - 3g_2^2 - \frac{16}{3}g_3^2 \IEEEyessubnumber\\
\beta(\log{y_b}) &=& |y_t|^2 + 6|y_b|^2 + |y_\tau|^2 - \frac{7}{15}g_1^2 - 3g_2^2 - \frac{16}{3}g_3^2 \IEEEyessubnumber\\
\beta(\log{y_\tau}) &=& 3|y_b|^2 + 4|y_\tau|^2 - \frac95g_1^2 - 3g_2^2 \IEEEyessubnumber
\end{IEEEeqnarray}
The terms in the $\beta$-functions proportional to $g_a^2$ can be eliminated by taking linear combinations with logarithms of gauge couplings, of which we can rewrite the $\beta$-functions as:
\begin{equation}
\beta(\log{g_a}) = b_ag_a^2 \qquad (a=1,2,3) \label{eq:loggaugebetafunctions}
\end{equation}
Hence $\mu$ can only appear in an RGI through a linear combination of $\log{\mu}$, $\log{y_t}$, $\log{y_b}$, $\log{y_\tau}$, $\log{g_1}$, $\log{g_2}$ and $\log{g_3}$.\footnote{We could also include logarithms of gaugino masses in these linear combinations, since their $\beta$-functions are also proportional to $g_a^2$. However, in a moment we will construct RGIs from the gauge couplings and gaugino masses only. Any RGI that contains both $\mu$ and the gaugino masses will be a function of those RGIs and the one we are constructing now.}
We have seven $\beta$-functions with six different terms to eliminate (namely terms linear in $|y_t|^2,|y_b|^2,|y_\tau|^2,g_1^2,g_2^2$ or $g_3^2$), so we can make one RG invariant linear combination of them.
Using elementary linear algebra we find that the linear combination
\begin{align}
&-\frac{27}{61}\log{y_t} - \frac{21}{61}\log{y_b} - \frac{10}{61}\log{y_\tau} + \log{\mu} - \frac{1}{61}\cdot\frac{73}{33}\log{g_1} \nonumber\\
&\qquad\qquad+ \frac{9}{61}\log{g_2} + \frac{1}{61}\cdot\frac{256}{3}\log{g_3} \nonumber \\
=& \log{\left(\mu\left[\frac{g_2^9g_3^{256/3}}{y_t^{27}y_b^{21}y_\tau^{10}g_1^{73/33}}\right]^{1/61}\right)}
\end{align}
has a vanishing $\beta$-function.
Thus we can choose the only independent RGI containing $\mu$ to be:
\begin{equation}
I_2 \equiv \mu\left[\frac{g_2^9g_3^{256/3}}{y_t^{27}y_b^{21}y_\tau^{10}g_1^{73/33}}\right]^{1/61}
\end{equation}
using the notation of \cite{Demir}.
To summarise, we have found a set of independent RGIs containing $\mu$ (in this case only one) by considering what terms in the MSSM $\beta$-functions could cancel each other.
This will be our general strategy for finding all RGIs of the MSSM, because the running parameters only enter the $\beta$-functions in a very limited number of combinations (e.g.\ the soft scalar masses only appear in the linear combinations $D_Y$, $X_t$, $X_b$ and $X_\tau$).

Now we turn to the parameter $B$.
It does not appear in any of the MSSM $\beta$-functions itself.
Its $\beta$-function contains only terms linear in $A_t|y_t|^2$, $A_b|y_b|^2$, $A_\tau|y_\tau|^2$, $g_1^2M_1$, $g_2^2M_2$ and $g_3^2M_3$.
The $\beta$-functions of $A_t$, $A_b$, $A_\tau$, $M_1$, $M_2$ and $M_3$ also contain only these terms, so $B$ should always appear in RGIs in a linear combination of these parameters.
This gives us seven $\beta$-functions with six different terms to eliminate, so again we can make one RG invariant linear combination.
Using elementary linear algebra this combination is found to be:
\begin{equation}
I_4 \equiv B - \frac{27}{61}A_t -\frac{21}{61}A_b -\frac{10}{61}A_\tau -\frac{256}{183}M_3 -\frac{9}{61}M_2 +\frac{73}{2013}M_1
\end{equation}

Indeed we have found only one independent RGI containing $\mu$ and one containing $B$.
As was argued at the end of section \ref{s:RGIs}, RGIs are only useful as long as their constituent running parameters also appear in other RGIs.
This is not the case for $I_2$ and $I_4$, so we are restricted to RGIs that contain neither $\mu$ nor $B$.
But in the above procedure, we needed their $\beta$-functions to eliminate the $|y_i|^2$ and $A_i|y_i|^2$ dependence respectively from the $\beta$-function of the RGI under construction!
If we wish to construct RGIs containing the Yukawa couplings without using $\mu$, we have to eliminate three different $|y_i|^2$ terms using three $\beta$-functions, so we cannot make any RG invariant combinations.
Similarly, we cannot make any RGIs containing the soft trilinear couplings without using $B$, because we have to eliminate three different $A_i|y_i|^2$ terms using three $\beta$-functions.

Thus if we want to construct RGIs without using $\mu$ and $B$, we cannot use the Yukawa and soft trilinear couplings either: we do not have enough equations to eliminate all terms from the $\beta$-function of the RGI under construction.
Therefore, from now on we will only consider RGIs that are functions of soft masses and/or gauge couplings.

Let us begin with RGIs constructed from the gauge couplings only.
First we rewrite their $\beta$-functions into the more convenient form:
\begin{IEEEeqnarray}{rCl}
\beta(g_1^{-2}) &=& -\frac{66}{5} \IEEEyessubnumber \\
\beta(g_2^{-2}) &=& -2 \IEEEyessubnumber \\
\beta(g_3^{-2}) &=& 6 \IEEEyessubnumber
\end{IEEEeqnarray}
This gives us three equations to eliminate a single term (namely a constant), hence we can make two independent RGIs out of them.
We choose them to be:
\begin{align}
I_{g_2} &\equiv g_1^{-2} - \frac{33}{5}g_2^{-2} \\
I_{g_3} &\equiv g_1^{-2} + \frac{11}{5}g_3^{-2}
\end{align}
Now we turn to the gaugino masses.
First we rewrite their $\beta$-functions as follows:
\begin{IEEEeqnarray}{rCl}
\beta(\log{M_1}) &=& \frac{66}{5}g_1^2 \IEEEyessubnumber \\
\beta(\log{M_2}) &=& 2g_2^2 \IEEEyessubnumber \\
\beta(\log{M_3}) &=& -6g_3^2 \IEEEyessubnumber
\end{IEEEeqnarray}
Together with \eqref{eq:loggaugebetafunctions} this gives six equations with three different terms (namely those proportional to $g_a^2$) to eliminate.
Hence we get three new RGIs by taking linear combinations of $\log{M_a}$ and $\log{g_a}$:
\begin{IEEEeqnarray}{rCl}
0 &=& \beta(\log{M_1} - 2\log{g_1}) = \beta(\log{\frac{M_1}{g_1^2}}) \IEEEyessubnumber \\
0 &=& \beta(\log{M_2} - 2\log{g_2}) = \beta(\log{\frac{M_2}{g_2^2}}) \IEEEyessubnumber \\
0 &=& \beta(\log{M_3} - 2\log{g_3}) = \beta(\log{\frac{M_3}{g_3^2}}) \IEEEyessubnumber
\end{IEEEeqnarray}
Thus we can choose the three independent RGIs to be:
\begin{align}
I_{B_1} &\equiv \frac{M_1}{g_1^2} \\
I_{B_2} &\equiv \frac{M_2}{g_2^2} \\
I_{B_3} &\equiv \frac{M_3}{g_3^2}
\end{align}
Now let us consider RGIs constructed solely from the twelve soft scalar masses.
First we eliminate the Yukawa terms $X_t$, $X_b$, $X_\tau$ and the gaugino mass terms $g_1^2M_1^2$, $g_2^2M_2^2$, $g_3^2M_3^2$ from the $\beta$-function.
Since we have to eliminate six terms using twelve equations, we can make six independent linear combinations of the soft scalar masses that have a $\beta$-function proportional to $g_1^2D_Y$.
Then we can make linear combinations of these quantities such that five of them have a vanishing $\beta$-function and the sixth quantity still runs with $g_1^2D_Y$.
In accordance with \cite{Carena1}, we choose the five RGIs to be:\footnote{The notation used for the RGIs may look odd here. In \cite{Carena1}, they are related to symmetries of the MSSM Lagrangian. In this context, the $D$-term $D_i$ of a charge $Q_i$ (which has nothing to do with the auxiliary component of a gauge supermultiplet!) is defined as $D_i\equiv \text{Tr}(Q_im^2)$, with the trace running over all chiral multiplets. Then one should interpret $D_{B_{13}}$ as \mbox{$D_{B_1}-D_{B_3}$,} where the subscripts 1 and 3 mean that the trace is restricted to the first and third generation sfermions respectively. See \cite{Carena1} for an explanation of the nomenclature for the remaining RGIs.}
\begin{align}
D_{B_{13}} &\equiv 2\left(\msqa-\msqc\right) - \msua + \msuc - \msda + \msdc \\
D_{L_{13}} &\equiv 2\left(\msla-\mslc\right) - \msea + \msec \\
D_{\chi_1} &\equiv 3\left(3\msda-2\left(\msqa-\msla\right)-\msua\right) - \msea \\
D_{Y_{13H}} &\equiv \msqa-2\msua+\msda-\msla+\msea \nonumber\\
&-\frac{10}{13}\left(\msqc-2\msuc+\msdc-\mslc+\msec+\mhu-\mhd\right) \\
D_Z &\equiv 3\left(\msdc-\msda\right) + 2\left(\mslc-\mhd\right)
\end{align}
The sixth quantity, which runs with $g_1^2D_Y$, can be chosen to be $D_Y$ itself because:
\begin{equation}
\beta(D_Y) = \frac{66}{5}g_1^2D_Y
\end{equation}
Note that $\log{D_Y}$ runs with $g_1^2$, so using \eqref{eq:loggaugebetafunctions} we find:
\begin{equation}
\beta(\log{D_Y} - 2\log{g_1}) = \beta(\log{\frac{D_Y}{g_1^2}}) = 0
\end{equation}
This gives us another independent RGI:
\begin{equation}
I_{Y_\alpha} \equiv \frac{D_Y}{g_1^2} = \frac{1}{g_1^2}\left(\mhu-\mhd + \sum_\text{gen}\left(\msq-2\msu+\msd-\msl+\mse\right)\right)
\end{equation}
Finally, we look for RGIs constructed from both scalar masses and gaugino masses.
Note that the gaugino mass $\beta$-functions can be rewritten as:
\begin{equation}
\beta(M_a^2) = 4b_ag_a^2M_a^2 \qquad (a=1,2,3)
\end{equation}
Combining the gaugino masses and scalar masses, we have fifteen $\beta$-functions with seven terms to eliminate, so we can construct eight RGIs by taking linear combinations of the gaugino masses squared and the scalar masses.
Five of them can be made from the scalar masses alone, so there must be three new RGIs.
In accordance with \cite{Carena1}, we take them to be:
\begin{align}
I_{M_1} &\equiv M_1^2 - \frac{33}{8}\left(\msda-\msua-\msea\right) \\
I_{M_2} &\equiv M_2^2 + \frac{1}{24}\left(9\left(\msda-\msua\right) + 16\msla-\msea\right) \\
I_{M_3} &\equiv M_3^2 - \frac{3}{16}\left(5\msda+\msua-\msea\right)
\end{align}
These complete the list of independent one-loop RGIs for the MSSM.

\end{document}